\documentclass[a4paper,fleqn,usenatbib]{mnras}

\usepackage{newtxtext,newtxmath}
\usepackage[T1]{fontenc}
\usepackage{ae,aecompl}

\usepackage{graphicx}	% Including figure files
\usepackage{amsmath}	% Advanced maths commands
\usepackage{amssymb}	% Extra maths symbols
\usepackage{paralist}

\usepackage{multirow}
\usepackage{subcaption}
\usepackage{diagbox}
\usepackage{todo}
\usepackage{xcolor}

\newcommand{\BOX}[1]{%
	{\color{red}%
		\fbox{\color{black}#1}}%
}											% highlighter box for tables

\title[Colour-dependent albedo of exoplanets with high-resolution spectroscopy]{Recovering the colour-dependent albedo of exoplanets with high-resolution spectroscopy: from ESPRESSO to the ELT.}

\author[J. H. C. Martins et al.]{
J. H. C. Martins$^{1,2,3}$,\thanks{E-mail: Jorge.Martins@astro.up.pt}
P. Figueira$^{2,1}$,
N. C. Santos$^{1,3}$,
C. Melo$^{2}$,
A. Garcia Mu\~noz$^{4}$,
\newauthor
J. Faria$^{1,3}$,
F. Pepe$^{5}$,
and C. Lovis$^{5}$
\\
$^{1}$Instituto de Astrof\'isica e Ci\^encias do Espa\c{c}o, Universidade do Porto, CAUP, Rua das Estrelas, 4150-762 Porto, Portugal\\
$^{2}$European Southern Observatory, Alonso de C\'ordova 3107, Vitacura, Regi\'on Metropolitana, Chile\\
$^{3}$Departamento de F\'isica e Astronomia, Faculdade de Ci\^encias, Universidade do Porto, Rua do Campo Alegre, 4169-007 Porto, Portugal\\
$^{4}$Zentrum f\"ur Astronomie und Astrophysik, Technische Universit\"at Berlin, D-10623 Berlin, Germany\\
$^{5}$Observatoire de Gen\`eve, Universit\'e de Gen\`eve, 51 ch. des Maillettes, CH-1290 Sauverny, Switzerland
}

\date{Accepted 2018 May 21. Received 2018 May 18; in original form 2017 September 8}

\pubyear{2018}

\begin{document}
\label{firstpage}
\pagerange{\pageref{firstpage}--\pageref{lastpage}}
\maketitle

\begin{abstract}
The characterization of planetary atmospheres is a daunting task, pushing current observing facilities to their limits. The next generation of high-resolution spectrographs mounted on large telescopes -- such as ESPRESSO@VLT and HIRES@ELT -- will allow {us} to probe and characterize exoplanetary atmospheres in greater detail than possible {to this point}. We present a method that permits the recovery of the colour-dependent reflectivity of exoplanets from high-resolution spectroscopic observations. Determining the wavelength-dependent albedo will provide insight into the chemical properties and weather of the exoplanet atmospheres.

For this work, we simulated ESPRESSO@VLT and HIRES@ELT high-resolution observations of known planetary systems with several albedo configurations. We demonstrate how {the cross correlation technique applied to} theses simulated observations can be used to successfully recover the geometric albedo {of exoplanets over a range of wavelengths}. In all cases, we were able to recover the wavelength dependent albedo of the simulated exoplanets and distinguish between several atmospheric models representing different atmospheric configurations.

In brief, we demonstrate that the cross correlation technique allows for the recovery of exoplanetary albedo functions from optical observations with the next generation of high-resolution spectrographs that will be mounted on large telescopes with reasonable exposure times. Its recovery will permit the characterization of exoplanetary atmospheres in terms of composition and dynamics and consolidates the cross correlation technique as a powerful tool for exoplanet characterization.

\end{abstract}


%%%%%%%%%%%%%%%%%%%%%%%%%%%%%%%%%%%%%%%%%%%%%%%%%%

%%%%%%%%%%%%%%%%% BODY OF PAPER %%%%%%%%%%%%%%%%%%

%-------------------------------------------------------------------------------
% Section 1 - Introduction
%-------------------------------------------------------------------------------
%
\section{Introduction}									\label{sec:intro}
The detection of what turned out to be the prototypical hot Jupiter planet around the solar twin 51 Pegasi \citep{Mayor1995} paved the way for one of the most prolific fields in current Astronomy. Currently over 3700 planets in around 2800 planetary systems have been discovered\footnote{From the \url{http://exoplanet.eu/} database \citep{Schneider2011}, as of March 2018.}, with more than 2600 additional candidates from the Kepler \citep{Borucki2010} and K2 \citep{Howell2014} missions still awaiting confirmation. 

Nowadays, one of the main research paths of exoplanetology is the study and characterization of exoplanetary atmospheres. Atmospheric characterization of exoplanets is a formidable task, at the limit of the capabilities of current generation observing facilities. However, in-depth knowledge of exoplanetary atmospheres is fundamental towards our understanding of the planet's chemistry, physics, origin and ultimately its {capacity for} sustaining life \citep[e.g.][]{Seager2010c}.

{A major challenge in for the study of exoplanetary atmospheres is how to detect the minute signal due to the extreme  planet-to-star contrast ratio required for the detection of even a hot Jupiter class planet. At infrared wavelengths, the planet-to-star flux ratio for a hot Jupiter at opposition can reach $10^{-3}$ \citep{Demory2011}, mainly due to {the thermal emission of the planet -- which peaks at these wavelengths --}}{and the lower stellar infrared flux}. At optical wavelengths, the planet-to-star flux ratio {at opposition} {is expected to reach up to $10^{-4}$ \citep[e.g.][]{Charbonneau1999} for giant planets in close-in orbits} {and is usually dominated by the reflected light component}\footnote{Not true for all cases, it has been shown that super close-in hot Jupiters have a non-negligible {thermal} contribution in red bands, e.g. \cite[][]{Demory2011}}. ! the Earth-to-Sun flux ratio is much lower, around $\rm 5\times10^{-10}$ {(computed from Equation \ref{eq:fluxRatioLambda}, assuming $R_{p}=6 371$km, $a=1 AU$ and $A_g=0.3$)}. {These extreme contrast ratios make} the detection of the optical reflected signal from exoplanets a challenging task with current generation facilities. 

{To overcome the low contrast problem, the most diverse -- yet complementary -- techniques have been developed. Phase curve measurements \citep[e.g.][]{Knutson2009, Stevenson2016} rely on the variation of the star+planet flux as the planet orbits the star. Transit \citep[e.g.][]{Charbonneau2002,Sedaghati2015,Sing2016} and occultation spectroscopy \citep[e.g.][]{2005Natur.434..740D,Alonso2009,2014arXiv1410.2241S} techniques measure the wavelength dependent variation in flux of the system as the planet partially occults and is occulted by the star, respectively.}
	
{Since both the planet and its host star orbit the system's centre of mass, their spectra will appear Doppler-shifted relative to each other, typically $\sim$100 km/s for a hot Jupiter. {This separation in the radial velocity space} allowed high-resolution spectroscopy techniques to spectroscopically resolve the planetary signature of the planet from the stellar one \citep[e.g.][]{Brogi2012, Snellen2015a, Martins2015} and to identify {the contribution from} {individual molecules on the atmosphere of exoplanets \citep{2012ApJ...753L..25R,DeKok2013a,Birkby2017}.} {In some cases -- young giant planets {on long} periods -- it has already been possible to spatially resolve the planet from the host star, and to obtain planetary spectra with negligible stellar light contamination \citep[e.g., $\beta$ Pictoris b with CRIRES - ][]{Snellen2014}}. For smaller planets and/or in close-in orbits, new techniques are being developed \citep[e.g., for Proxima b][]{lovis2016}. The combination of high-resolution spectroscopy with polarimetry provides added value to brightness-only measurements in the characterization of the atmospheres and orbits of exoplanets \citep[e.g.][]{Munoz2018}.For a more in-depth review of the available techniques and their results, {the reader is referred} to e.g. \cite{Deming2017}.}

{As done in the past with the planets in our Solar System, the direct detection of reflected light from exoplanets can be used as a tool towards the comprehension of their atmospheres. When the light from a star is reflected on a planet's atmosphere, it will be modulated by the planet's reflectance or albedo which are highly dependent on the atmospheric chemical and physical characteristics. Different gases and condensates will {scatter} {(and absorb)} light at different wavelengths with different intensities \citep[e.g.][]{GarciaMunoz2015}. The detection of absorption features in an atmosphere depends greatly on the presence of clouds and hazes for a variety of planets \citep{Sing2016}.} 
Thus, the recovery of the colour dependence of exoplanet albedos {provides unique insight into the atmospheric properties of the planet}. At optical wavelengths, the reflected spectrum from exoplanets is expected to be a copy -- albeit scaled-down by several orders of magnitude -- of the stellar spectrum modulated by the planetary albedo plus the contribution of absorption bands by the gases in the exoplanet atmosphere.
	
{The first attempts to detect reflected optical light from exoplanets -- such as the pioneering works of \citet{Charbonneau1999} and \citet[][]{Cameron1999} -- were unfruitful and only yielded upper limits for the planetary albedo of some exoplanets -- e.g., $\upsilon$ And b with $A_g < 0.39$ \citep[at a 99.9\% confidence level --][]{2002MNRAS.330..187C} or {HD 75289A b with $A_g < 0.46$ \citep[at a 99.9\% confidence level assuming $R_p=1.2\;R_{Jup}$ --][]{2008A&A...485..859R}.} }

Currently, the number of exoplanets with measured albedos is still low. \citet{Demory2014} presents constrains on the albedo of 27 Super-Earth class Kepler planet candidates. In terms of larger planets, {only about 20 hot Jupiters \citep[see e.g.][]{Angerhausen2015,Schwartz2015a,Esteves2015} have had their albedo {constrained},} which is quite a small sample. The general trend seems to be that hot Jupiters are dark in the optical, with a few exceptions such as Kepler-7b. The disparity of results in terms of measured albedos does not offer an unifying picture on the properties of exoplanet atmospheres, {with values ranging from $0.0253 \pm 0.0072$ for TrES2-b \citep{2011MNRAS.417L..88K-eprintv2} to} $0.35 \pm 0.02$ for Kepler-7b \citep{Demory2013}. Therefore, the statistics of exoplanetary albedos are still a field with much to explore. 

{Using a technique that cross-correlates high-resolution (R \textgreater 100 000) spectral observations of planetary systems with a numerical spectral mask, \citet[][]{Martins2015} presented the first direct detection of the reflected spectrum of 51 Pegasi on its orbiting planet. {Simply put, the cross-correlation technique sums up all spectral features that are present in both the spectrum and the numerical mask, effectively merging them into one single average line. This process enhances the signal-to-noise ratio of the spectra, thereby making the planetary reflected spectrum to stand above the spectral noise.} With this detection, they inferred that this planet is most likely a highly-inflated hot-Jupiter type planet with a high albedo ($A_g= 0.5$ for $R_p = 1.9 R_{Jup}$).}

{{The work described herein} focuses on the direct detection of the high-resolution optical spectrum from the host star reflected on an orbiting planet with the next generation of observing facilities, using {the cross-correlation technique presented} in \citet{Martins2015}}. Next generation observing facilities (e.g ESPRESSO@VLT, HIRES@ELT) will permit to increase the expected number of photon counts at the detector, and consequently move our observations into a higher S/N domain. This increase is accomplished both from an increased sensitivity of the whole optical systems (e.g., from $\sim$4 per cent for HARPS\footnote{From HARPS Exposure Time Calculator (ETC) version P100.2.} to $\sim$10 per cent for ESPRESSO\footnote{Lovis, Pepe and Sosnowska, private communication.}) and increased collecting areas (from a 3.6-m telescope for HARPS@ESO3.6 to up to 4 $\times$ 8.4-m with ESPRESSO@VLT and to a 39-m telescope for HIRES@ELT).

With these observing facilities in mind, we expanded the cross-correlation function technique presented in \cite{Martins2015} to allow it to recover not only the reflected light signal from the planet, but also its colour dependency. This should make it a powerful tool for exoplanet characterization as the colour dependency of a planetary reflected signal is highly modulated by the gases and condensates in its atmosphere {\cite[e.g.][]{GarciaMunoz2015}}. In this work, we apply the proposed technique to simulated star+planet observations with both ESPRESSO@VLT and HIRES@ELT. Our main goals are to estimate to which level we can: $a)$ recover the colour dependency of the reflected light spectrum from selected exoplanets and $b)$ distinguish between possible albedo function models.

{In Section \ref{sec:method}, we describe the cross-correlation technique in detail. How the simulated observations were created is the object of Section \ref{sec:simulations}. The results from our simulations are presented in Section \ref{sec:results} and discussed in Section \ref{sec:discussion}. We {summarize our conclusions} in Section \ref{sec:conclusions}.}

%-------------------------------------------------------------------------------
% Section 2 - The method
%-------------------------------------------------------------------------------
%
\section{The cross correlation Function Technique}			\label{sec:method}

In this work, we propose and test on simulated observations (see Section \ref{sec:simulations} for details on the creation of the simulated observations) a {modified version of the cross correlation Function Technique} used in \cite{Martins2013,Martins2015} to recover the reflected spectrum from exoplanets and infer their albedo functions. 

The {CCF} of high-resolution spectra with binary masks has been used very successfully for many years to determine with high-precision the radial velocity of astronomical objects \citep[e.g.][]{Zinn1984}. In particular for exoplanets, astronomers cross-correlate high-resolution spectra of stellar systems with numerical masks over a range around the host's radial velocity \citep[e.g.][]{Baranne1996,Mayor2003}.

{These numerical masks are fundamentally lists of spectral lines identified on the particular spectral type of the host. Each spectral line on the mask is represented by a theoretical initial and final wavelengths, as well as the weight for each line \citep[e.g.][]{Pepe2002}. These parameters are computed assuming that the line can be represented by a rectangle with the same area as the real spectral line, with the continuum normalized to one and the weight corresponding to the relative depth of the line, centred at the rest wavelength of the spectral line. {A detailed description of the CCF is beyond the scope of this work and as such we refer the reader to \citet[][]{Baranne1996} and \cite{Pepe2000} for the detailed mathematical formulation.}}

{By cross-correlating high-resolution spectra obtained with instruments such as HARPS \citep{Pepe2000} with these numerical masks, researchers are effectively constructing an average spectral line of the spectrum. This average line will be centred at the radial velocity of the object being observed and will benefit of a signal-to noise ratio (S/N) increase given approximately by }
\begin{equation}
	{S/N}_{ccf} \approx N_{lines} \times {S/N}_{sp}
\end{equation}
where ${S/N}_{ccf}$ and $ {S/N}_{sp}$ are respectively the signal-to-noise ratio (S/N) of the resulting CCF and of the original spectrum, and $N_{lines}$ is the number of spectral lines used in the cross-correlation. Since a typical spectral line mask for a solar type star has several thousands of spectral lines (e.g. the HARPS G2 mask has around 4000 lines), the CCF provides an increase in S/N of one to two orders of magnitude (for example, in the case of the standard HARPS G2 mask, this yields an increase in S/N over of 60 times). {Also, moving the observed spectrum into a much higher S/N domain allows to greatly increase the precision at which the radial velocity of the object is measured.} 

When searching for the reflected signal of a planet, we expect that the planetary signal will, to a large extend, mimic the stellar {spectrum}, but at a different radial velocity and scaled-down by several orders of magnitude. {The exoplanet will also contribute through absorption by the gases in its atmosphere.} For a given orbital phase $\phi$ and wavelength $\lambda$, the fraction of stellar light reaching the planetary disk that will be reflected back to us is given by 
\begin{equation}
\frac{F_{\rm{p}}\left(\lambda, \phi\right)}{F_{\rm{*}}\left(\lambda\right)} = \; A_{g}\left(\lambda\right) \;g(\alpha)\left(\frac{R_{\rm{p}}}{a}\right)^2
\label{eq:fluxRatioLambda}
\end{equation}
where $\rm A_g $ is the geometric albedo, $R_{\rm{p}}$ the planetary radius, $g(\alpha)$ the {phase function} and $a$ the semi-major axis of the planetary orbit.{ By cross-correlating the star+planet spectrum with a numerical mask representing the star will produce a CCF with two peaks with different amplitudes at different radial velocities, one representing the star and the other -- with a much lower amplitude -- representing the planet. In brief, the spectrum will resemble the one of a spectroscopic binary, with one component several orders of magnitude fainter. Inverting Eq. \ref{eq:fluxRatioLambda}, we can use the ratio between both signals to recover the planetary albedo function {(times the square of the planet radius if the planet size is not known)}. Note that albeit powerful, this technique is limited in terms of planetary characterization by the photon noise from the star. }

\subsection{Recovery of the planetary albedo} 

{To recover the albedo from our simulated observations, we implemented a modified version of the cross correlation recovery technique from \cite{Martins2015} as a Python tool which can be downloaded at \url{https://github.com/jorgehumberto/albedo_recovery_tool.git}. We will now proceed by summarizing each step of the albedo recovery process.}

\subsection*{Computation of the CCFs}
{Since HARPS data was used as the spectral template and standard type of our simulations, for simplicity we used the \textit{HARPS {Data Reduction Software (DRS)} vs 3.6.} to compute the CCF for each spectral order of each simulated observation. {This recipe cross-correlates high-resolution spectra with a numerical mask that contains the spectral information for thousands of spectral lines that where identified for each stellar spectral type. Since each spectral order will have different S/N levels due to the wavelength dependency of the instrumental setup transmission, the recipe uses an optimally weighted cross-correlation function which accounts for the individual S/N of each individual spectral line\footnote{similarly to what was done before with the CORALIE spectrograph \cite{Pepe2002}.}. The parameters for each spectral line of the mask {(initial and final wavelength, and line depth)} are computed assuming that the line can be represented by a rectangle with the same area as the real spectral line, with the continuum normalized to one and the weight corresponding the relative depth of the line, centred at the theoretical wavelength line if the radial velocity was equal to zero. We refer the reader to \citet{Baranne1996} for a detailed description of the computation of the CCF and to \citet[][]{Pepe2002} for details on the spectral mask.}

	\subsection*{Wavelength binning of the CCFs}
	{We intend to determine the exoplanet reflectivity over a number of bins spanning the entire spectral range available to the spectrograph.} {Thus, the wavelength} coverage of the spectrograph is broken into $\rm N_{bins}$ wavelength bins {of equal size}, where $\rm N_{bins}$ is the desired sampling for the recovered albedo function. 
	Then, the CCFs of the orders covered by the wavelength range of each bin are merged together, effectively re-sampling them for the wavelengths defined by each {bin} {as the wavelength region of each individual bin can span several spectral orders}. This can be done as it is mathematically equivalent to compute the CCF for a wavelength bin or merge the CCFs of the orders covered by the same bin\footnote{This can be demonstrated from Eq. 8 of \cite[e.g.][]{Baranne1996} by splitting the total number of orders into $\rm N_{bins}$ wavelength bins.}. Note that in most orders, there is an overlap in wavelength between adjacent orders (see Figure \ref{fig:harpsOrders}). In those cases, by binning the orders, we are effectively increasing the photon information in the overlapping ranges as some spectral lines will contribute twice to the binned CCFs.
	
	\begin{figure}
		\includegraphics[width = \hsize]{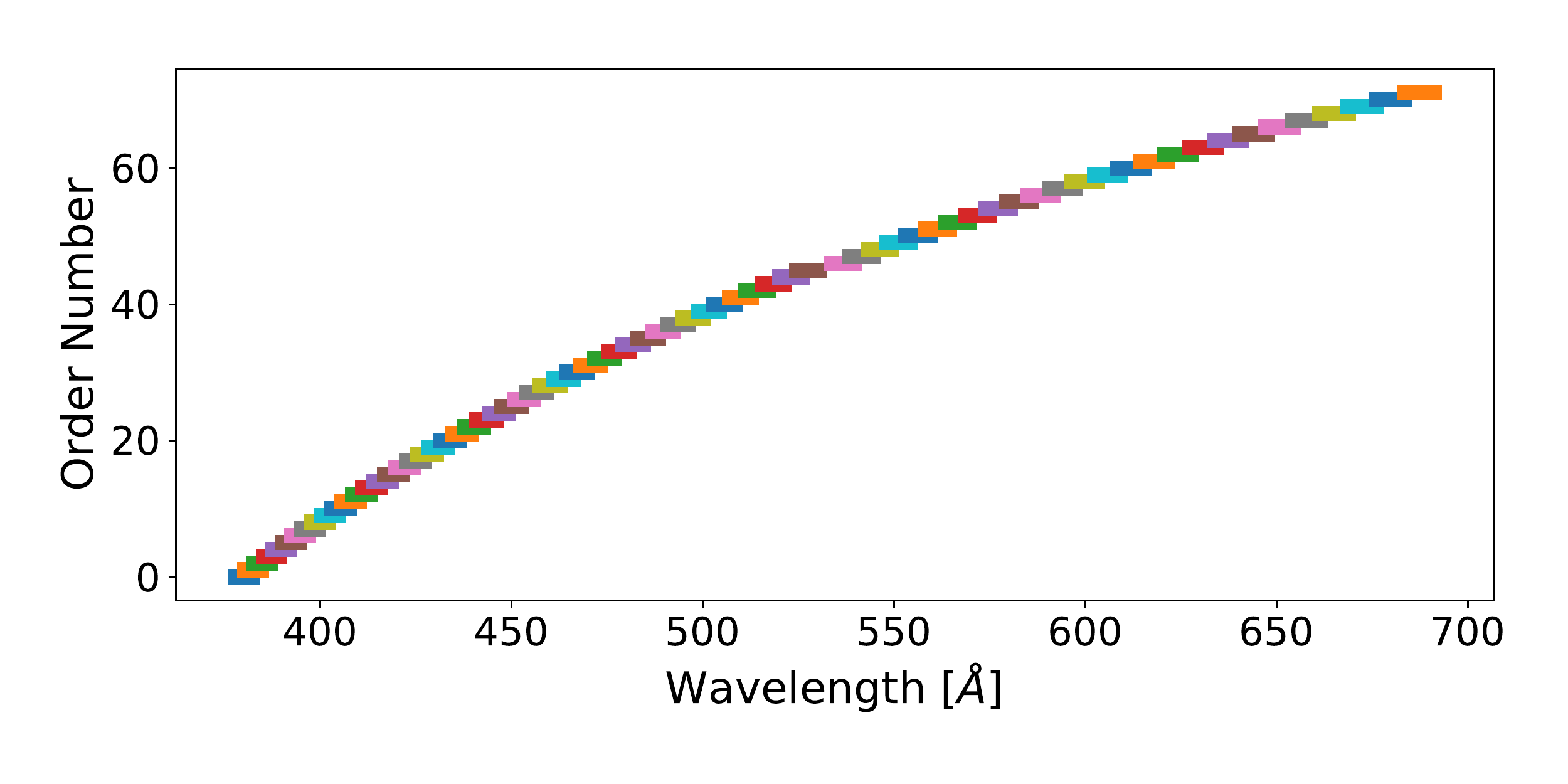}
		\caption{Comparative wavelength coverage of all HARPS spectral orders. Each colour represents an order.}
		\label{fig:harpsOrders}
	\end{figure}

	{Note that although arbitrary and highly dependent on the target planetary system properties and observing facility, the value of  $\rm N_{bins}$ needs to balance the required exposure time and level of precision required for the recovered albedo function. Selecting a higher value for $\rm N_{bins}$ will increase the level of detail of the recovered albedo function, thereby enabling a more detailed characterization of the planetary atmosphere. However, the required exposure time for a targeted accuracy varies linearly with $\rm N_{bins}$, and as $\rm N_{bins}$ increases, the required exposure time increase as well (see Fig. \ref{fig:nBins_vs_time}) and might become prohibitive for too high a value of $\rm N_{bins}$. More details on the selection of the value of  $\rm N_{bins}$ can be found in Section \ref{sec:targets}.}

	\subsection*{Removal of the stellar signal}	\label{sec:removal}
	{For each wavelength bin, the stellar CCF template is constructed by stacking together the CCFs corresponding to each wavelength bin on all individual observations}, after correction of the stellar radial velocity drift {that results from the reflex motion of the star (i.e., the CCFs of all individual observations are re-centered to the rest frame of the star)}. This dilutes the planetary signal amidst the stellar noise, while increasing the strength of the stellar signal. {Once the stellar template has been constructed, we divide the CCF of each individual observation by it, effectively removing the stellar signature. Figure \ref{fig:ccfExample} shows an example of a: i) the CCF for one of the simulated observations (\textit{Top Panel}) ii) the stellar template for one of the simulated observing runs (\textit{Middle Panel}) and iii) the same CCF presented in the Top Panel, but after removal of the stellar signal (\textit{Bottom Panel}).} 	Note that as a rule of thumb, the S/N of the stellar template should be at least over 10 times that of the individual CCFs to avoid the introduction of non-negligible noise onto the planetary CCF \citep[see][Appendix A]{Martins2013} during the removal process. As such, it should be built from over 100 individual CCFs.
	
	\begin{figure}
		\includegraphics[width=\linewidth]{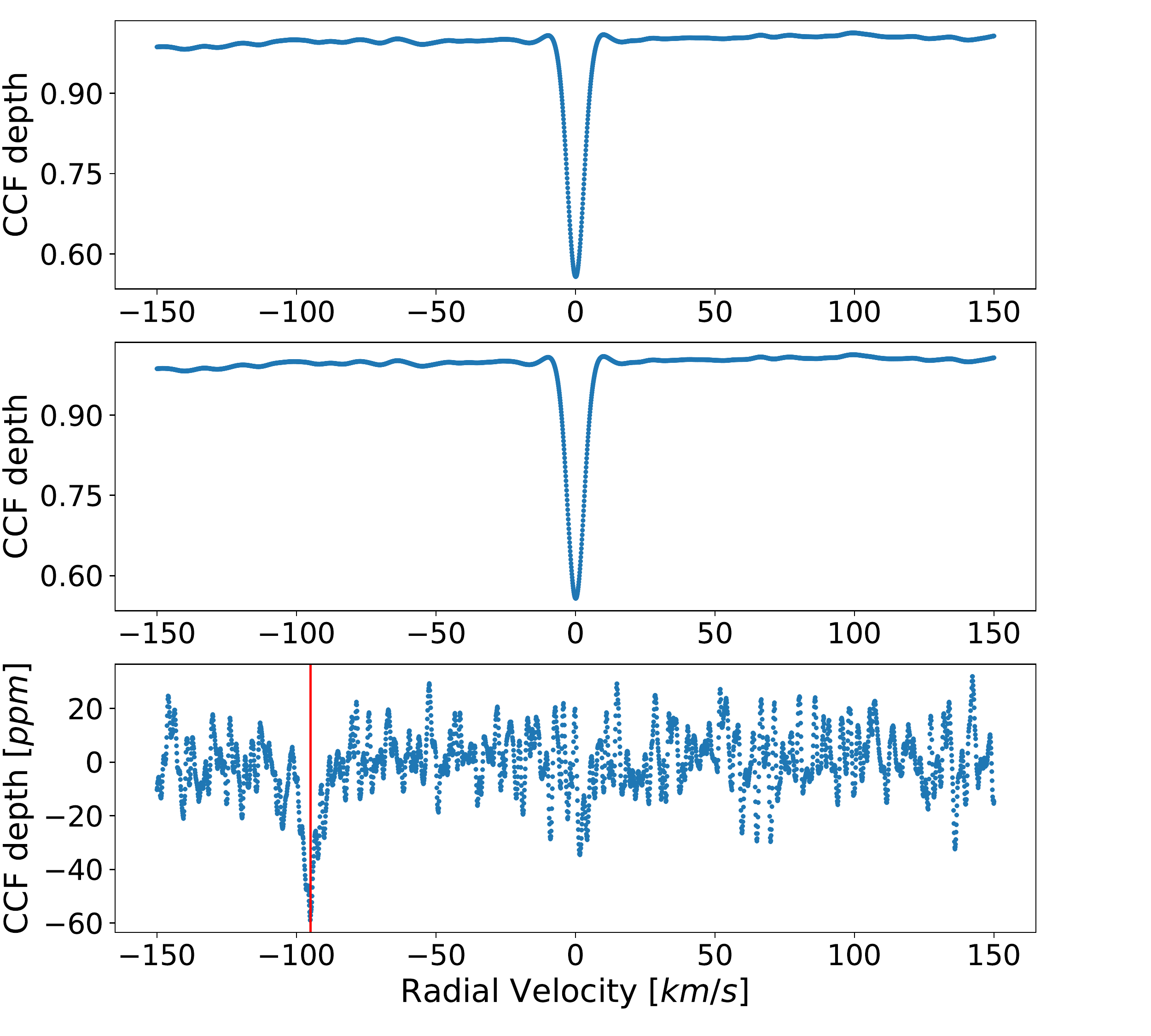}
		\caption{{\textit{Top Panel:} Example of the CCF for one of the simulated observations of the HD209458 system with ESPRESSO.\\
			\textit{Middle Panel:} Example of the CCF template for one of the simulated observations of the HD209458 system with ESPRESSO, constructed by adding 100 different CCFs centred on the stellar rest frame.\\
			\textit{Bottom Panel:} Example of the CCF for one of the simulated observations of the HD209458 system with ESPRESSO, after removal of the stellar signal. The red line corresponds to the expected radial velocity of the planet.\\
		{Note that although both the \textit{top} and \textit{middle} panels look identical, this is just a scale effect and there are minute differences between them, as can be seen on the \textit{bottom} panel.}}}
		\label{fig:ccfExample}
	\end{figure}
	
	We further increase the S/N of the planetary signal  by stacking together all planetary CCFs, after correction of the planet radial velocity for each wavelength bin and observation. 
	
	{For each observation, the radial velocity of the star ($RV_*$) and planet ($RV_P$) can be computed from the Radial Velocity Equation} 
	\begin{equation}
			RV_{*} = RV_{0} - k_{*} \left(cos(\omega+f) + e \,cos(\omega)\right)
			\label{eq:RVEquationStar}
	\end{equation}
	for the star and 
	\begin{equation}
		RV_{P} = RV_{0} + k_{P} \left(cos(\omega+f) + e \,cos(\omega)\right)
		\label{eq:RVEquationPlanet}
	\end{equation}
	{for the planet. $RV_{0}$ corresponds to the radial velocity of the system's barycentre relatively to the Sun, $f$ is the true anomaly\footnote{The angle between the periastron direction and the star or planet position on its orbit. For circular orbits the $\omega=0$ and $f$ is replaced by the orbital phase $\phi$, defined from the point of maximum radial velocity.} and $\omega$ is the argument of periastron from the orbit and $k_{*}$ and $k_{*}$ are the radial velocity semi-amplitude of respectively the stellar and planetary orbits \citep[see][]{Paddock1913,Lovis2010}.}
	
	{Both the stellar and planetary semi-amplitudes can be computed from the orbital parameters through}
	\begin{equation}
		k_{*} = \left(\frac{2 \,\pi \,G}{P}\right)^{{1}/{3}} \frac{m_{p} \,sin(i)}{m_{*}^{{2}/{3}}} \frac{1}{\sqrt{1-e^{2}}}
		\label{eq:KStar}
	\end{equation}
	and
	\begin{equation}
			k_{P} = \frac{K_*}{q} 
		\label{eq:KPlanet}
	\end{equation}
	{where $G$ is the universal gravitational constant, $P$ is the orbital period, $m_{*}$ and $m_{p}$ are respectively the stellar and planetary masses, $i$ is the orbital inclination relatively to the sk plane, $e$ eccentricity of the orbit, and $q=m_P/m_*$ is the planet-to-star mass ratio..}
	
	{When dealing with real observations, $K_{*}$ can be computed from the observed radial velocity of the star}
	\begin{equation}
		k_{*} = \frac{max\left(RV_{*}\right) - min\left(RV_{*}\right)}{2} 
		\label{eq:KStar}
	\end{equation}
	{However, the radial velocity method only allows to recover the minimum mass of the planet ($m_P\;\sin\left(i\right)$) and for most non-transiting planets, the orbital inclination and planetary mass are degenerate. As such, the precise radial velocity of the planet is unknown. In those cases, for each observation, the radial velocity for the planet is computed over a range of possible radial velocity semi-amplitudes. The best fit model will permit to infer the correct $K_P$ and break the mass/inclination degeneracy for the planet by solving the system defined by Equations \ref{eq:KStar} and \ref{eq:KPlanet}. For more details on this analysis, we refer the reader to \cite{Martins2015}.}	
	
	\subsection*{Recovery of stellar signal reflected on the planet signal and the planetary albedo function.}	
	
	{After removal of the stellar contribution to the CCF, for each wavelength bin we are left with the planetary CCF plus noise. Figure \ref{fig:planetCCFs} shows the recovered planetary CCFs from the simulated observations of HD 209458 b for each wavelength bin (\textit{Left Panels}), as well as the recovered albedo for each bin (\textit{Right Panel}).}
	
	\begin{figure}
		\includegraphics[width=\hsize]{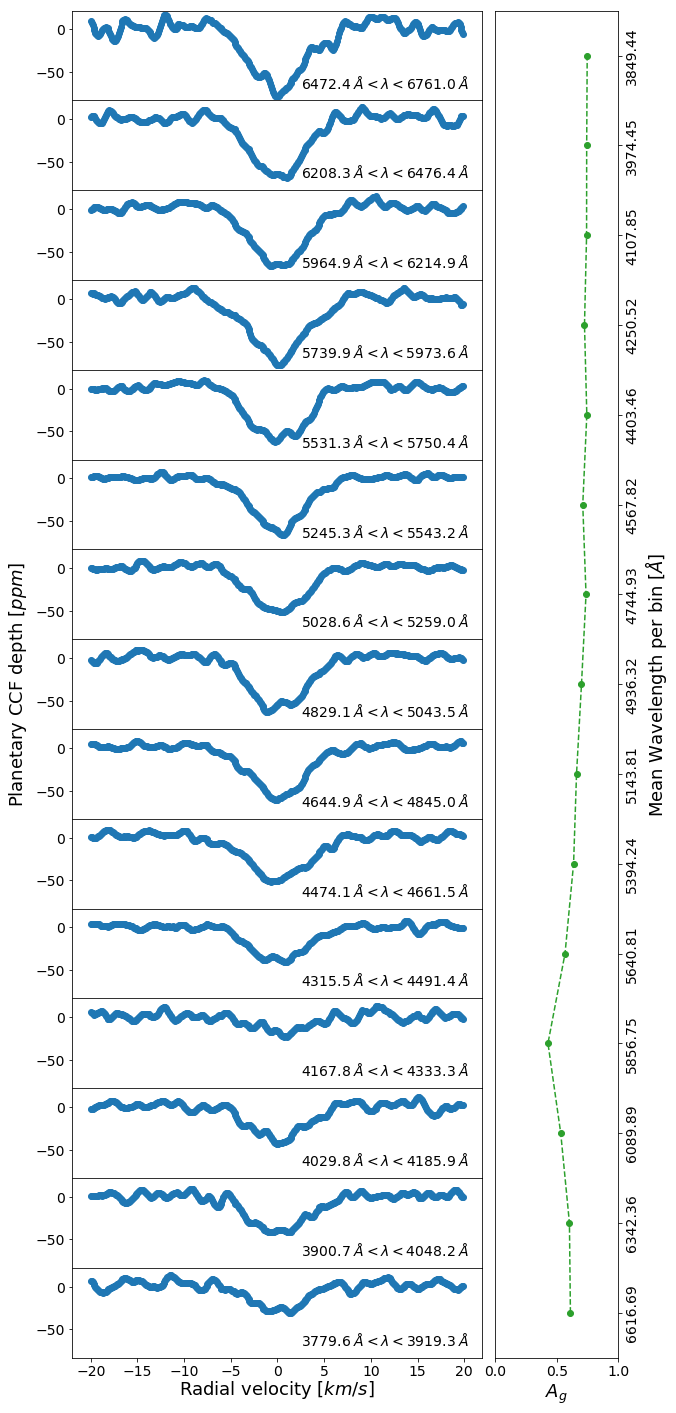}
		\caption{\textit{[Left Panels:]} Planetary CCFs recovered over 15 wavelength bins from simulated ESPRESSO observations of the HD 209458 system. {$RV=0$ corresponds to the radial velocity of the planet, and thus, the expected centre of the planetary CCF}. \textit{[Right Panel]}: Recovered albedo for each wavelength bin. Note that the albedo can be recovered as the radius is known. {For cases where the radius is unknown, only the value of $Ag \times R_p^2$ is recovered}.}
		\label{fig:planetCCFs}
	\end{figure}
	
	{To infer the albedo for a given wavelength bin, we fitted Gaussian functions to both the resulting planetary CCF and {the stellar template CCF}. The ratio of the area of the Gaussian fit to the planet CCF by the Gaussian fit to the stellar template CCF is the planet-to-star flux ratio per bin $\left(\frac{F_{\rm{p}}}{F_{\rm{*}}}\right)_{bin}$. Note that we are not including line broadening effects planetary rotation or the fact that the planet sees the star with a different $\rm v \sin(i)$ due to orbital motion. However, these effects maintain the area of the spectral lines and as such this approach is still valid. Using Equation \ref{eq:fluxRatioLambda}, the planetary geometric albedo for each bin can be inferred from}
	\begin{equation}
	A_{g, bin} = \frac{1}{<g\left(\alpha\right)>} \times \left(\frac{a}{R_{\rm{p}}}\right)^2 \times \left(\frac{F_{\rm{p}}}{F_{\rm{*}}}\right)_{bin}
	\end{equation}
	{Where $<g\left(\alpha\right)>$ is the mean phase function over all observations stacked to recover the planetary albedo %(See Section \ref{sec:removal})
		, $a$ is the planetary orbit semi-major axis and $R_{\rm{p}}$ is the planetary radius. For the case depicted in Figure \ref{fig:planetCCFs}, we generated observations with random orbital phases between 15 and 45 degrees on both sides of superior conjunction {(i.e., orbital phases with $0.37 \lessapprox \phi \lessapprox 0.46$ or $0.54 \lessapprox \phi \lessapprox 0.63$, assuming that the inferior conjunction/transit phase corresponds to $\phi=0$ and one full orbit corresponds to $\Delta \phi = 1$)}. Assuming a Lambert phase function for an edge-on orbit (Equation \ref{eq:Phasefunction}), this yields an average $<g\left(\alpha\right)> \approx 0.87$.} {We refer the reader to Appendix \ref{app:optimalWindows} for a detailed description regarding this choice of orbital phases.}
	
	{This process is repeated for all wavelength bins to recover the wavelength dependent albedo function of the planet, shown in the \textit{Right Panel} of Figure \ref{fig:planetCCFs}.} We will now proceed {to describe} the creation of the simulated observations that will be used for this work.

%-------------------------------------------------------------------------------
% Section 2 - The simulations
%-------------------------------------------------------------------------------
%
\section{The simulations}			\label{sec:simulations}

\subsection{The instruments}	\label{sec:instruments}

\paragraph*{{ESPRESSO}} \citep[Echelle Spectrograph for Rocky Exoplanets and Stable Spectroscopic Observations, ][]{Pepe2010} is a new high-resolution spectrograph currently being commissioned for the VLT at ESO's LaSilla-Paranal Observatory.

In terms of resolution, ESPRESSO will have 3 available modes: 
\begin{inparaenum}[i)]
	\item Multi UT Mode (MR) with a resolving power of 60 000;
	\item High-Resolution (HR) with a resolving power of 130 000; and
	\item Ultra High-Resolution (UHR) with a resolving power of 200 000.
\end{inparaenum}
In MR mode, ESPRESSO will be able to combine incoherently the light from up to all 4 VLT Unit Telescopes (UT), effectively increasing the collecting area up to the equivalent of a 16-m telescope. In both HR and UHR modes, it will only be able to receive the light {from one of the four UT telescopes.}. {For both MR and HR modes, ESPRESSO will be fed by a {1 arc sec} fibre, while on UHR mode, the fibre diameter drops to {0.5 arc sec}. While allowing for a large increase in the resolution for the UHR mode, it comes at the cost of a lower photon collecting power}. 

Figure \ref{fig:espressoEff} shows the efficiency curve of all 3 resolution modes of ESPRESSO (Pepe, Lovis and Sosnowska - private communication), as well as the efficiency curve of HARPS (from HARPS ETC). Both HARPS and ESPRESSO (all 3 modes) efficiency curves were computed for a seeing of 0.65 arc sec and air mass of 1.0 (i.e. pointing towards the zenith). For all of them, the efficiency includes the transmission {factors of the atmosphere, telescope, and spectrograph, as well as slit losses}.
\begin{figure}
	\includegraphics[width = \hsize]{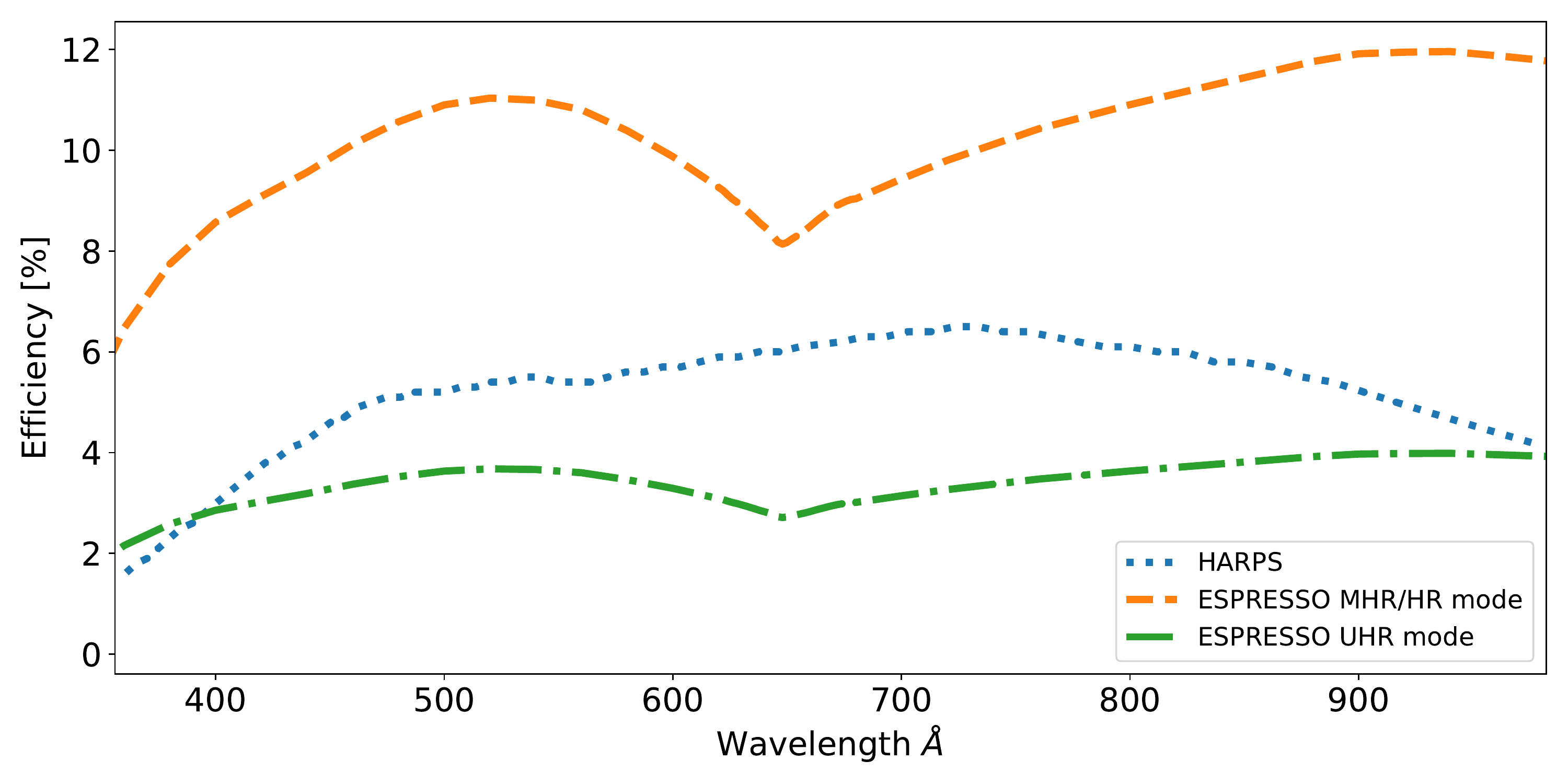}
	\caption{Total efficiency of ESPRESSO (in all resolution modes) and HARPS/HIRES as a function of wavelength, {which includes the efficiency of the atmosphere, telescope, optical components and detector.}}
	\label{fig:espressoEff}
\end{figure}

\paragraph*{{HIRES}} \citep[][]{Maiolino2013}
	is a high-resolution spectrograph proposed for ESO's ELT that will operate simultaneously at visible and infrared wavelengths. Recently, the HIRES@ELT consortium has successfully published the phase A study \citep[][]{Maiolino2013}, where a technical solution and a clear science case is proposed. The detection of exoplanet atmospheres is considered the top science driver of the instrument. Since the instrument is still in the planning phase and the baseline design was not finalised at the time this paper was conceived, we considered for our simulations a high-resolution spectrograph for the ELT which is a scaled up version of HARPS. Therefore, we assumed a HARPS-like profile with a resolving power of around 110 000 and a HARPS-like efficiency.

\paragraph*{{HARPS}} \citep[][]{Pepe2000,Mayor2003}
{is a fibre-fed and cross-dispersed high-resolution echelle spectrograph installed on ESO's 3.6-m telescope at the LaSilla-Paranal observatory. The main scientific goal that led to its development was the detection of exoplanets from extremely precise radial velocity measurements. The combination of its resolving power ($R \approx 115 000)$ with precise wavelength calibrations allows it to achieve a long term radial velocity precision better than 80 cm s\textsuperscript{-1} \citep{LoCurto2012}. For more details, we refer you to HARPS User Manual\footnote{\url{https://www.eso.org/sci/facilities/lasilla/instruments/harps/doc.html}} Issue 2.1, October 1st 2011.}

{Throughout this work we used HARPS as the spectral template for both ESPRESSO and HIRES as these instruments are still not available. {First of all, we restricted the spectral coverage for both ESPRESSO and HIRES to the same wavelength range as HARPS (378--691 nm). The reason behind this choice is that this is also the spectral range covered by the spectral masks used by the HARPS cross correlation recipe, the same we apply in this work. Note that ESPRESSO will have a larger spectral coverage than HARPS, and as such we should be able to cover a broader wavelength region of the planetary atmospheres. A broader wavelength coverage in principle also implies a larger amount of spectral lines to construct the CCF and consequently a larger increase in S/N.}}

Additionally, the collected flux and S/N for both instruments was extrapolated from the values estimated with HARPS ETC\footnote{\url{https://www.eso.org/observing/etc/}} version 100.2. Since we are looking at bright targets, to avoid saturation and ensure linearity of the instrument, we set for all targets the individual exposure time in the ETC to 300s. E.g., for our brightest star -- 51 Pegasi with $\rm mag_V = 5.49$ -- in 300s we only reach about 65 per cent of the detector's saturation limit per pixel, a regime where the detector is still linear. The stellar parameters for each target were obtained from the exoplanet.eu database. The target independent parameters are presented in Table \ref{tab:ETC}. {Note that for all observations we assumed the Moon to be far from our simulated targets. In this case, the sky brightness has a visual magnitude in the range $20.0 < mag_V\; \/ arc sec^2 < 21.8$ \citep[e.g. see][]{Cunha2013}. Therefore the Moon phase can be ignored for our simulations and was left at its default value in the ETC.}

\begin{table}
	\caption{HARPS ETC settings. The settings not in this table were left to their default values.}
	\begin{tabular}{l l l}
		\hline\hline\\[-1em]
		\bf Target Input Flux Distribution 	&							\\
		Template spectrum					&	G2V (Kurucz)			\\			
		\bf Sky Conditions					&							\\
		Moon phase							&	3 days 	\\
		Airmass								&	1.2						\\
		Seeing								&	0.8						\\[.5em]
		\bf Instrument Setup				&							\\
		Exposure Time						&	300 s					\\				
		\hline
	\end{tabular}
	\label{tab:ETC}
\end{table}

To extrapolate the S/N that can be obtained with either ESPRESSO or HIRES we used
\begin{equation}
SN_{\rm{ins}}\left(\lambda, t\right) = \sqrt{A \times E\left(\lambda\right)} \; SN_{HARPS}\left(\lambda, t\right)
\label{eq:efficienciesSN}
\end{equation}
where $SN_{\rm{ins}}\left(\lambda\right)$ and $SN_{HARPS}\left(\lambda\right)$ are respectively the expected wavelength dependent S/N for the simulated instrument and for HARPS. $A$ is the ratio of the collecting areas for the selected instrument and {HARPS 3.6-m telescope} ($A \approx 5.4$ for ESPRESSO@VLT,assuming observations done with one single UT, and $A \approx 117.4$ for HIRES@ELT). $E\left(\lambda\right)$ is a scaling factor to account for the different wavelength dependent efficiencies between the ESPRESSO/HIRES and HARPS. For ESPRESSO, this scaling factor is defined as
\begin{equation}
E\left(\lambda\right) = \frac{Eff_{mode}\left(\lambda\right)}{Eff_{HARPS}\left(\lambda\right)}
\label{eq:efficiency}
\end{equation}
where ${Eff_{mode}\left(\lambda\right)}$ and ${Eff_{HARPS}\left(\lambda\right)}$ are respectively the efficiency of the selected ESPRESSO mode (see Figure \ref{fig:espressoEff}) and of the HARPS spectrograph. {Note that in this case, the efficiency curve of ESPRESSO was computed for a seeing of 0.65 arcsec. As such, ${Eff_{HARPS}\left(\lambda\right)}$ was computed for the same seeing of 0.65 arcsec, independently of $SN_{HARPS}\left(\lambda, t\right)$.} As discussed above, for the case of HIRES we assumed a HARPS-like efficiency and thus $E=1$.

Similarly, the expected stellar flux {at the spectrograph} $F_{\rm{*}}\left(\lambda\right)$ can be extrapolated for both ESPRESSO and HIRES with
\begin{equation}
F_{\rm{*}}\left(\lambda\right) = A \times E\left(\lambda\right) \times F_{HARPS}\left(\lambda\right)
\label{eq:efficienciesFlux}
\end{equation}
where $F_{HARPS}\left(\lambda\right)$ corresponds to the collected flux from the star collected by HARPS\footnote{Column \emph{Obj} for the central column for each order of the results estimated with HARPS ETC.}.

\subsection{Target selection}	\label{sec:targets}
{As {test targets we selected} known exoplanets representative of three different planet categories:
	\begin{inparaenum}[i)]
		\item hot Jupiters, 
		\item hot Neptunes and
		\item {short-period} super-Earths.
\end{inparaenum}}
Our initial sample included all planets from the exoplanet.eu database \citep{Schneider2011} that 
	\begin{inparaenum}[i)]
		\item orbit FGK dwarfs with $\rm mag_V \le 10$, 
		\item have orbital periods shorter than 14 days and
		\item have measured radii.
\end{inparaenum}
{This choice ensures that
	\begin{inparaenum}[i)]
	\item high-S/N data can be obtained within reasonable exposure times\footnote{e.g., for a $mag_V = 8$ G2 star, the average S/N for 1-h of exposure time will be $\sim$120 as per HARPS ETC.}, 
	\item the planet-to-star flux ratio will not be too low for detection and
	\item we can break the $A_g\;R_p^2$ degeneracy coming from Eq. \ref{eq:fluxRatioLambda} and recover the albedo directly.
\end{inparaenum}}

For all the planets in our initial sample, we estimated the required exposure time (see Section \ref{sec:expTime} for a detailed description on how this was computed) to recover the albedo function from the planet assuming a grey albedo -- i.e. wavelength independent -- with $\rm A_g=0.2$. {This value is somewhat arbitrary and was selected as it {corresponds to} about 70 per cent of the lowest mean albedo for the albedo functions we simulated (see Section \ref*{sec:atmospheres}) over HARPS wavelength coverage.} {Furthermore, to claim a detection, the albedo function should be recovered for $\rm N_{bins}$ wavelength bins over the whole spectral range of the selected instrument with a $3\sigma$ {confidence} per bin. Note that the number of wavelength bins selected for each planet type needs to balance the required exposure time and level of precision on the recovered albedo function.} Figure \ref{fig:nBins_vs_time} shows on the top panel the estimated required times to recover the albedo function of {a prototypical} hot Jupiter ($\rm p = 3 days$; $\rm R_{\rm{p}} = 1.2 \; R_{\rm{Jup}}$; grey albedo with $\rm A_g=0.2$) with ESPRESSO (HR mode), as a function of $\rm N_{bins}$ and different visual magnitudes of the host star. The bottom panel shows the same, but from HIRES observations.
	\begin{figure}
		\includegraphics[width = \hsize]{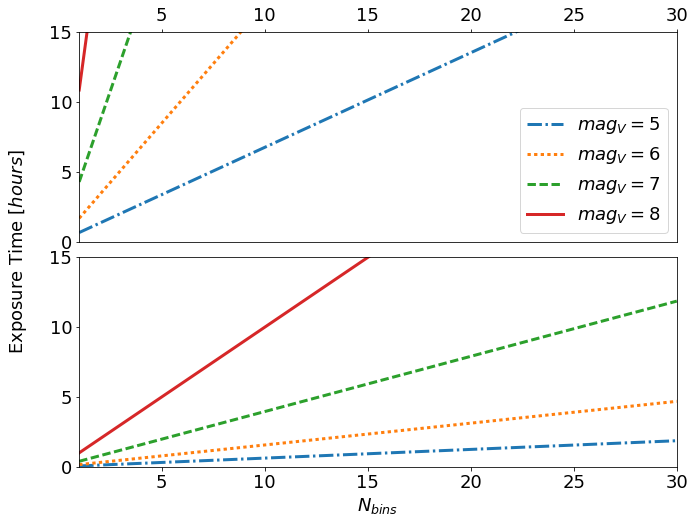}
		\caption{\textit{Top Panel:} Estimated exposure times for the recovery of the grey albedo function ($\rm A_g=0.2$) for a 3-day period hot Jupiter with a radius of 1.2 $\rm R_{\rm{Jup}}$ from {ESPRESSO (HR mode) observations} as a function of $\rm N_{bins}$ and of the host's magnitude. \textit{Bottom Panel:} Same as top panel, but from HIRES observations. {In both cases, these exposure times correspond to a 3$\sigma$ detection of the planetary signal.}}
		\label{fig:nBins_vs_time}
	\end{figure}
Note that $\rm N_{bins}=1$ corresponds to the work presented in \citet{Martins2015}.
	A high number for $\rm N_{bins}$ and the exposure time cost will be prohibitively high (e.g., for $\rm N_{bins}=25$ the required S/N {to recover the reflected light spectrum }is close to 800 000. For a $\rm mag_V=8$ star this requires over 100 hours {of ESPRESSO time}. Too low a value for $\rm N_{bins}$ and the level of detail of the recovered albedo function will be too low to distinguish between albedo models as no spectral features can be recovered (e.g., the sodium doublet at 589-nm).
As such, for simulated observations of hot Jupiters with ESPRESSO we settled on 15 wavelength bins for the recovered albedo function. For simulated observations with HIRES we selected
	\begin{inparaenum}[i)]
		\item 70 wavelength bins for hot Jupiters;
		\item 6 wavelength bins for hot Neptunes and 
		\item 5 wavelength bins for Super-Earths.
	\end{inparaenum}
	We believe these values to yield a good balance between exposure time and the level of detail in the recovered albedo function. Note that the values we propose for $\rm N_{bins}$ were arbitrarily selected for this exploratory work. For real observations we would attempt different values for $\rm N_{bins}$ as a trade-off between the level of detail of the recovered albedo function and the associated uncertainty. Once the signal of an exoplanet is detected at $3\sigma$, we passed the detection threshold, and the significance for a detection on a different $\rm N_{bins}$ value or integration time can be calculated by scaling from the presented values.

Of the resulting planet sample we selected a few targets which we consider representative of the different exoplanet types: 
	\begin{inparaenum}[i)]
		\item two hot Jupiters -- 51 Pegasi b, HD 209458 b -- to be observed with both instruments\footnote{Selecting the same targets for instruments allows to compare the effect of an increased telescope size on sampling of the recovered albedo function.};
		\item two hot Neptunes -- HD 109749 b, HD 76700 b -- to be observed with HIRES and 
		\item one Super-Earth -- 55 Cnc e -- to be observed with HIRES.
	\end{inparaenum} Table \ref{tab:paramsPlanets} shows the selected sample of target planetary systems and their relevant parameters, {where:}
	 \begin{inparaenum}[i)]
	 	\item {$\rm M_{\rm{p}}$ and $\rm M_{\rm{*}}$ are respectively the planet and stellar mass,}
	 	\item {$\rm R_{\rm{p}}$ and $\rm R_{\rm{*}}$ are respectively the planet and stellar radii,}
	 	\item {$P$ is the orbital period,}
	 	\item {$a$ is the semi-major axis of the planetary orbit,}
	 	\item {$k_p$ and $k_\star$ are respectively the radial velocity semi-amplitude of the planetary and stellar orbits,}
	 	\item {$i$ is the inclination of the orbit relatively to the plane of the sky,}
	 	\item {$T_{Eff}$ is the effective temperature of the star and}
	 	\item {$mag_V$ is the visual magnitude of the star.}
	 \end{inparaenum}
	
	The parameters for the simulated observations of our sample are presented in Table \ref{tab:paramsObservations}. For each target, we define one observing run as a set of simulated observations whose cumulative exposure times is given by $t_{\rm{total}}$. Note that we arbitrarily set a minimum observing time of 1h, which corresponds to the time span of one observing block with HARPS instruments.
4
\begin{table*}
	\caption{Parameters for simulated planets and hosts \citep[see the exoplanet.eu database,][]{Schneider2011}. The radius for 51 Pegb was not measured directly \citep[see ][]{Martins2015}. All other radii were obtained from transit measurements.}
	\label{tab:paramsPlanets}
	\centering\resizebox{\linewidth}{!}{\begin{tabular}{l | c c c c c c c | c c c c c }
			\hline\hline
			Planet name &	$\rm M_{\rm{p}}$&	$\rm R_{\rm{p}}$&	 P &	 a &	$\rm k_{\rm{p}}$&	I &	$\rm \left(\frac{R_{\rm{p}}}{a} \right)^{2}$&	$\rm M_{\star}$&	$\rm k_{\star}$&	$\rm T_{eff}$&	 Spectral &	\rm $\rm mag_V$ \\
			&	$\rm[M_{Jup}]$ &	$\rm[R_{Jup}]$ &	 $\rm[days]$ &	 $\rm[AU]$ &	 $\rm[\rm km \; s^{-1}]$ &	$\rm[degrees]$ &	 $\rm[ppm]$ &	$\rm[M_{\sun}]$&	 $\rm[\rm m \; s^{-1}]$ &	 $\rm[K]$ &	 Type &	 \\
			\hline
			\rule{0pt}{3ex}{\textbf{ESPRESSO + HIRES}} &&&&&&&&\\
			51 Peg b &	 0.47 &	 1.90 &	 4.2 &	 0.051 &	 130.8 &	 80.0 &	 315 &	 1.11 &	 52.9 &	 5793 &	 G2 IV &	 5.5 \\
			HD 209458 b &	 0.69 &	 1.38 &	 3.5 &	 0.045 &	 137.2 &	 86.6 &	 212 &	 1.15 &	 78.7 &	 6092 &	 G0 V &	 7.7 \\
			\rule{0pt}{3ex}{\textbf{HIRES}} &&&&&&&\\
			HD 109749 b &	 0.28 &	 0.99 &	 5.2 &	 0.059 &	 109.7 &	 90.0 &	 64 &	 1.20 &	 24.4 &	 5610 &	 G3 IV &	 8.1 \\
			HD 76700 b &	 0.23 &	 0.99 &	 4.0 &	 0.049 &	 120.4 &	 90.0 &	 93 &	 1.00 &	 26.4 &	 5726 &	 G6 V &	 8.1 \\
			55 Cnc e &	 0.03 &	 0.18 &	 0.7 &	 0.016 &	 128.8 &	 85.4 &	 28 &	 0.91 &	 3.6 &	 5196 &	 K0IV-V &	 6.0 \\
			\hline\hline
	\end{tabular}}
\end{table*}

\begin{table}
	\caption{Parameters for simulated observing runs.}
	\label{tab:paramsObservations}
	\centering\resizebox{\linewidth}{!}{\begin{tabular}{l | c | c c c | c }
			\hline\hline
			&	\multirow{2}{*}{$\rm \left(\frac{R_{\rm{p}}}{a} \right)^{2}$}	&	\multirow{2}{*}{$\rm t_{exp}$} 	&	\multirow{2}{*}{$\left<S/N\right>$}	& \multirow{2}{*}{$\rm t_{\rm{total}}$} 	&	 \multirow{2}{*}{$\rm N_{bins}$}			\\
			Planet name	&			& 					&			&					&					\\
			& $[ppm]$	&	$[seconds]$ 	&			& $[hours]$ 	& 		\\ \hline
			\rule{0pt}{3ex}{\textbf{ESPRESSO}} &	& 			& 		&					& 	\\
			51 Peg b & 315 & 		36			&	330	& 1.0 		& 15 \\
			HD 209458 b & 212 & 		432			&	430		& 12.0 		&		15 \\[.5em]
			\rule{0pt}{3ex}{\textbf{HIRES}} 	 & & 			&			& 		&					\\
			51 Peg b & 315 & 		36			&	1100	& 1.0 		& 70 \\
			HD 209458 b & 212 & 		 188			&	930		& 5.2 		&		70 \\[.5em]
			HD 109749 b & 64 & 		 263		&	890	& 7.3 		& 6 \\
			HD 76700 b & 93 & 		 130		&	610	& 3.6 		& 6 \\ [.5em]
			55 Cnc e & 28 & 		 153		&	1850	& 4.25 		& 5 \\\hline\hline
	\end{tabular}}
\end{table}

\subsection{Computing the exposure time}	\label{sec:expTime}

{To estimate the required exposure time, we begin by estimating the average planet-to-star flux ratio (${F_{\rm{*}}}/{F_{\rm{p}}}$) from Eq. \ref{eq:fluxRatioLambda} assuming:
	\begin{inparaenum}[i)]
		\item{ an average orbital phase yielding an average Lambert phase function} $g(\alpha) \approxeq 0.87$;
		\item a grey albedo function with $\rm A_g=0.2$;
		\item the planetary radius $R_{\rm{p}}$ and semi-major axis $a$ from the exoplanet.eu database.
\end{inparaenum}}
{Note that in most works \citep[e.g.][]{Esteves2015}, Lambert scattering is often selected for simplicity -- as this scattering model is wavelength independent -- but other more complex to model scattering functions could have been used \citep[see e.g.][]{GarciaMunoz2015,Munoz2018}}. 

{Under the above conditions, the required signal-to-noise ratio ($SN_{\rm{req}}$) required to {recover the planetary signal} at a $3\sigma_{\rm{noise}}$ significance is given by
	\begin{equation}
	SN_{\rm{req}} = \frac{3}{\frac{F_{\rm{p}}}{F_{\rm{*}}}}
	\end{equation}
}

As mentioned in Section \ref{sec:instruments}, for both instruments we estimated their S/N by correcting the S/N computed with HARPS ETC for the different collecting areas of each telescope and efficiencies of the instruments.

{The technique we propose makes use of the cross correlation of high-resolution spectra with numerical masks to increase the S/N of the observations (see Section \ref{sec:method}). This increase is proportional to the square root of the number of spectra lines used in the correlation. Since we intend a $3\sigma$  detection per bin, we assume that the increase in SN per bin is proportional to the average number of spectral lines per bin and as such}
\begin{equation}
\left<SN_{\rm{ins, bin}}\left(300s\right)\right> = \left<SN_{\rm{ins}}\left(300s\right)\right> \sqrt{\frac{N_{lines}}{N_{bins}}}
\end{equation}
{where $\left<SN_{\rm{ins, bin}}\left(300s\right)\right>$ is the average SN per wavelength bin that can be achieved with the selected instrument after a 300s exposure with the cross correlation technique (Equation \ref{eq:efficienciesSN}).}

{Finally the total exposure time $t_{\rm{total}}$ {to achieve the required SN} is given by:}
\begin{equation}
t_{\rm{total}} = 300s \times \left(\frac{SN_{\rm{req}}}{\left<SN_{\rm{ins, bin}}\left(300s\right)\right>}\right)^2
\end{equation}

Note that the simulation of observations and the computation of the cross correlation function (CCF) is computationally expensive. Therefore, to {optimize} the computational effort, it is ideal to keep the number of simulation to a minimum. However, the cross correlation technique applied in this work requires the construction of a stellar template (constructed by summing the simulated observations) with a S/N at least 10 times higher that of the individual observations, even assuming a white noise component. As such a minimum of 100 individual exposures is required, a value adopted for the number of observations for each simulated observing run\footnote{{In some cases, the proposed individual exposure times will most likely saturate the simulated instrument. However, we can assume that each individual simulated observation can be constructed by stacking individual observations with lower exposure times until the required S/N is attained.}}.

For our simulations, we assumed that we are working in a photon noise limited domain, and that instrumental noise can be neglected.  

\subsection{Constructing the simulated observations}		\label{sec:simulation}

Our simulations were constructed from the extremely high-resolution solar spectrum\footnote{This spectrum is freely available at \url{http://kurucz.harvard.edu/sun.html}.} from \citet[][]{2006astro.ph..5029K}. This spectrum -- with a resolution of 500 000 -- was degraded to the intended resolutions for both ESPRESSO and HIRES (see Section \ref{sec:instruments}).{ To do so, we convolved the spectrum with a Gaussian profile of unit area and FWHM corresponding to the intended resolution using the algorithm presented in \citep{Neal2017}.}

Since this template spectrum is normalized in flux, its continuum $S(\lambda, RV = 0)$ needs to be scaled in flux to match the expected flux of the host star ($F_{\rm{*}}\left(\lambda\right)$). As mentioned in Section \ref{sec:instruments}, this flux was estimated from HARPS ETC and extrapolated for each instrument with Equation \ref{eq:efficienciesFlux} assuming the following settings per target and instrument: 
\begin{inparaenum}[i)]
\item Table \ref{tab:paramsPlanets} for the stellar parameters;
\item Table \ref{tab:paramsObservations} for the exposure time and
\item Table \ref{tab:ETC} for the remaining parameters.
\end{inparaenum} 

	For any given orbital phase $\phi$, each simulated spectrum has three main components: 
	\paragraph*{\bf The stellar spectrum} $S_{\rm{*}}\left(\lambda, RV_{\rm{*}}\left(\phi\right)\right)$ with a continuum flux $F_{\rm{*}}\left(\lambda\right)$ given by Eq. \ref{eq:efficienciesFlux}.

	\paragraph*{\bf The planetary spectrum} $S_{\rm{p}}\left(\lambda, \phi\right)$ with flux $F_{\rm{p}}\left(\lambda\right)$ and radial velocity $RV_{\rm{p}}\left(\phi\right)$. Since we are working at optical wavelengths {-- and assumeing tha   thermal contribution from the planet is negligible --} {the planetary spectrum will basically be a scaled-down copy of the stellar spectrum modulated by the planetary atmospheric absorption profile}, given by
	\begin{equation}
	S_{\rm{p}}\left(\lambda, \phi\right) = \frac{F_{\rm{p}}\left(\lambda, \phi\right)}{F_{\rm{*}}\left(\lambda\right)} \times S\left(\lambda, RV_{\rm{p}}\left(\phi\right)\right)
	\end{equation}
	where $\frac{F_{\rm{p}}\left(\lambda, \phi\right)}{F_{\rm{*}}\left(\lambda\right)}$ comes from Eq. \ref{eq:fluxRatioLambda}. We assume for our observations a Lambert phase function $g(\alpha )$ \citep[see][]{Langford2011}
	\begin{equation}
	g(\alpha ) = \frac{[\sin(\alpha ) + (\pi - \alpha ) \, \cos(\alpha)]}{\pi}
	\label{eq:Phasefunction}
	\end{equation}
	where $\alpha$ is the phase angle, given by
	\begin{equation}
		\cos(\alpha ) = \sin(I) \, \cos(2 \pi \phi )
		\label{eq:PhaseAngle}
	\end{equation}
	{where $I$ is the inclination of the orbit.}
	
	\paragraph*{\bf The noise component} assumed to be Gaussian noise with a mean of zero and a standard deviation of $\sigma = \sqrt{F_{\rm{*}}\left(\lambda_c\right)}$
	as we are simulating extremely high signal-to-noise spectra and thus working in a photon noise limited domain. 
	
	These three components are all summed together.{ We assume that HARPS instrumental profile defines the spectral template for each simulated instrument in terms of spectral order, coverage and in-order characteristic efficiency (i.e., the blaze function).  As such, the computed spectra is projected on a wavelength grid defined by the HARPS spectrograph instrumental profile. Likewise, each order is multiplied by HARPS blaze function $B\left(\lambda\right)$ {to} account for in-order sensitivity variations.} 
	
	{Each resulting spectrum was then saved in the standard fits format used for HARPS observations, which permitted us to use \emph{HARPS Data Reduction Software (DRS) vs 3.6} to compute the CCF for each observation.}
	
	{Note that for these simulations we have assumed a photon limited S/N. When it comes to real observations, the situation might be more complex as the S/N will probably be limited by the instrument itself.}

	\subsection{Simulated atmospheric models}		\label{sec:atmospheres}
			
	The main purpose of this exercise is to test our method to recover the colour dependence of the reflected optical spectrum from an exoplanet. In particular, we are interested in the ability of the {CCF technique} to distinguish between possible atmospheric models and thus hint at the composition of the simulated planetary atmosphere. Thus, for this study, we decided to simulate our observations assuming albedo function models with real physical meaning.
	
	{As stated before, the reflectivity of a planet is highly dependent on its atmospheric composition and physics, and in particular cloud coverage. Unlike their gas giant Solar System counterparts (with albedos of about 0.5 due to the presence of ices in the atmosphere), close-in giant exoplanets were expected to be dark at optical wavelengths, due to heavy absorption from TiO and VO and/or alkali metals \citep[e.g][]{Marley1999,2000ApJ...538..885S,Madhusudhan2012}. In particular, atmospheric models predict a strong absorption feature at 589-nm due to the presence of Na in exoplanetary atmospheres which has already been identified in several planets \citep[e.g.][]{Redfield2008,Wyttenbach2015}. The technique presented in Setion \ref{sec:method} should be able to identify this feature.}
		
	{However, several exoplanets have been detected with albedo values above 0.3, \citep[e.g ][]{2010A&A...514A..23R,Demory2011,Esteves2015,Martins2015}, with these high values usually attributed to the presence of high altitude clouds and hazes \cite[e.g.][]{Sing2013a} that would mask the deeper absorption features. These would present flat reflection spectra as the clouds/hazes would mask the deeper absorbing molecules. However, the recovered albedos should allow to infer the particle size and density of the clouds/hazes \citep[e.g.][]{GarciaMunoz2015,Webber2015a}} 
	
	For this particular exploratory study, for both the hot Jupiter and hot Neptune type planets, the atmospheric models we used were produced by solving the problem of multiple scattering of starlight in the planet atmosphere. For integration over the planetary disk, we utilized the algorithm described in \cite{Horak1950}. Partial solutions at the integration points were obtained with a discrete-ordinate method \citep{Stamnes1988}. The line lists for molecular absorption were taken from HITEMP \citep{Rothman2010}. For the absorption by the alkalis, we incorporated the parameterization given in \cite{Iro2004}. The albedo function for each planetary model was created at full illumination (zero phase angle and zero inclination), and then adapted to each simulated illumination configuration by assuming a Lambert phase function (see Eq. \ref{eq:Phasefunction}). {Additional details on the model can be found in e.g. \citet{Munoz2012} and \citet{Nielsen2016}.}
	
	{We assumed hydrostatic balance in the atmosphere, and a temperature profile dictated by a planet equilibrium temperature of 1150 K. The temperature profile plays a second order effect in the scattering of reflected starlight.} {We also assumed that gas absorption at wavelengths shorter than 1 $\rm \mu m$ is dominated by water and alkalis (Na and K), thereby ignoring the potential contribution from other gases such as VO or TiO.} In terms of composition, we selected somewhat arbitrary values for the altitude-independent Volume Mixing Ratios (VMR's) for water and the {alkalis} (see Table \ref{tab:models}). 
	
	\begin{table}
		\caption{Atmospheric models composition.}
		\centering\begin{tabular}{l c c c c c c c }
			\hline\hline\\[-.5em]
			&	$Model \; A$	&	$Model \; B$	\\	
			\hline\\[-.5em]
			\multicolumn{1}{l}{{Volume Mixing Ratios}}				\\[.5em]
			$\rm H_2 O$	&	\multicolumn{2}{c}{$5 \times 10^{-4}$}				\\[.5em]
			$\rm Na$		&	$1.5 \times 10^{-6}$	&	$1.5 \times 10^{-8}$	\\[.5em]
			$\rm K$		&	$1.2 \times 10^{-7}$	&	$1.2 \times 10^{-9}$	\\[.5em]
			\multicolumn{1}{l}{{Configurations}}				\\[.5em]
			Aerosols		&	$\times 1, \;\times 100$	&	$\times 1, \;\times 100$	\\[.5em]
			\hline\hline					
		\end{tabular}
		\label{tab:models}
	\end{table}
	
	For the 'Model A' family of atmospheric models we assumed VMR's of about half of the concentration in solar composition atmospheres. In the 'Model B' family of atmospheric models we assume the same VMR for water, but the alkali concentrations {were} reduced by a factor of 100. The rationale for these cases is to see if a more structured spectrum (diminishing the Na/K impact enhances the relative impact of $\rm H_2 O$) may affect the recovery of the albedo function. The approximation of well-mixed gases is probably more realistic for water than for the alkalis. The alkalis potentially ionize {high-up} in the atmosphere, which will effectively result in the removal of their neutral forms at pressures less than $10^{-4}$ bar according to photo-chemical models of hot Jupiters \citep[e.g.][]{Lavvas2014}. 
	
	We include Rayleigh scattering by the $\rm{H_2/He}$ background atmospheric gas. For each model family, we also consider different aerosol scattering configurations ($\times 1$, $\times 100$) by adding extra scattering opacity. This extra opacity is equal to the gas opacity in our baseline $\times 1$ configuration, and one hundred times larger in the $\times 100$ configuration. {The aerosols are assumed to scatter without absorbing, i.e. their single scattering albedo is identically one.} Figure \ref{fig:models} shows the different planet geometric albedos obtained for each simulated configuration. In spectral regions away from the water and alkali bands, our treatment produces a geometric albedo of 0.75, as expected for a semi-infinite, conservative Rayleigh atmosphere. In the regions where gas absorption is significant (e.g around the 589-nm neutral sodium line), the planetary reflectivity depends strongly on the adopted aerosol {and gas} content.
	
	\begin{figure}
		\centering
		\includegraphics[width=.9\hsize]{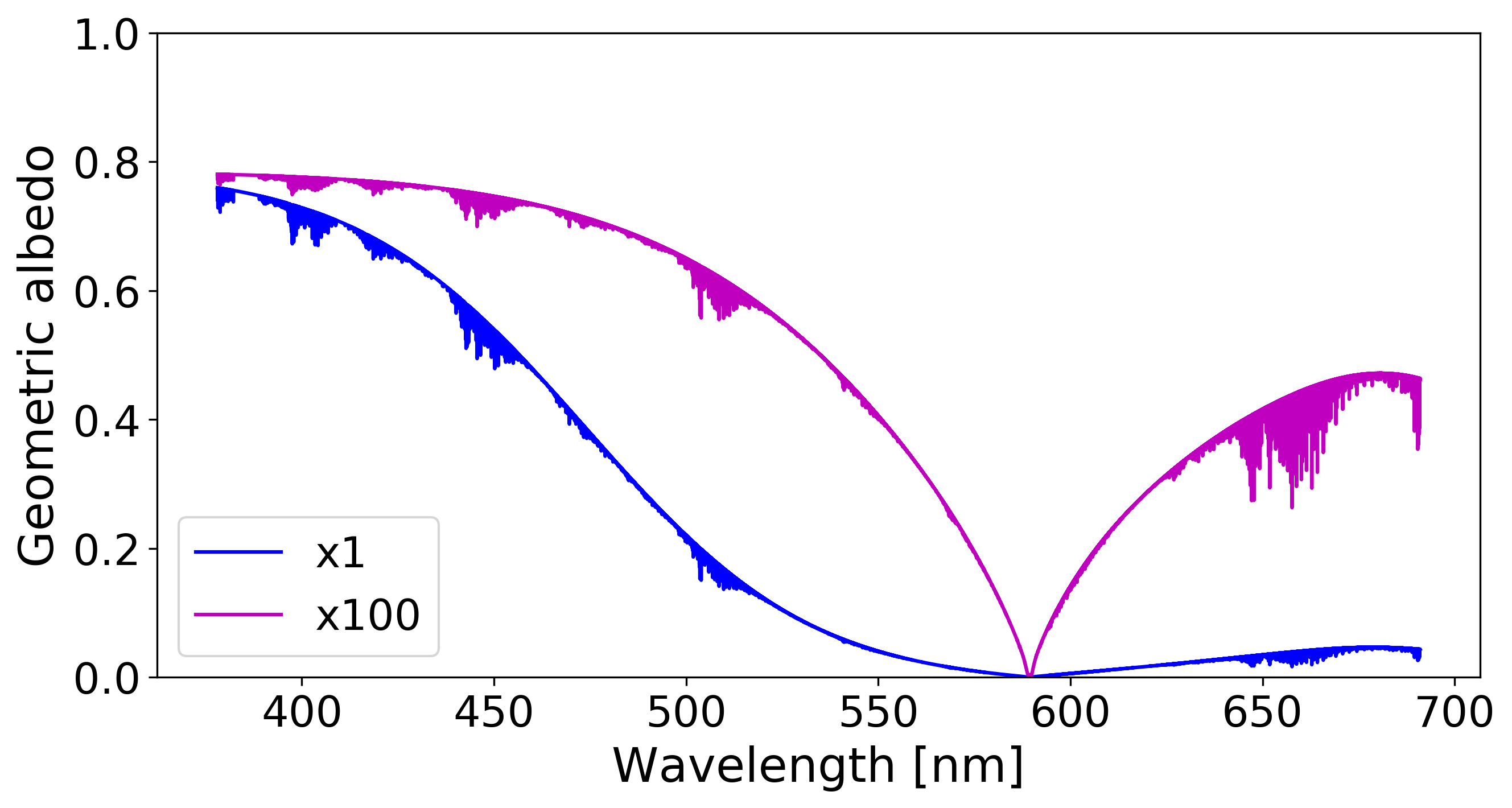}
		
		\includegraphics[width=.9\hsize]{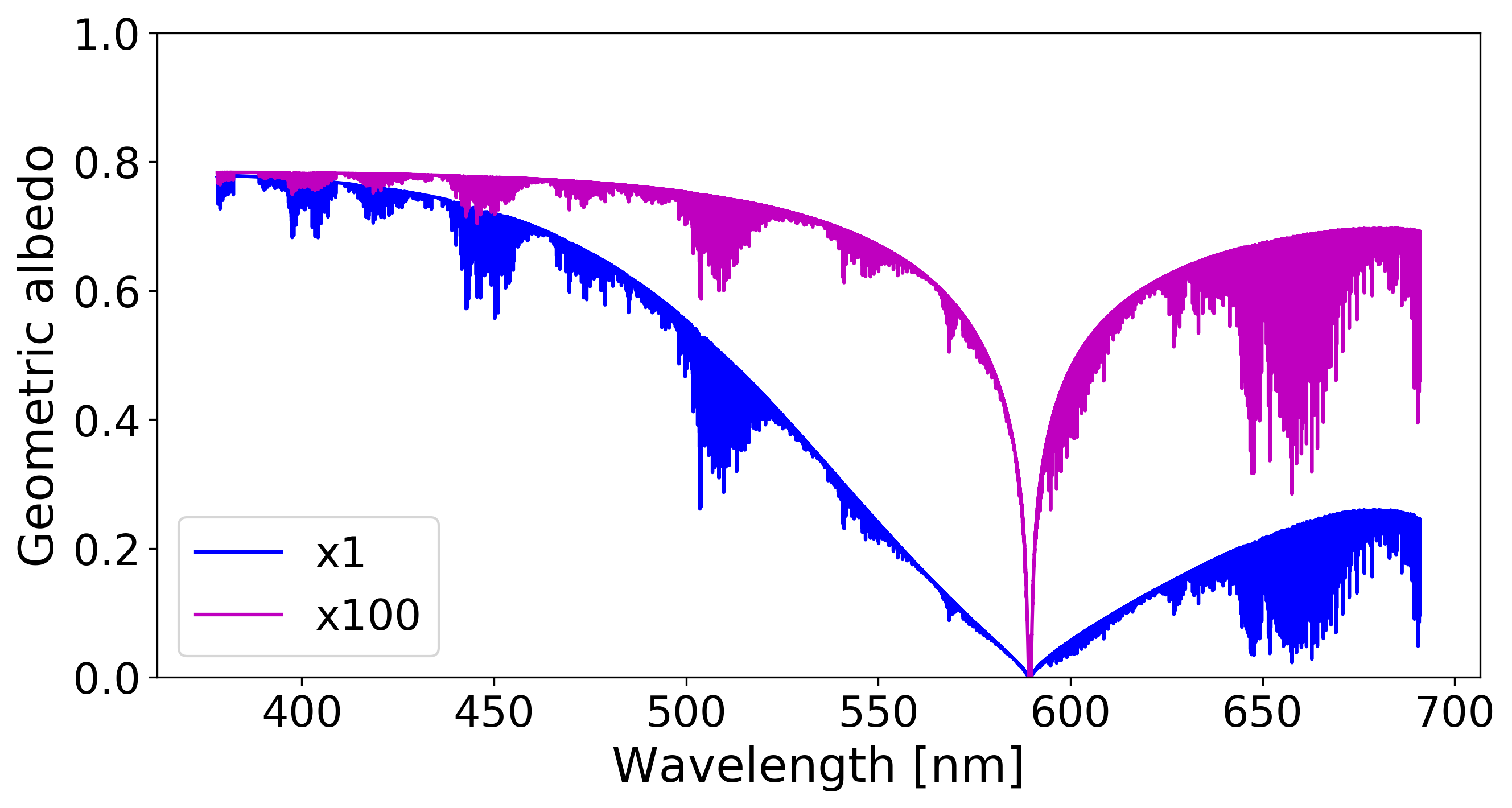}
		\caption{Albedo function for the different configurations of the atmospheric models in Table \ref{tab:models}. {\textit{Top Panel:} Model A ($\times 1, \times 100$ configurations); \textit{Bottom Panel:} Model B ($\times 1, \times 100$ configurations).} The $\times 1, \times 100$ configurations correspond respectively to different scattering configurations, where the opacity in the $\times 100$ configuration is 100 times larger that on the baseline configuration.}
		\label{fig:models}
	\end{figure}
	
	{For all targets, we also decided to simulate a grey albedo function with a conservative fixed albedo $\rm A_g=0.2$ with the goal of testing {whether} the proposed technique allows to distinguish between grey albedo models (e.g., as caused by clouds/hazes) and the {chromatic} models described above.} 
	
	In the case of 55 Cnc e, although an atmosphere has been detected around the planet \citep[see ][]{Tsiaras2016}, there is no information about its composition. Thus, we simulate a grey albedo function with a fixed albedo of 0.2.

%-------------------------------------------------------------------------------
% Section 4 - Results and discussion
%-------------------------------------------------------------------------------
%
\section{Results}										\label{sec:results}

We simulated ESPRESSO@VLT (in all 3 modes -- resolutions: MHR=60 000, HR=130 000 and UHR=200 000) and HIRES@ELT (resolution=110 000) observational runs of the star+planet systems presented in Table
\ref{tab:paramsPlanets}. {Table \ref{tab:paramsObservations} shows the parameters for each observing run of {all} star+planet systems, {as observed with} both ESPRESSO and HIRES. The total exposure time for each star+planet system with ESPRESSO {was set to be} the same for all modes (see Table \ref{tab:paramsObservations}). {For each instrument+star+planet+albedo configuration, we simulated 100 observing runs -- each with the total exposure time presented in Table \ref{tab:paramsPlanets} -- of 100 observations each. In total, we simulated close to $12000$ hours of ESPRESSO observations of 24 planet+albedo configurations, with an average S/N around $300-400$. For HIRES, we created about $6000$ hours of telescope time of 13 planet+albedo configurations, with an average S/N varying between $600$ and $1800$.}}

We attempted the recovery of the wavelength dependence of the planetary albedo function from all simulated observations. For each planet+albedo configuration we simulated 100 independent observational runs, for which we recovered the colour dependent albedo function independently. {From the recovered value distribution, we estimate }the $1\sigma$ error inherent to the method . 

The results are presented in Figures \ref{fig:espresso_hr} to \ref{fig:hires}. The green diamonds represent the mean of the recovered albedos over all simulated observing runs {with $1\sigma$ error bars}, for each wavelength bin. The blue horizontal lines represent the mean albedo from the simulated model over each wavelength bin. The horizontal orange lines represent the superior detection limit with a $3\sigma$  confidence for the recovered albedo.

{We defined the upper limit for non-detection (or minimum detectable albedo) on each wavelength bin as 3 times the average noise across all 100 observing runs simulated per instrument+star+planet+albedo configuration}. For the bins where the mean albedo from the model is lower than the minimum detectable albedo, we assume the value for the recovered albedo for those bins as the {upper} detection limit. Those cases are represented in the plots by the orange arrows.

{In all simulated cases, we tested our recovered albedo function against all atmospheric models presented {in} Section \ref{sec:atmospheres}. To do so, we computed the reduced chi-squared ($\chi^2_\nu$) value for the recovered data against the albedo models for all planet+albedo configurations. $\chi^2_\nu$ was computed from:}

\begin{equation}
\chi^2_\nu = \sum_{i = 1}^{N_{bins}} \frac{1}{N_{bins}} \frac{\left(<A_{measured, i}> - {<{A_{model, i}}>}\right)^2}{\sigma_i^2}
\end{equation}

where $\rm N_{bins}$ is the number of wavelength bins for which the wavelength was recovered; 
$<{A_{model, i}}>$ represents the mean of the albedo function over the wavelength range covered by bin $i$; 
$<A_{measured, i}>$ is the {average} measured albedo for wavelength bin $i$ {over all independent runs};
$\sigma_i$ is the standard deviation of the recovered albedos {for all runs}. In the sense of identifying the correct model, note that a different analysis using {Akaike Information Criterion (AIC)} or {Bayesian Inference Criterion (BIC)} could have been used with no change in the results as we are {assuming a constant number of degrees of freedom}. {The assumption of non-correlated noise makes is that} the distribution of the recovered values for the albedo follows a normal distribution and consequently so does the likelihood. Furthermore, each model has the same number of parameters (in this case zero). As a result the AIC/BIC values are only proportional to the likelihood and therefore to the $\chi^2_\nu$ value.

{The $\chi^2_\nu$ results are presented in Tables \ref{tab:results_espresso_hr} to \ref{tab:results_hires}. For each simulated configuration, the lowest reduced $\chi^2_\nu$ value is represented in bold.}
\begin{table*}
	\caption{Results of the recovery of the albedo function for each batch of simulated ESPRESSO (HR mode) observations. The first column shows the simulated instrument+planet configurations, the second one the simulated albedo for each configuration, and the remaining columns the reduced $\chi^2$ comparison between the simulated model and the models from Table \ref{tab:models}. For each configuration, the lowest $\chi^2$ value is highlighted with a red box. Note that in all cases it occurs when the simulated model matches the comparison model, indicating that the correct albedo function has been recovered.}
	\label{tab:results_espresso_hr}
	\resizebox{\hsize}{!}{\begin{tabular}{l c | c c c c c c c}
	\hline
	  &      &  \multicolumn{5}{c}{$\chi^2$ value}   \\[.5em]
	  &      &     $\rm A_g=0.2$&Model A ($\times 1$)&Model A ($\times 100$)&Model B ($\times 1$)&Model B ($\times 100$)   \\[.5em]
	\hline\\[.25em]
	\multicolumn{2}{l|}{\textbf{ESPRESSO (HR mode)}}&&&&& \\[.25em]
51 Peg b	&	 $\rm A_g=0.2$	&		\BOX{0.56}	&		1252.09	&		2883.25	&		2380.86	&		3750.83\\[.25em]
	&	Model A ($\times 1$)	&		1497.32	&		\BOX{1.71}	&		833.82	&		487.17	&		1324.44\\[.25em]
	&	Model A ($\times 100$)	&		3166.72	&		850.43	&		\BOX{4.37}	&		84.88	&		307.53\\[.25em]
&&&&&& \\[-.5em]
HD 209458 b	&	 $\rm A_g=0.2$	&		\BOX{0.34}	&		630.85	&		1415.97	&		1165.54	&		1855.49\\[.25em]
	&	Model B ($\times 1$)	&		1672.73	&		244.56	&		74.97	&		\BOX{1.78}	&		383.86\\[.25em]
	&	Model B ($\times 100$)	&		2362.25	&		937.77	&		111.94	&		308.75	&		\BOX{3.62}\\[.25em]
&&&&&& \\[-.5em]
\hline
\end{tabular}}
\end{table*}
\par

\begin{table*}
	\caption{Results of the recovery of the albedo function for each batch of simulated ESPRESSO (MHR mode - 1UT) observations. The first column shows the simulated instrument+planet configurations, the second one the simulated albedo for each configuration, and the remaining columns the reduced $\chi^2$ comparison between the simulated model and the models from Table \ref{tab:models}. For each configuration, the lowest $\chi^2$ value is highlighted with a red box. Note that in all cases it occurs when the simulated model matches the comparison model, indicating that the correct wavelength-dependent albedo has been recovered.}
	\label{tab:results}
	\resizebox{\hsize}{!}{\begin{tabular}{l c | c c c c c c c}
	\hline
	  &      &  \multicolumn{5}{c}{$\chi^2$ value}   \\[.5em]
	  &      &     $\rm A_g=0.2$&Model A ($\times 1$)&Model A ($\times 100$)&Model B ($\times 1$)&Model B ($\times 100$)   \\[.5em]
	\hline\\[.25em]
	\multicolumn{2}{l|}{\textbf{ESPRESSO (MHR mode - 1UT)}}&&&&& \\[.25em]
51 Peg b	&	 $\rm A_g=0.2$	&		\BOX{0.59}	&		862.38	&		1992.87	&		1633.98	&		2618.26\\[.25em]
	&	Model A ($\times 1$)	&		1240.68	&		\BOX{1.94}	&		707.1	&		413.69	&		1143.95\\[.25em]
	&	Model A ($\times 100$)	&		2511.01	&		688.48	&		\BOX{3.97}	&		74.61	&		285.29\\[.25em]
&&&&&& \\[-.5em]
HD 209458 b	&	 $\rm A_g=0.2$	&		\BOX{0.29}	&		342.57	&		846.75	&		682.27	&		1129.48\\[.25em]
	&	Model B ($\times 1$)	&		1338.98	&		193.76	&		60.34	&		\BOX{1.72}	&		303.29\\[.25em]
	&	Model B ($\times 100$)	&		1933.61	&		831.58	&		103.41	&		279.58	&		\BOX{3.51}\\[.25em]
&&&&&& \\[-.5em]
\hline
\end{tabular}}
\end{table*}
\par

\begin{table*}
	\caption{Results of the recovery of the albedo function for each batch of simulated ESPRESSO (MHR mode - 2UT) observations. The first column shows the simulated instrument+planet configurations, the second one the simulated albedo for each configuration, and the remaining columns the reduced $\chi^2$ comparison between the simulated model and the models from Table \ref{tab:models}. For each configuration, the lowest $\chi^2$ value is highlighted with a red box. Note that in all cases it occurs when the simulated model matches the comparison model, indicating that the correct albedo function has been recovered.}
	\label{tab:results}
	\resizebox{\hsize}{!}{\begin{tabular}{l c | c c c c c c c}
	\hline
	  &      &  \multicolumn{5}{c}{$\chi^2$ value}   \\[.5em]
	  &      &     $\rm A_g=0.2$&Model A ($\times 1$)&Model A ($\times 100$)&Model B ($\times 1$)&Model B ($\times 100$)   \\[.5em]
	\hline\\[.25em]
	\multicolumn{2}{l|}{\textbf{ESPRESSO (MHR mode - 2UT)}}&&&&& \\[.25em]
51 Peg b	&	 $\rm A_g=0.2$	&		\BOX{0.89}	&		1819.82	&		3823.3	&		3176.37	&		5028.64\\[.25em]
	&	Model A ($\times 1$)	&		2101.37	&		\BOX{3.45}	&		1424.64	&		817.0	&		2315.4\\[.25em]
	&	Model A ($\times 100$)	&		4268.15	&		1154.99	&		\BOX{6.75}	&		133.3	&		508.01\\[.25em]
&&&&&& \\[-.5em]
HD 209458 b	&	 $\rm A_g=0.2$	&		\BOX{0.54}	&		758.16	&		1742.24	&		1425.48	&		2325.58\\[.25em]
	&	Model B ($\times 1$)	&		2268.51	&		349.04	&		115.93	&		\BOX{3.44}	&		589.1\\[.25em]
	&	Model B ($\times 100$)	&		3312.67	&		1511.6	&		205.49	&		540.77	&		\BOX{6.68}\\[.25em]
&&&&&& \\[-.5em]
\hline
\end{tabular}}
\end{table*}
\par

\begin{table*}
	\caption{Results of the recovery of the albedo function for each batch of simulated ESPRESSO (MHR mode - 4UT) observations. The first column shows the simulated instrument+planet configurations, the second one the simulated albedo for each configuration, and the remaining columns the reduced $\chi^2$ comparison between the simulated model and the models from Table \ref{tab:models}. For each configuration, the lowest $\chi^2$ value is highlighted with a red box. Note that in all cases it occurs when the simulated model matches the comparison model, indicating that the correct albedo function has been recovered.}
	\label{tab:results}
	\resizebox{\hsize}{!}{\begin{tabular}{l c | c c c c c c c}
	\hline
	  &      &  \multicolumn{5}{c}{$\chi^2$ value}   \\[.5em]
	  &      &     $\rm A_g=0.2$&Model A ($\times 1$)&Model A ($\times 100$)&Model B ($\times 1$)&Model B ($\times 100$)   \\[.5em]
	\hline\\[.25em]
	\multicolumn{2}{l|}{\textbf{ESPRESSO (MHR mode - 4UT)}}&&&&& \\[.25em]
51 Peg b	&	 $\rm A_g=0.2$	&		\BOX{1.91}	&		3074.13	&		6962.03	&		5675.48	&		9370.81\\[.25em]
	&	Model A ($\times 1$)	&		4951.37	&		\BOX{8.11}	&		6415.28	&		3127.65	&		12758.4\\[.25em]
	&	Model A ($\times 100$)	&		8253.72	&		3034.02	&		\BOX{14.87}	&		453.84	&		1874.56\\[.25em]
&&&&&& \\[-.5em]
HD 209458 b	&	 $\rm A_g=0.2$	&		\BOX{1.42}	&		2213.95	&		5323.42	&		4298.97	&		7102.65\\[.25em]
	&	Model B ($\times 1$)	&		4455.57	&		665.3	&		259.87	&		\BOX{5.98}	&		1476.43\\[.25em]
	&	Model B ($\times 100$)	&		6442.4	&		3093.26	&		493.54	&		1244.16	&		\BOX{13.79}\\[.25em]
&&&&&& \\[-.5em]
\hline
\end{tabular}}
\end{table*}
\par

\begin{table*}
	\caption{Results of the recovery of the albedo function for each batch of simulated ESPRESSO (UHR mode) observations. The first column shows the simulated instrument+planet configurations, the second one the simulated albedo for each configuration, and the remaining columns the reduced $\chi^2$ comparison between the simulated model and the models from Table \ref{tab:models}. For each configuration, the lowest $\chi^2$ value is highlighted with a red box. Note that in all cases it occurs when the simulated model matches the comparison model, indicating that the correct albedo function has been recovered.}
	\label{tab:results}
	\resizebox{\hsize}{!}{\begin{tabular}{l c | c c c c c c c}
	\hline
	  &      &  \multicolumn{5}{c}{$\chi^2$ value}   \\[.5em]
	  &      &     $\rm A_g=0.2$&Model A ($\times 1$)&Model A ($\times 100$)&Model B ($\times 1$)&Model B ($\times 100$)   \\[.5em]
	\hline
\multicolumn{2}{l|}{\textbf{ESPRESSO (UHR mode)}}&&&&& \\[.25em]
51 Peg b	&	 $\rm A_g=0.2$	&		\BOX{0.73}	&		1422.0	&		3200.04	&		2636.73	&		4198.8\\[.25em]
	&	Model A ($\times 1$)	&		1558.19	&		\BOX{1.61}	&		866.93	&		502.27	&		1394.81\\[.25em]
	&	Model A ($\times 100$)	&		3247.76	&		765.05	&		\BOX{3.6}	&		77.2	&		271.94\\[.25em]
&&&&&& \\[-.5em]
HD 209458 b	&	 $\rm A_g=0.2$	&		\BOX{0.41}	&		927.23	&		2207.22	&		1789.34	&		2935.23\\[.25em]
	&	Model B ($\times 1$)	&		2060.49	&		303.4	&		95.82	&		\BOX{1.99}	&		501.21\\[.25em]
	&	Model B ($\times 100$)	&		2919.3	&		1195.16	&		158.4	&		422.0	&		\BOX{3.96}\\[.25em]
&&&&&& \\[-.5em]
\hline
\end{tabular}}
\end{table*}
\par

\begin{table*}
	\caption{Results of the recovery of the albedo function for each batch of simulated HIRES observations. The first column shows the simulated instrument+planet configurations, the second one the simulated albedo for each configuration, and the remaining columns the reduced $\chi^2$ comparison between the simulated model and the models from Table \ref{tab:models}. For each configuration, the lowest $\chi^2$ value is highlighted with a red box. Note that in all cases it occurs when the simulated model matches the comparison model, indicating that the correct albedo function has been recovered.}
	\label{tab:results_hires}
	\resizebox{\hsize}{!}{\begin{tabular}{l c | c c c c c c c}
	\hline
	  &      &  \multicolumn{5}{c}{$\chi^2$ value}   \\[.5em]
	  &      &     $\rm A_g=0.2$&Model A ($\times 1$)&Model A ($\times 100$)&Model B ($\times 1$)&Model B ($\times 100$)   \\[.5em]
	\hline\\[.25em]
	\multicolumn{2}{l|}{\textbf{HIRES}}&&&&& \\[.25em]
51 Peg b	&	 $\rm A_g=0.2$	&		\BOX{5.38}	&		7150.26	&		17168.62	&		13772.93	&		23739.76\\[.25em]
	&	Model A ($\times 1$)	&		4881.94	&		\BOX{7.84}	&		6924.96	&		3480.13	&		13008.05\\[.25em]
	&	Model A ($\times 100$)	&		7725.97	&		2878.74	&		\BOX{19.77}	&		391.53	&		1771.47\\[.25em]
&&&&&& \\[-.5em]
HD 209458 b	&	 $\rm A_g=0.2$	&		\BOX{2.04}	&		2440.33	&		6088.13	&		4835.52	&		8452.54\\[.25em]
&	Model B ($\times 1$)	&		3911.57	&		801.72	&		317.0	&		\BOX{8.17}	&		1730.91\\[.25em]
&	Model B ($\times 100$)	&		6073.89	&		2987.44	&		393.93	&		1055.57	&		\BOX{15.94}\\[.25em]
&&&&&& \\[-.5em]
HD 109749 b	&	 $\rm A_g=0.2$	&		\BOX{1.05}	&		645.38	&		2006.96	&		1478.67	&		2928.78\\[.25em]
	&	Model A ($\times 100$)	&		1897.37	&		662.68	&		\BOX{8.24}	&		70.7	&		241.85\\[.25em]
	&	Model B ($\times 100$)	&		3610.79	&		1798.4	&		258.4	&		647.7	&		\BOX{10.29}\\[.25em]
&&&&&& \\[-.5em]
HD 76700 b	&	 $\rm A_g=0.2$	&		\BOX{2.13}	&		2623.96	&		6485.17	&		5179.69	&		8967.27\\[.25em]
	&	Model A ($\times 1$)	&		3007.09	&		\BOX{3.0}	&		2699.31	&		1508.58	&		4400.09\\[.25em]
	&	Model B ($\times 1$)	&		4379.23	&		834.28	&		318.99	&		\BOX{10.91}	&		1777.48\\[.25em]
&&&&&& \\[-.5em]
55 Cnc e	&	 $\rm A_g=0.2$	&		\BOX{0.67}	&		1081.99	&		2078.01	&		1732.81	&		2942.96\\[.25em]
&&&&&& \\[-.5em]
\hline
\end{tabular}}
\end{table*}
\par

\section{Discussion} \label{sec:discussion}

{Figures \ref{fig:espresso_hr} to \ref{fig:hires} show for each star+planet+albedo configuration the recovered (green points) versus the simulated (grey line) albedo functions. It can be seen that for each wavelength bin, the mean simulated albedo (horizontal blue lines) is within the $3\sigma$  error bars of the recovered albedo functions.} In all cases, the recovered albedo function is close to the simulated one. {Note the unusually large error bar on panel h of Figure \ref{fig:hires}, which results from the low number of spectral lines (3) in the numerical mask for that particular order (order \textit{103} of the HARPS spectrograph, with {590.7-nm <$\lambda$< 597.4-nm}). Albeit not as evident, the same effect can be seen on panel a of the same figure for the same spectral order.}

{Furthermore, from Tables \ref{tab:results_espresso_hr} to \ref{tab:results_hires}, it can be seen that for each star+planet+albedo configuration -- regardless of the simulated instrument or atmosphere -- the lowest reduced $\chi^2$ value ({highlighted with a red box}) is obtained when comparing with the correct simulated albedo model for the configuration. }

{{These results are a clear indication that the cross-correlation function of high-resolution spectra with numerical masks can be used to not only recover the reflected optical signal} from an exoplanet but also recover the colour dependency of the planetary albedo function. {With HARPS, we were able to recover the optical reflected signature of 51 Pegasi b and infer the planetary albedo from about 2.5h worth of observing time \citep{Martins2015}}. From 2018 -- due to the increased collecting power and higher efficiency of the spectrograph -- ESPRESSO should permit us to recover the albedo function from hot Jupiters with just a few hours of observing time (e.g 1h for 51 Pegasi b and $\sim$ 8h for HD 209458 b assuming 20 wavelength bins). HIRES -- and following similar stable high-resolution optical spectrographs to be mounted on 30m class telescopes \citep[e.g G-CLEF@GMT][]{Szentgyorgyi2014} -- will be able to probe the same targets but with an increased number of wavelength bins, effectively increasing the level of detail of the albedo function. {For both instruments, lowering the number of wavelength bins and/or increasing the total exposure time should permit to recover the albedo function from smaller and/or with larger period planets.}} 

{Note that an alternative approach can be followed. Instead of finding the required times for a fixed amount of wavelength bins, for each target, we can estimate for how many wavelength bins can the albedo be recovered at a $3\sigma$  significance for a fixed amount of time (e.g. one night, $\sim$ 11h). Table \ref{tab:fixedTime} shows -- for the same targets we selected -- 
	the estimated number wavelength bins for which the albedo can be recovered at a $3\sigma$  significance for a total observing time of 11 hours (about one observing night) for a grey albedo model with different $\left<A_g\right>$. We limited $\rm N_{bins}$ at 70, which is the number of orders of the HARPS spectrograph. }

\begin{table}
\caption{Estimated number of wavelength bins for which the albedo could recovered at an average $3\sigma$  significance for a total observing time of 11 hours (about one observing night) for different grey albedos models. We limited $\rm N_{bins}$ at 70, which is the number of orders of the HARPS spectrograph.}
\label{tab:fixedTime}
\centering
{\begin{tabular}{l | c c c c c c }
		\hline\hline
		&	\multicolumn{5}{c}{$\rm N_{bins}$ for $\left<A_g\right>$}\\
		&	0.1	& 0.2 & 0.3 & 0.4 & 0.5 \\
		Planet name	&			& 					&			&					&					\\
		\hline
		\rule{0pt}{3ex}{\textbf{ESPRESSO}} &	& 			& 		&					& 	\\
		51 Peg b &55 &70 &70 &70 &70\\
		HD 209458 b &3 &13 &31 &55 &70\\[.5em]
		\rule{0pt}{3ex}{\textbf{HIRES}} 	 & & 			&			& 		&					\\
		51 Peg b &70 &70 &70 &70 &70\\
		HD 209458 b &37 &70 &70 &70 &70\\[.5em]
		HD 109749 b &2 &9 &20 &36 &56\\
		HD 76700 b &4 &18 &41 &70 &70\\[.5em]				
		55 Cnc e &3 &12 &29 &51 &70\\
		\hline\hline
\end{tabular}}
\end{table}

In conclusion, as soon as 2018, ESPRESSO will make it possible to probe the atmospheres of hot Jupiter class planets {in reflected light} with a few hours worth of observing time. Increasing the exposure time or decreasing the level of detail, ESPRESSO will surely be able to probe smaller planets, but for those HIRES will be much more efficient due to the sheer increase of collecting power that will allow it to reach much higher S/N domains.

{In the study, for all cases except 51 Pegasi b (for which we assumed the radius proposed in \citet{Martins2015}) the planetary radius is known. In the cases where the radius is not known, the radius and the albedo are degenerate and as such we will recover $A_{g}\left(\lambda\right) \; R_{\rm{p}}^2$ instead of the wavelength-dependent albedo.}

\subsection{Observing strategies}

{In order to apply the CCF technique it is necessary to carefully choose the orbital phases at which the planet is being observed. As such we have defined \textit{optimal phases} as the orbital phases within 15 and 45 degrees on each side of superior conjunction {(i.e., phases with $0.37 \lessapprox \phi \lessapprox 0.46$ or $0.54 \lessapprox \phi \lessapprox 0.63$}, see Appendix \ref*{app:optimalWindows} for details). Given the total exposure times presented in Table \ref{tab:paramsObservations} and the time each planet spends at optimal phases\footnote{51 Pegasi b: 16.8h; HD 209458 b: 14h; HD 109749 b: 20.8h; HD 76700 b: 16h; 55 Cnc e: 2.8h}, it is clear that apart from the proposed observations of 55 Cnc e with HIRES all the proposed observations can be completed during the same orbit.}

{However, it is important to consider that to remove the stellar signal, this technique relies on the construction {of} stellar template built from adding at least 100 observations. Should these observations be taken too close together in the RV domain, the template will also remove the planetary signal. To avoid this issue, three different approaches can be followed.}

{The most intuitive approach consists in acquiring  100+ additional spectra at orbital phases close to inferior conjunction. In those cases, {and assuming that the orbit is not too far from edge on,} not only the planetary signal will be minimal (or {close to zero }for transiting planets), but also the planetary and stellar signals will be aligned in the RV domain. This should allow for the construction of the purest stellar template, with (almost) no planetary contamination. The downside of this approach is that it requires additional telescope time. While for bright targets such as 51 Peg b this might not be a problem, faint targets such as HD 209458 b could easily greatly increase the required exposure time or even double it.}

{Another approach is to divide the observations in two sets, defined by their orbital phases. Assuming that the orbital phase of superior conjunction is $\phi_0$, one set will have orbital phases in the optimal window (See Appendix \ref{app:optimalWindows} for details) \textit{before}  superior conjunction ($\phi_0-45^\circ < \phi< \phi_0-15^\circ$) and the other set will have orbital phases in the orbital windows \textit{after} ($\phi_0+15^\circ < \phi< \phi_0+45^\circ$) superior conjunction. Then for each set, the template will be constructed from observations from the other set. In this case, we guarantee that when removing the stellar signal from a given CCF, the stellar template will be built from CCFs where the planet will have a RV which is symmetric to the one of the observation. }

{A different approach, and the one we used for this work, is to make sure that the observations are randomly spread across the optimal phases window. This dilutes the planetary signal on the template across a large range of planetary RVs, increasing the S/N of the template by a factor equal to the square root of the number of spectra used in its construction.}	

\subsection{{Identification of atmospheric spectral absorption features}}

{One of the main goals of atmospheric characterization is the detection of spectral features in the recovered albedo functions. The feature that dominates the albedo functions from our models is the optical sodium doublet at 589-nm. In the best case scenario -- where $\rm N_{bins}$ is equal to the number of orders of the spectrograph -- each wavelength bin will be over {4-nm} wide, clearly insufficient to resolve both lines of the doublet. Yet, this feature is extremely wide -- typically over {10-nm overall} -- and as such we should be able to sample it with over 2 wavelength bins and as such resolve this spectral feature. Note that the amount of aerosols will impact the width of the sodium doublet. The lower the amount of aerosols, the wider the line will be {and the easier will be to resolve and sample it.} On the other hand, a wider spectral feature will mean more absorption and as such a lower albedo.}

{Narrower spectral features -- with widths smaller than the spectral range covered by 2-3 wavelength bins -- will not be possible to resolve. However, in wavelengths regions where a significant number of absorption lines can be identified, the mean albedo will drop relatively to the 'continuum' of the albedo function and dips can be identified. This is clear where albedos from the Model B family were simulated. Due to the higher water to alkali metals ratio, these albedos are more structured, i.e. have 'forests' of water absorption lines. This causes the mean albedo in those regions to drop significantly relatively to the 'continuum' of the albedo function (see panels e and f of Figures 9 to 11, panel g, i, j and m of Figure 12). {This the equivalent of a water band as seen in a low-resolution spectrum}.}

{An interesting feature from the model albedo functions {presented in this work }is the impact of aerosols on the spectral dependency of the albedo function. The higher the aerosol content of the simulated albedo function, the smaller the width of the dominating sodium doublet absorption line and the higher the average geometric albedo. {In fact, higher-altitude aerosols prevent the access of the starlight to the lower altitudes responsible for the line broadening.} In the simulated atmospheric scenarios, as the amount of aerosols diminishes, the average geometric albedo will drop  because absorption by the alkalis becomes dominating. However, {in the models presented here }this drop is not colour independent: although the albedo towards lower wavelengths will remain almost unchanged, it will be quite significant towards longer wavelengths. As such, even in cases where only a small number of bins is possible, the ratio of the recovered albedo function between redder and bluer wavelengths should at least allow to estimate {basic information on the aerosols such as the dependence of their cross sections and/or single scattering albedo with wavelength.}}

{It is intuitive that atmospheres with high reflectivity/albedos will be easier to detect. Fundamentally, it means that atmospheres with a high content of {poorly-absorbing and high-altitude} aerosols will be easier to detect. Also, the detectability of an atmosphere will also depend on the alkali metal content. Atmospheres with low alkali content (e.g. the B family of models we present) will be much less difficult to detect than atmospheres with high alkali content (e.g. the A family of models we present). In particular, a high sodium content, will lead to an extremely large drop in the reflective of the planet due to the impact of the sodium doublet at 589-nm.}

\subsection{ESPRESSO resolution dependence}

One of the hypotheses we wanted to test was if the resolving power of the instrument {can} affect the results. To test for this, we constructed additional simulated observations of the HD209458 planet+star system by ESPRESSO in all 3 modes (resolving power: MHR=60 000, HR=130 000 and UHR=200 000). {Furthermore, since the MHR mode of ESPRESSO allows to combine the light from up to 4 of the telescope's UTs, we simulated observations with 1, 2 and 4 UTs}. For the sake of simplicity we assumed for all observations a grey albedo model with $\rm A_g=0.2$.
For each star+planet+mode configuration, we simulated 100 observing runs. Each observing run has a total exposure time of 10h -- independent of the simulated resolution mode -- split into 100 spectra (as before to allow for a good stellar template correction). This makes sure that in all cases we are working in a S/N domain much higher than the required for an individual $3\sigma$  recovery of the colour dependency of the reflected planetary signal -- and consequently of the albedo function -- over 10 wavelength bins. For the low resolving power mode of ESPRESSO, we only considered observations with one telescope UT. As stated before, both HR and MHR modes will have higher efficiency than the UHR mode (see Figure \ref{fig:espressoEff})

\begin{figure}
\centering
\includegraphics[width=\hsize]{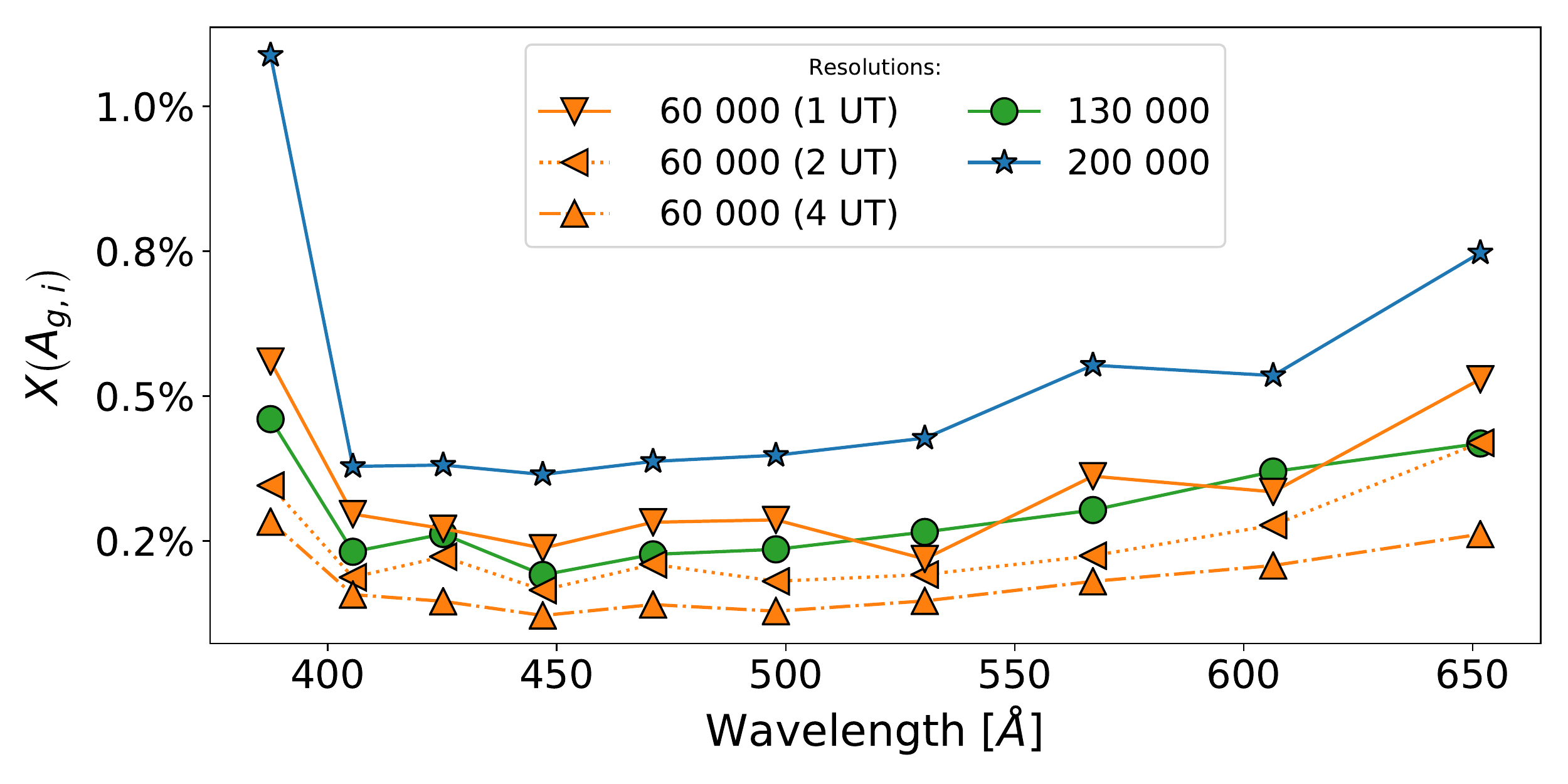}
\caption{Relative errors for the recovered albedo function from simulated observations of HD209458 b with ESPRESSO in all 3 resolution modes. {$X\left(A_{g,i}\right)$ represents the relative {error} on the geometric albedo $A_{g,i}$ recovered for each wavelength bin $i$, as defined in Equation \ref{eq:mean_albedo_dist}.} For the MHR mode, we simulated observations with 1, 2 and 4 UT telescopes.}
\label{fig:resolutionEffect}
\end{figure}

{Figure \ref{fig:resolutionEffect} shows the relative {error} $X\left(A_{g,i}\right)$ on the geometric albedo $A_{g,i}$ recovered for each wavelength bin $i$, defined as}
\begin{equation}
	\label{eq:mean_albedo_dist}
	X\left(A_{g,i}\right) = \frac{\left< \sigma_{A_{g,i}} \right>}{A_{g,i}}
\end{equation}
{where $A_{g, i}$ is the mean measured geometric albedo for bin $i$ and $\sigma\left(A_{g, i}\right)$ {is} the dispersion over all simulated runs of $A_{g, i}$.}

For the UHR mode ($R = 200 000$), it is clear that the error in the recovered albedos is significantly {larger} than with the two other modes. This can easily be {explained} if we consider that in this mode the instrument is being fed by a much smaller fiber (0.5 instead of 1 arcsec){, leading to a much lower number of photons reaching the detector. As such,} the observation will be conducted on a lower S/N domain. For the other two modes -- HR ($R = 130 000$) and MHR ($R = 60 000$) -- there is no evidence that might suggest an impact of the resolution on the error for the recovered albedos. However, an increased resolution is of great importance to be able to obtain precise orbital parameters for both host and planet. Thus, the choice of resolution needs to balance the required high spectral resolution, but not at the loss of collecting area (e.g. as with ESPRESSO's UHR mode). 

{It is also clear the effect of an increased collecting area. On the MHR mode observations, as we move from 1 UT to 2 and then 4 UTs, the error on the recovered albedo decreases.} As such, should only one UT be available to ESPRESSO, HR mode is the best choice, as it combines the highest resolution for the same collecting area. However, should 2 or more UT be available to ESPRESSO, the sheer increase in collecting power will make MHR mode a better choice. Despite the highest resolution, the UHR mode of ESPRESSO does not present any advantage (for this technique) over the other modes due to the much lower collecting power.

{An interesting case is the case of Figure \ref{fig:espresso_uhr}, panel d. When we computed the total exposure times for each observing run, we assume ESPRESSO to be in the HR configuration. Since the fiber diameter for the UHR mode is half of the fiber for the other modes, the planetary CCFs will be significantly noisier. Therefore, in most wavelength bins, the S/N of the recovered planetary CCF is lower than the {upper limit} for the albedo with a $3\sigma$  confidence. }

\subsection{Noise variation with wavelength}

It can be seen from the red horizontal bars on Figures \ref{fig:espresso_hr} to \ref{fig:hires} that the noise on the recovered CCF is wavelength dependent. This dependence is the combination of {several} factors. 

First, the efficiency of the instruments has a colour dependence. Figure \ref{fig:espressoEff} shows respectively the efficiency curves of ESPRESSO (in all 3 resolution modes) and HARPS (used as a template for HIRES). 
Secondly, the distribution of spectral lines along the spectral coverage of the instrument is not uniform. Figure \ref{fig:nLinesPerOrder} shows the distribution of spectral lines for the different number of wavelength bins simulated in this document (5, 6, 20 and 70 wavelength bins). The number of spectral lines per bins is indicated on each bar from the histogram, except for $\rm N_{bins} = 70$ as it would be unreadable. It can be seen that the number of spectral lines decreases towards redder wavelengths, yielding a lower increase in S/N due to the CCF technique in redder bins. For $\rm N_{bins} = 70$, each bin corresponds to a spectral order of the HARPS spectrograph. In that case several gaps can be seen. The gap around 530-nm corresponds to a missing order that falls between the two chips of the HARPS spectrograph. The other 3 gaps (around 590, 630 and 645 nm) correspond to orders where the number of spectral lines is inferior to 5. Those are orders known for a large telluric contamination {and as such larges wavelength ranges are not considered for the CCF calculation.} Those gaps do not show on lower numbers of bins as these orders will be merged with adjacent orders.

\begin{figure}
	\begin{subfigure}[t]{\hsize}
		\includegraphics[width=\hsize]{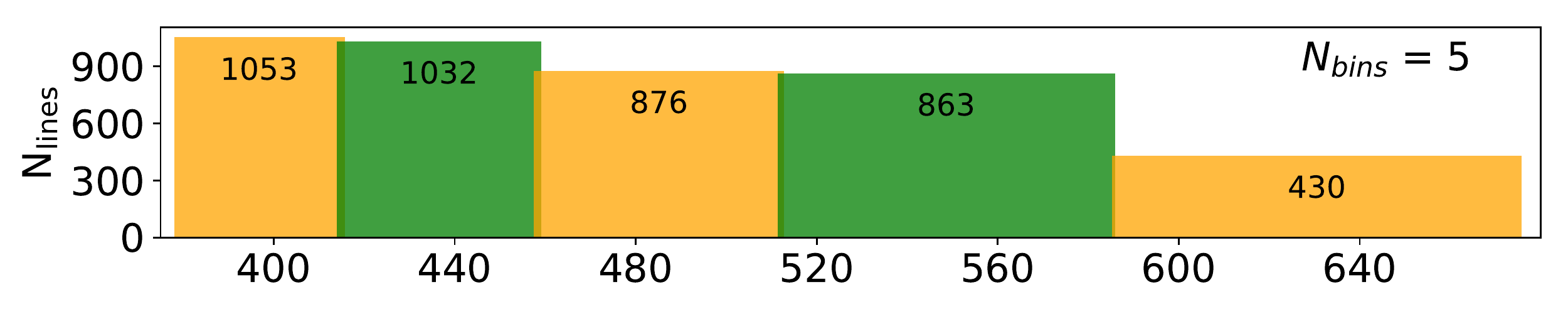}
	\end{subfigure}
	\begin{subfigure}[t]{\hsize}
		\includegraphics[width=\hsize]{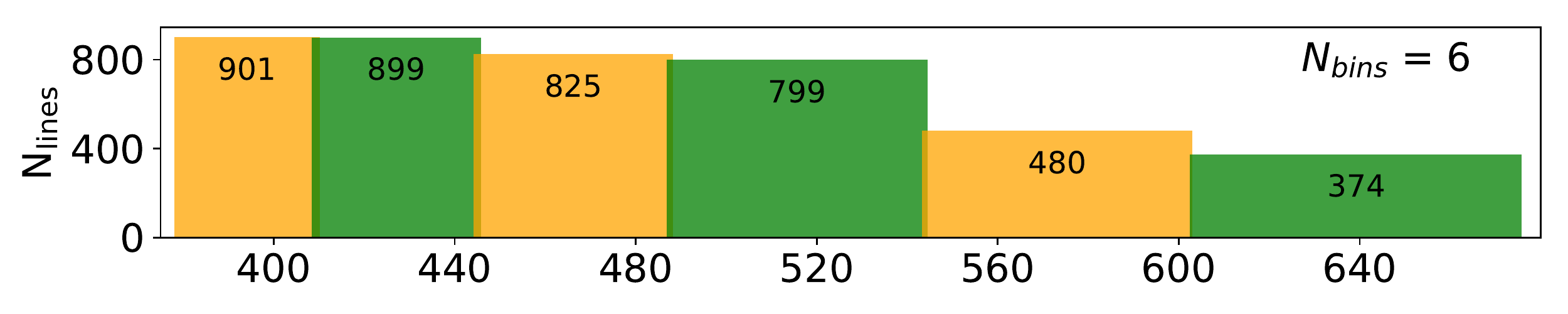}
	\end{subfigure}
	\begin{subfigure}[t]{\hsize}
		\includegraphics[width=\hsize]{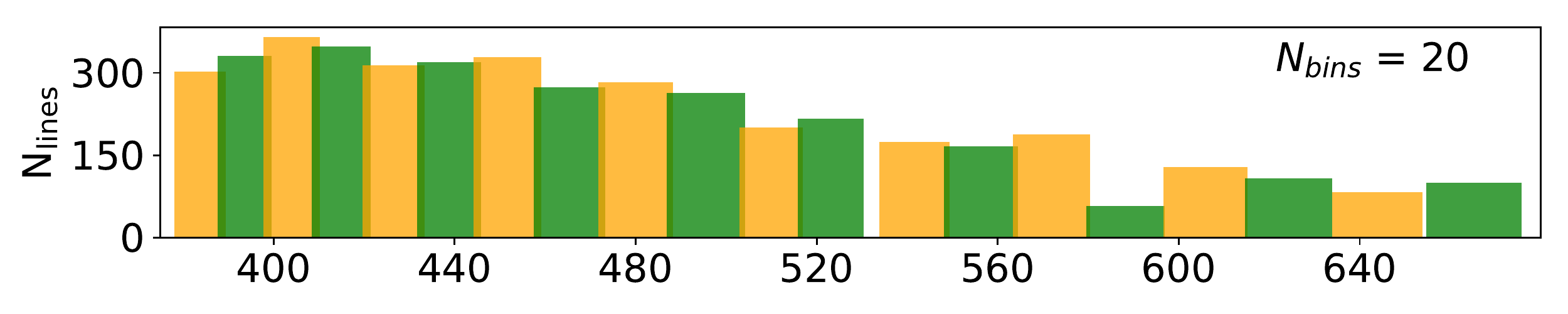}
	\end{subfigure}
	\begin{subfigure}[t]{\hsize}
		\includegraphics[width=\hsize]{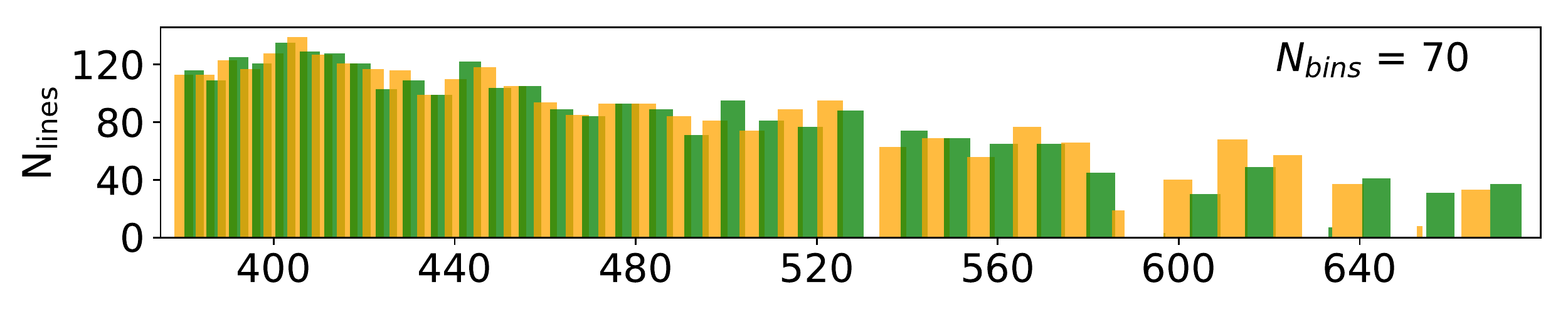}
	\end{subfigure}
	\centering{$\rm wavelength \; [nm]$}
	\caption{Number of spectral lines per wavelength bin for the HARPS CCF mask of a G2 star, for 5, 6, 20 and 70 bins.}
	\label{fig:nLinesPerOrder}
\end{figure}

{Finally, the spectral type of the star: different spectral types will have different spectral energy distributions. Figure \ref{fig:specTypeEffect} shows the expected S/N from a 300s exposure with HARPS of $\rm mag_V = 6.0$ F0, G2  K2 and M2 spectral type stars as computed with HARPS ETC. As expected, the 3 curves are different: M stars will attain higher S/N towards redder wavelengths, while F stars will reach higher S/N than G to M stars towards bluer wavelengths, with G stars somewhere in the middle of both.}

\begin{figure}
\centering
\includegraphics[width=\hsize]{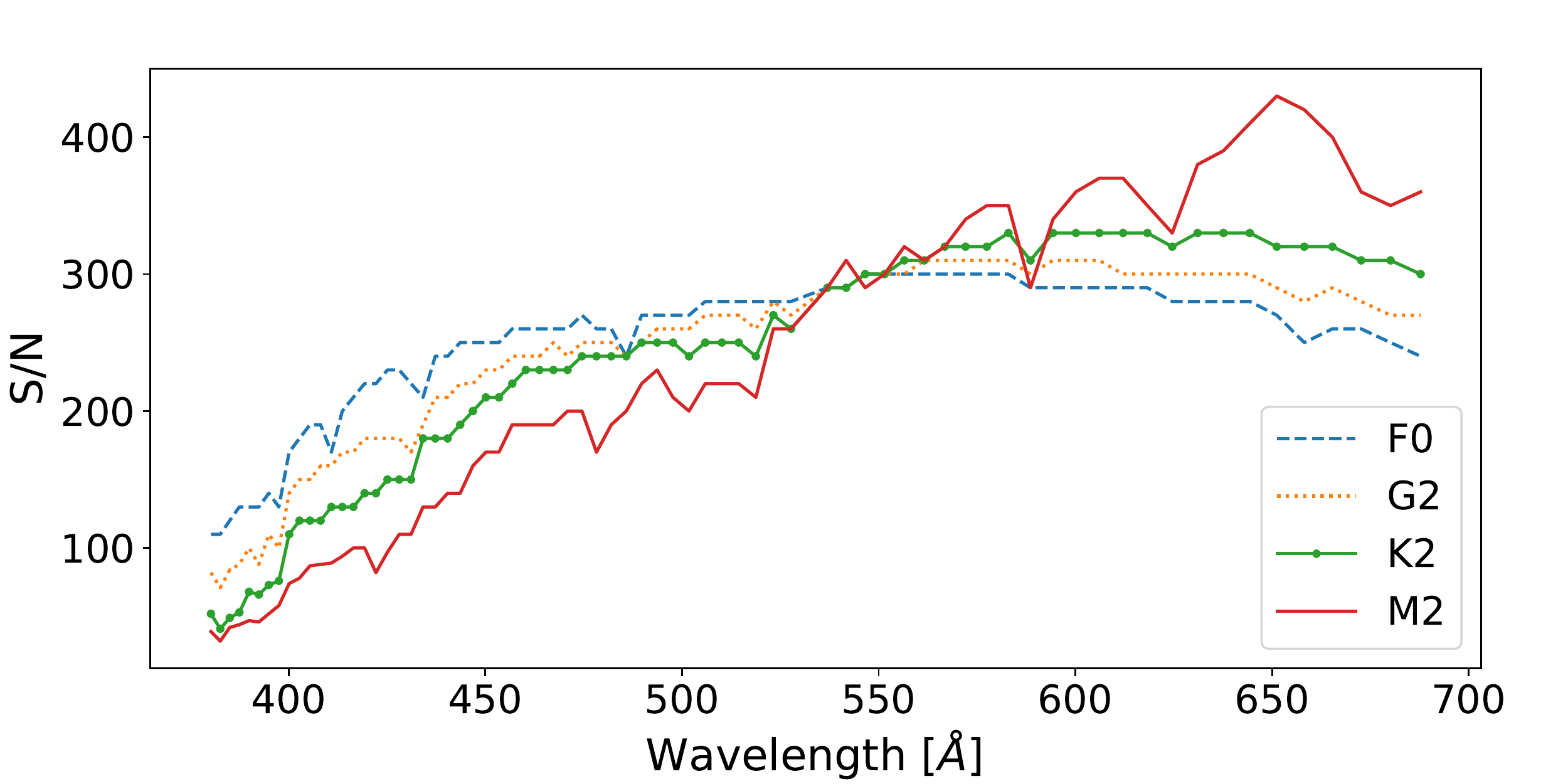}
\caption{{Comparison of the expected S/N from a 300s exposure with HARPS of $\rm mag_V = 6.0$ F0, G2  K2 and M2 spectral type stars (from HARPS ETC).}}
\label{fig:specTypeEffect}
\end{figure}

\subsection{{Spectral Types}}

For this work, all the spectra we have simulated were constructed from a template G2 spectrum. A question that arises is if the spectral type of the star will affect the recovery of the albedo function from a planet. From Equation \ref{eq:fluxRatioLambda}, it can be seen that the stellar type has no impact on the albedo function itself as we are dealing with the planet-to-star flux ratio and not the fluxes directly.

As discussed on the previous section, spectral type affects the S/N levels registered on the detectors -- through the number of counts recorded by the detector -- as well as the number of spectral lines over a specified spectral range.  Figure \ref{fig:specTypeEffect} shows the spectral energy distribution for F0, G2  K2 and M2 spectral type stars: M stars will attain higher S/N towards redder wavelengths, while F stars will reach higher S/N than G to M stars towards bluer wavelengths, with G stars somewhere in the middle of both.

As a result, detectability levels will depend on the stellar type. Assuming the same magnitude and exposure times:
\begin{itemize}
	\item for redder wavelengths, the detection limit for the albedo function will decrease from F to M stars;
	\item for bluer wavelengths, the detection limit for the albedo function will increase from F to M stars;
\end{itemize}

It is interesting to compare this result with the albedo models presented in this work (see Figure \ref{fig:models}). These models show higher albedos for {bluer} wavelengths, with most of the absorption (depending on the alkali metal concentration) occurring for redder wavelengths. As such, it is realistic to assume that cooler spectral type stars will be better targets than warmer stars as lower albedo detection limits can be attained for the same exposure time. Note that this assumption is only valid for the models we present here (or similar) and is not universal. It is not unusual in the solar system planets that some absorber affects the ultraviolet wavelengths and diminishes the planet reflectivity towards bluer wavelengths \citep[e.g. the case Venus -- ][]{Lee2017,Perez-Hoyos2018}.

\subsection{Advantages of the method relative to transit spectroscopy}		\label{sec:restrictions}

Currently, one of the most used and powerful techniques to characterize planetary atmospheres is transit spectroscopy. Albeit a powerful technique, transit spectroscopy is limited on the range of orbital parameters that can be accessed as it cannot be applied to non-transiting planets. With the technique we propose in this work, this restriction is lifted as it can access the same parameter range as the radial velocity detection method, i.e., most planets that do not have face-on (or close to) orbits. However, in those cases, since no radii measurements are available, the radius and albedo will be degenerate.

An additional strong point from the CCF technique relatively to transit spectroscopy is that it detects reflected light from the planet. {Transit spectroscopy is significantly affected by the presence of clouds and/or hazes that will flatten -- totally or partially -- the absorption features from the transmission spectrum of an exoplanet \citep[e.g.][]{Kreidberg2014}. This is not the case when observing the light reflected on existing clouds, which may scatter the incident light and make a planet much brighter, if the scattering particles are poorly absorbing. Moreover, any absorption feature from the atmosphere above the clouds will be imprinted onto the reflected light spectrum, which can be used as a proxy to study the composition of cloudy and hazy planetary atmospheres \citep[e.g.][]{Morley2015}. In the absence of clouds, reflected light spectra probe deep into the atmosphere. In summary, reflected light studies complement the information gained from transit spectroscopy, and are an extremely powerful tool to understand the bulk composition of exoplanets, their energy budget and meteorology, in particular for cloudy and hazy atmospheres. }

Furthermore, the technique we propose allows for the direct detection of the planet, and thus permits to {break down the mass-inclination degeneracy} resulting from using radial velocities as a detection method. {{Also}, when compared to methods that involve a combination of high spectral and spatial resolution \citep{lovis2016,Snellen2015a}, this method is insensitive to the target distance from the observer as it is not required to spatially resolve the star and planet.}

%-------------------------------------------------------------------------------
% Section 5 - Conclusions
%-------------------------------------------------------------------------------
%
\section{Conclusions}									\label{sec:conclusions}
For this work we simulated ESPRESSO@VLT (in all 3 modes) and HIRES@ELT observational runs of the star+planet systems presented in Table \ref{tab:paramsPlanets}, selected from the exoplanet.eu database \citep{Schneider2011} as discussed in Section \ref{sec:simulations}. For each of the selected planets, we simulated three different atmospheric models. 

We attempted the recovery of the wavelength dependence of the planetary albedo function from all simulated observations. On all cases we performed a reduced $\chi^2$ comparison of the recovered colour dependent albedo functions against the atmospheric models presented Section \ref{sec:atmospheres} plus a grey/flat albedo model with $\rm A_g=0.2$. The results for $\chi^2$ are summarized in Tables \ref{tab:results_espresso_hr} to \ref{tab:results_hires}. Regardless of the simulated instrument or atmosphere, the lowest reduced $\chi^2$ value (in bold) corresponds to the simulated model, {which confirms that the technique is viable}. 

The results we present of this work are encouraging, as they exemplify the power of the cross correlation function technique \citep{Martins2015} as a powerful tool to characterize exoplanetary atmospheres from the optical reflected spectra from exoplanets. {The major results of this work can be summarized as follows:}
{\begin{itemize}
		\item  the CCF technique permits the recovery of the wavelength dependence of reflected spectra of exoplanets, and consequently their albedo function;
		\item  the CCF technique is able to distinguish between possible albedo configurations and hint at the planetary atmosphere composition;
		\item  it will be possible to probe at planetary atmospheres for planets in a vastly accessible orbital parameter range;
		\item  even in cases where a low number of wavelength bins is possible, the ratio between the albedo at redder and bluer wavelengths should help constrain possible chemical and physical atmospheric configurations.
		\item as close as 2018, ESPRESSO should already be able to recover the albedo function from hot Jupiter class exoplanets with a few hours of exposure time (e.g. 51 Pegasi b with 1h);
		\item the advent of HIRES will allow to probe smaller planets (e.g. hot Neptunes and Super Earths) with just with a few hours of exposure time (e.g. HD 76700 b with around 5h and 55 Cnc e with about 4h)
\end{itemize}}

However, the albedo from an exoplanet is not the only factor that modulates the reflected spectrum from an exoplanet. An exciting prospect to further characterize the planet comes from the observation of the phase function of the planet and how it varies with wavelength. This function describes the light curve of a planet as a function of orbital phase. It depends not only on the fraction of the planet illuminated by the host star but also on the {general} scattering properties of the planet's atmosphere.

The detection of the reflected optical from exoplanets is a daunting task which requires a broad spectral coverage and a extremely high-resolution to achieve the level of S/N we require, as well as an extremely high PSF stability for long periods of time. This can be achieved with ESPRESSO@VLT, HIRES@ELT and similar observing facilities, which will lead the way towards a more detailed characterization of exoplanets and their atmospheres. The sheer collecting power of giant telescopes, combined with the precision and stability of next generation high-resolution spectrographs, will put researchers in a position to not only look at the reflected light from exoplanets, but use it to characterize their atmospheres in great detail. A very bright future awaits exoplanetology indeed.

%-----------------------------------------------------------------
% acknowledgements
\section*{Acknowledgements}
This work was supported by Funda\c{c}\~ao para a Ci\^encia e a Tecnologia (FCT) through national funds and by FEDER through COMPETE2020 by these grants UID/FIS/04434/2013 \& POCI-01-0145-FEDER-007672 and PTDC/FIS-AST/1526/2014 \& POCI-01-0145-FEDER-016886.
NCS acknowledge support by FCT through Investigador FCT contract of reference IF/00169/2012/CP0150/CT0002.
PF acknowledge support by FCT through Investigador FCT contract and an exploratory project, both with the reference IF/01037/2013/CP1191/CT0001. J.F. acknowledges support by the fellowship SFRH/BD/93848/2013 funded by FCT (Portugal) and POPH/FSE (EC).
{We also would like to thank the referee for his/her careful revision of the manuscript.}

\clearpage\begin{figure*}
	\begin{subfigure}[t]{0.5\hsize}
	\includegraphics[width=\hsize]{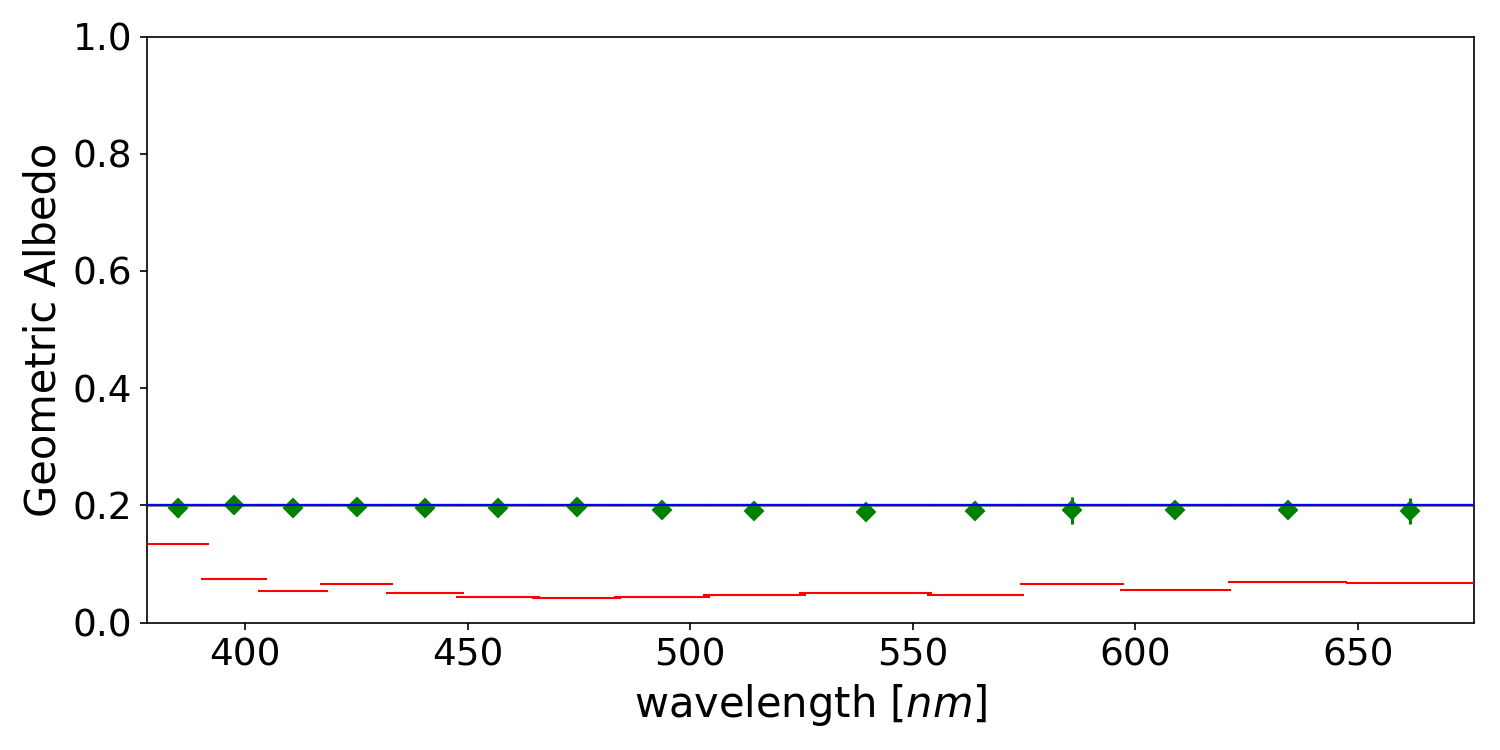}
	\caption{51 Peg b; $A_g$ = 0.2; $\chi^2 = 0.56$}
\end{subfigure}%
\begin{subfigure}[t]{0.5\hsize}
	\includegraphics[width=\hsize]{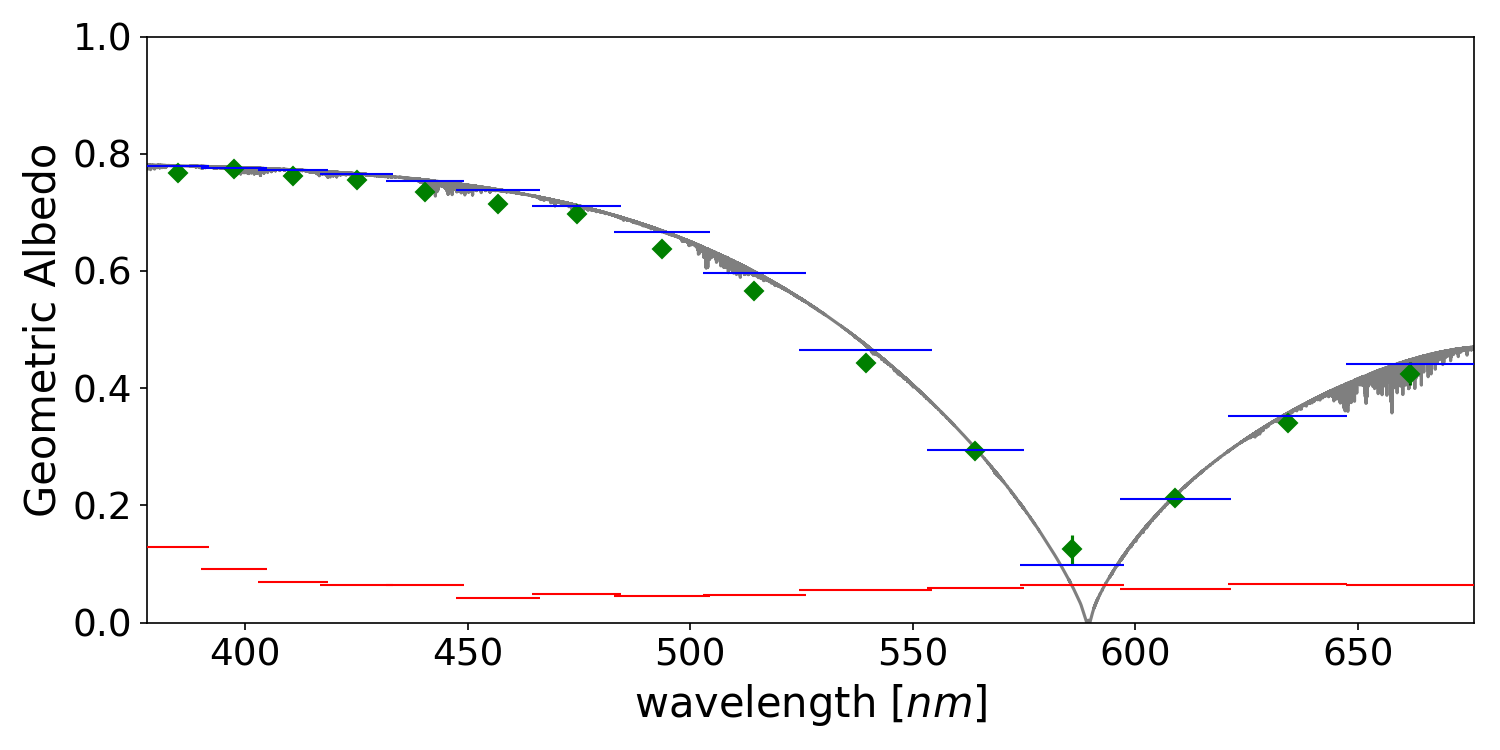}
	\caption{51 Peg b; Model A ($\times 100$); $\chi^2 = 4.37$}
\end{subfigure}%

\begin{subfigure}[t]{0.5\hsize}
	\includegraphics[width=\hsize]{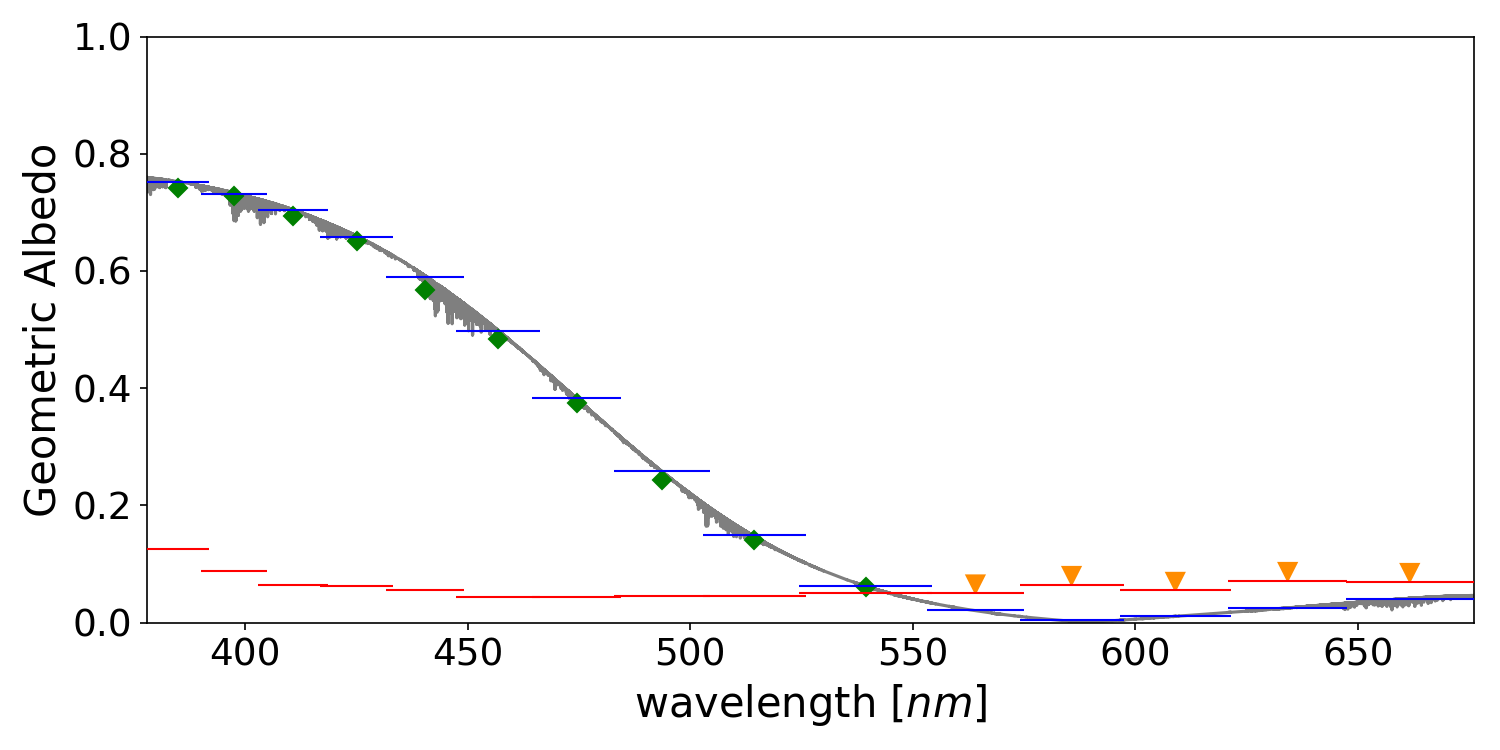}
	\caption{51 Peg b; Model A ($\times 1$); $\chi^2 = 1.71$}
\end{subfigure}%
\begin{subfigure}[t]{0.5\hsize}
	\includegraphics[width=\hsize]{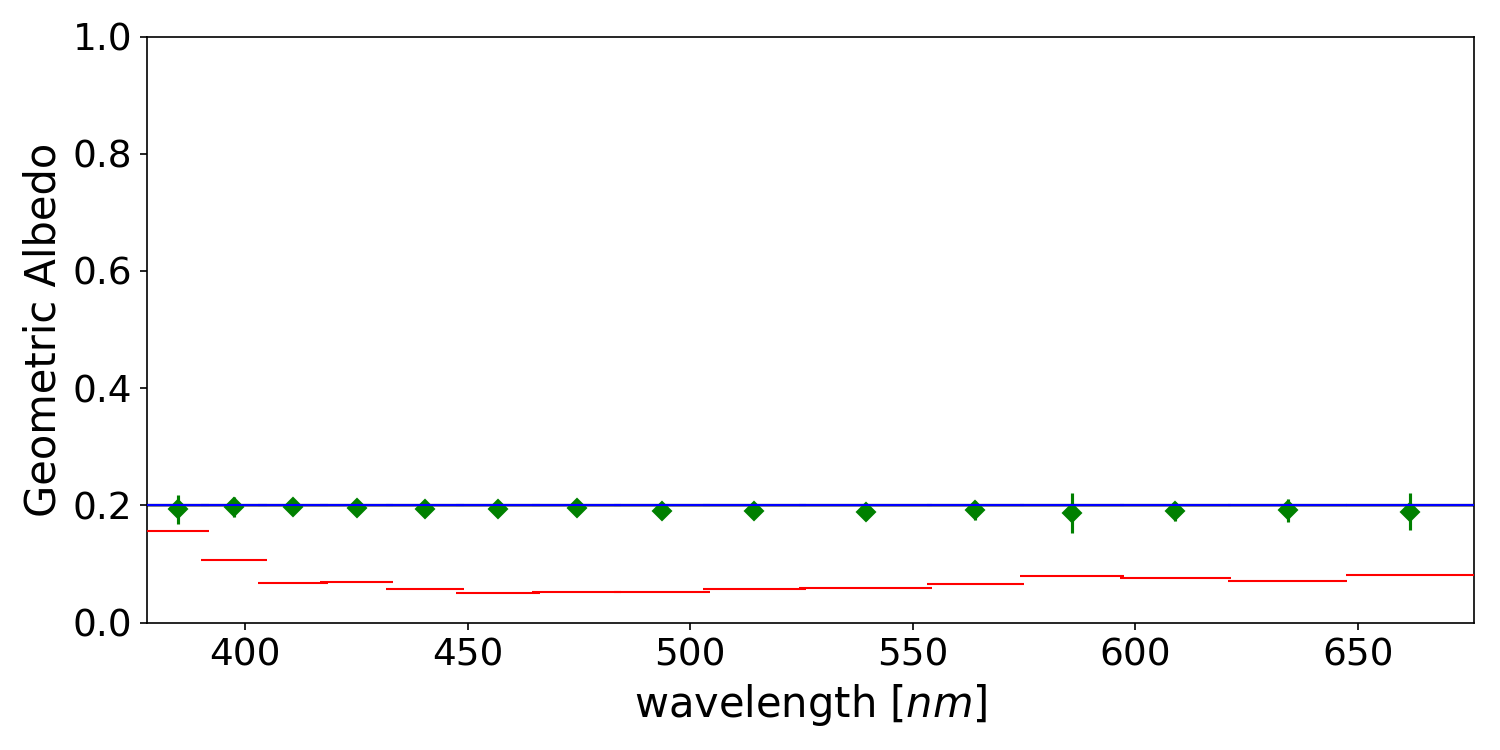}
	\caption{HD 209458 b; $A_g$ = 0.2; $\chi^2 = 0.34$}
\end{subfigure}%

\begin{subfigure}[t]{0.5\hsize}
	\includegraphics[width=\hsize]{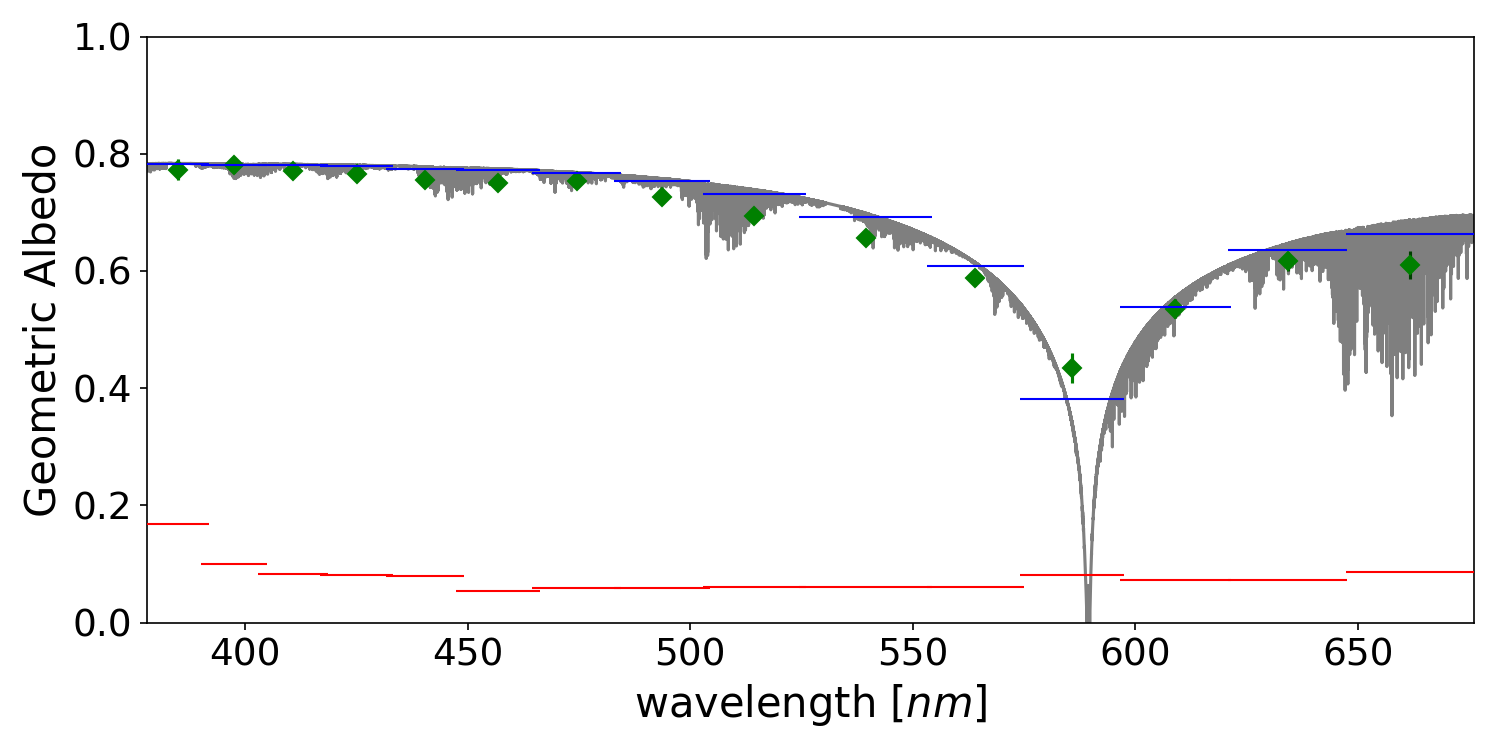}
	\caption{HD 209458 b; Model B ($\times 100$); $\chi^2 = 3.62$}
\end{subfigure}%
\begin{subfigure}[t]{0.5\hsize}
	\includegraphics[width=\hsize]{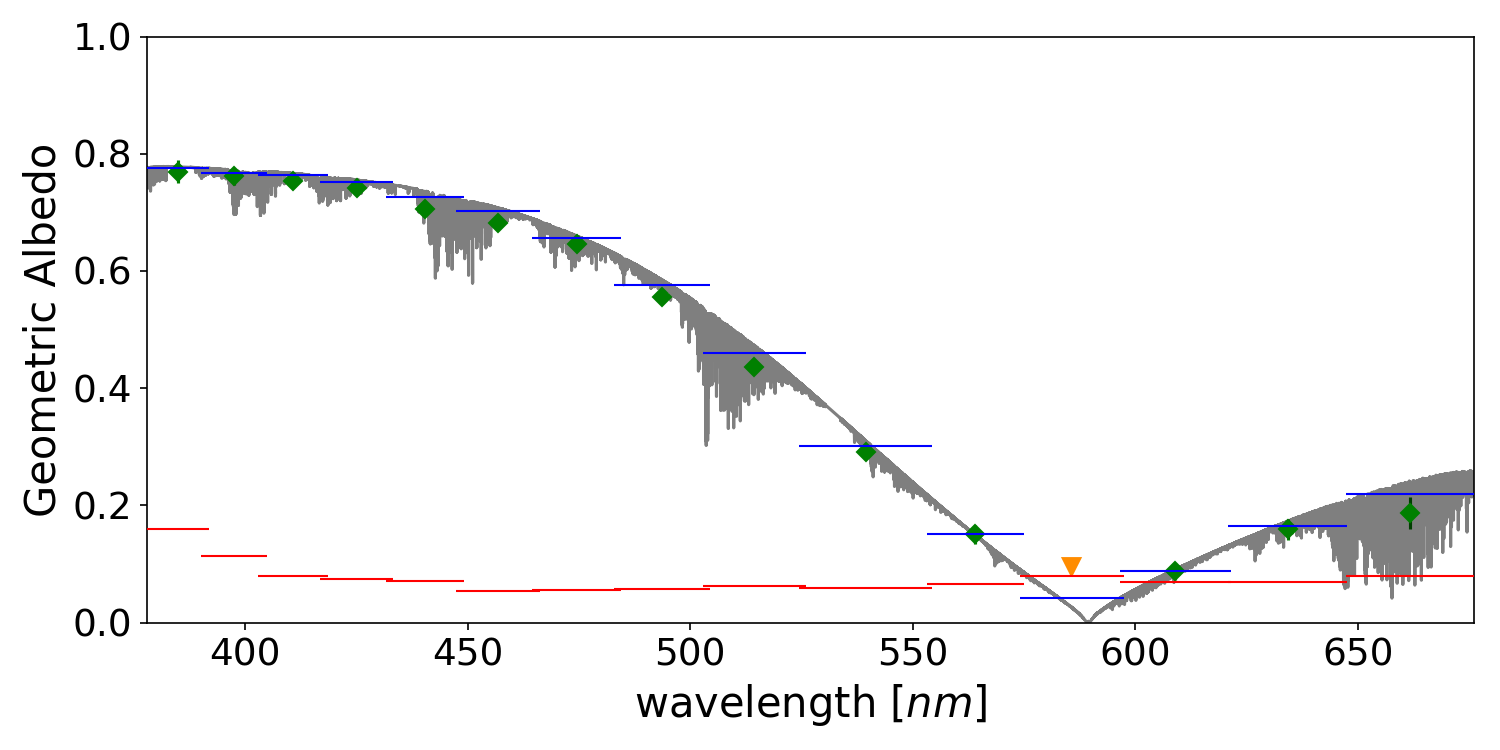}
	\caption{HD 209458 b; Model B ($\times 1$); $\chi^2 = 1.78$}
\end{subfigure}%

\caption{Distribution of the recovered albedo functions from the simulated ESPRESSO (HR mode) observations. For each wavelength bin: \textit{i)} the green dots represent the mean recovered albedo over the 100 simulated runs with error bars given by the 3 times the standard deviation; \textit{ii)}the blue horizontal lines represent the mean albedo of the simulated model over the bin; \textit{iii)} the red horizontal bar represents the $3\sigma$ detection limit of the albedo.}
	\label{fig:espresso_hr}
	\end{figure*}

\clearpage\begin{figure*}
	\begin{subfigure}[t]{0.5\hsize}
	\includegraphics[width=\hsize]{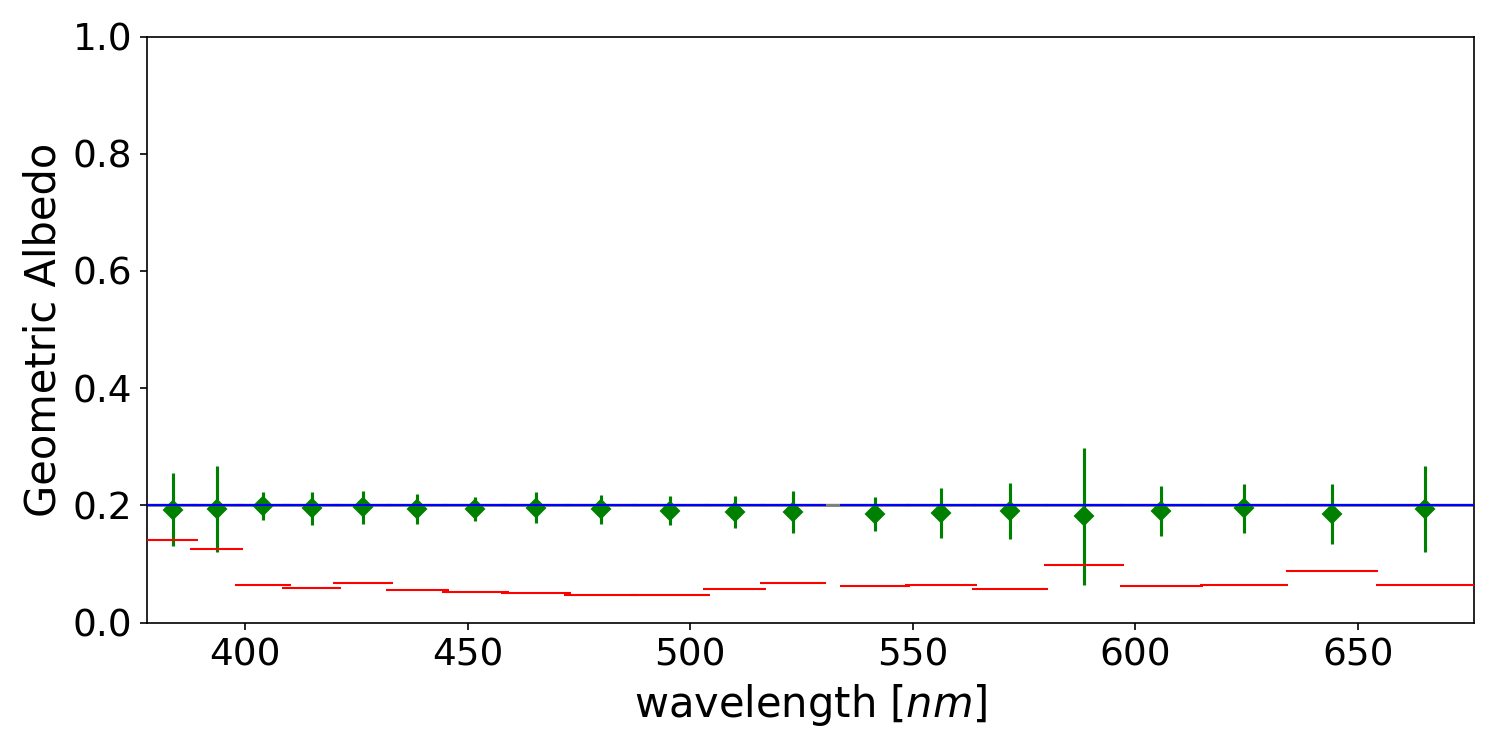}
	\caption{51 Peg b; $A_g$ = 0.2; $\chi^2 = 0.59$}
\end{subfigure}%
\begin{subfigure}[t]{0.5\hsize}
	\includegraphics[width=\hsize]{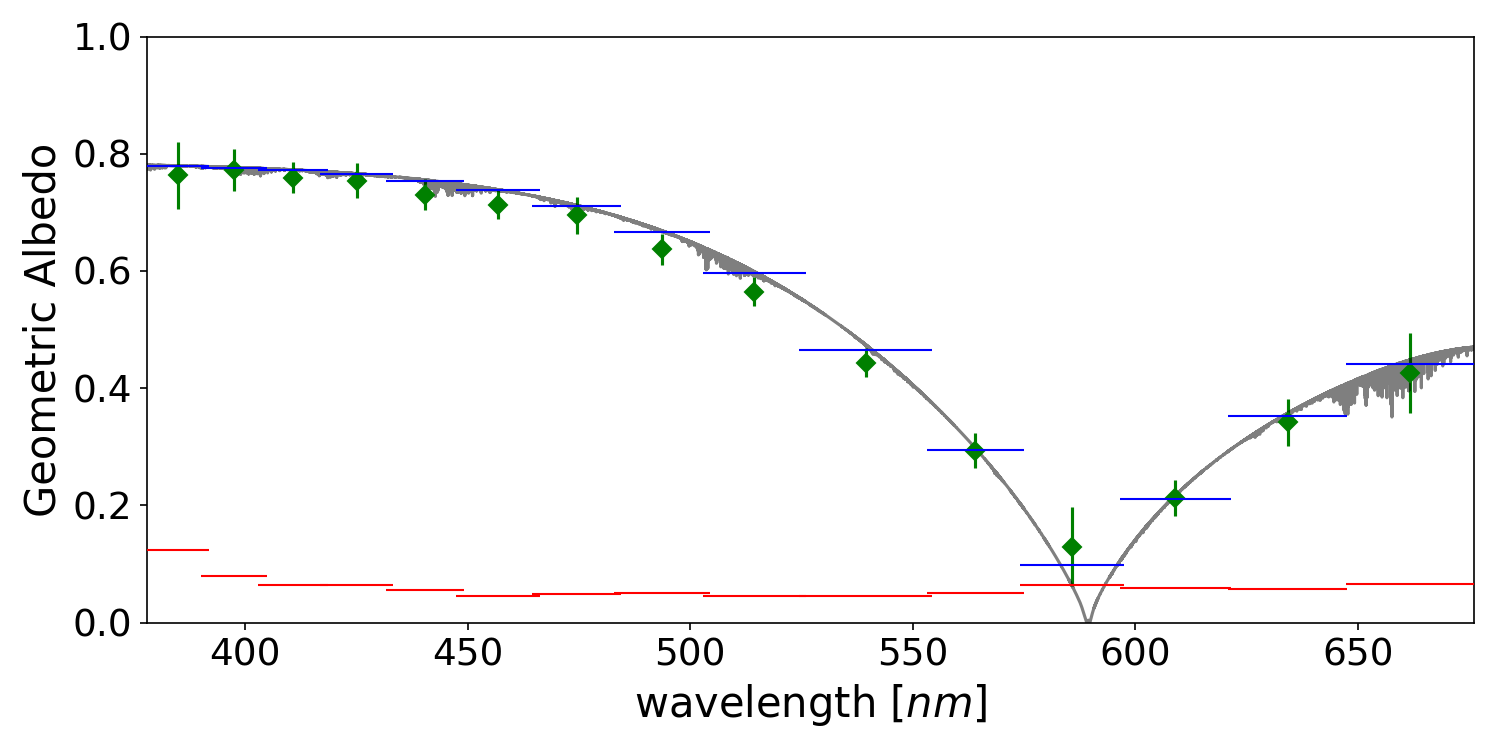}
	\caption{51 Peg b; Model A ($\times 100$); $\chi^2 = 3.97$}
\end{subfigure}%

\begin{subfigure}[t]{0.5\hsize}
	\includegraphics[width=\hsize]{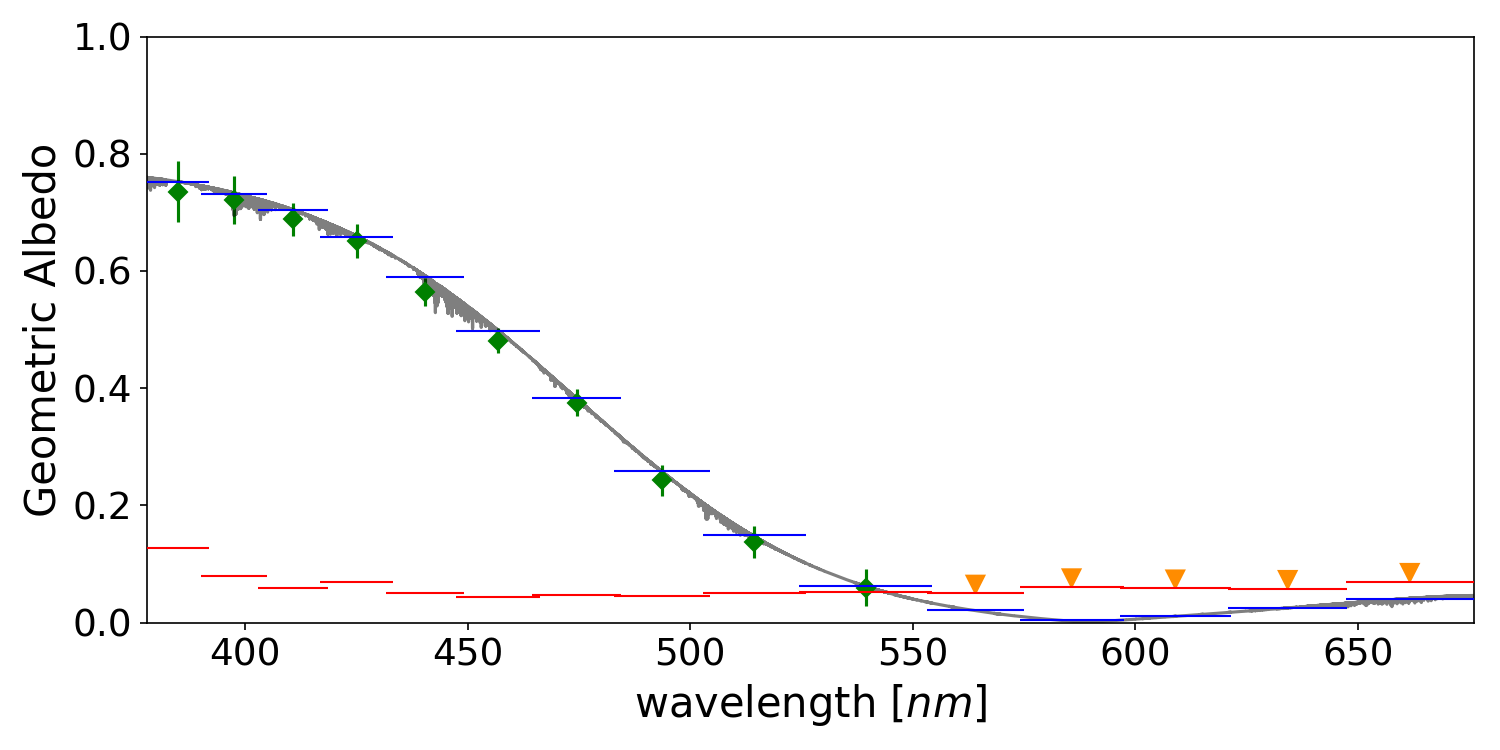}
	\caption{51 Peg b; Model A ($\times 1$); $\chi^2 = 1.94$}
\end{subfigure}%
\begin{subfigure}[t]{0.5\hsize}
	\includegraphics[width=\hsize]{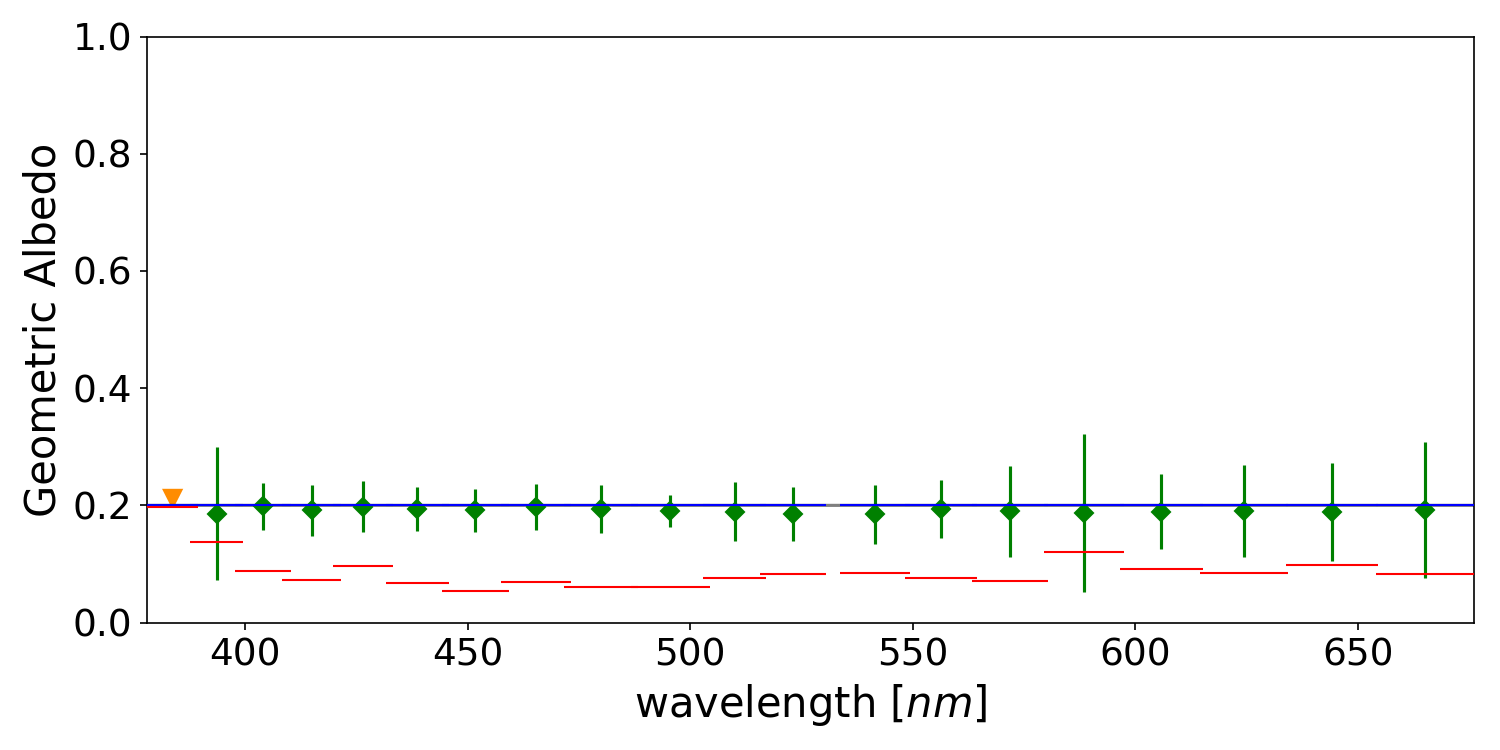}
	\caption{HD 209458 b; $A_g$ = 0.2; $\chi^2 = 0.29$}
\end{subfigure}%

\begin{subfigure}[t]{0.5\hsize}
	\includegraphics[width=\hsize]{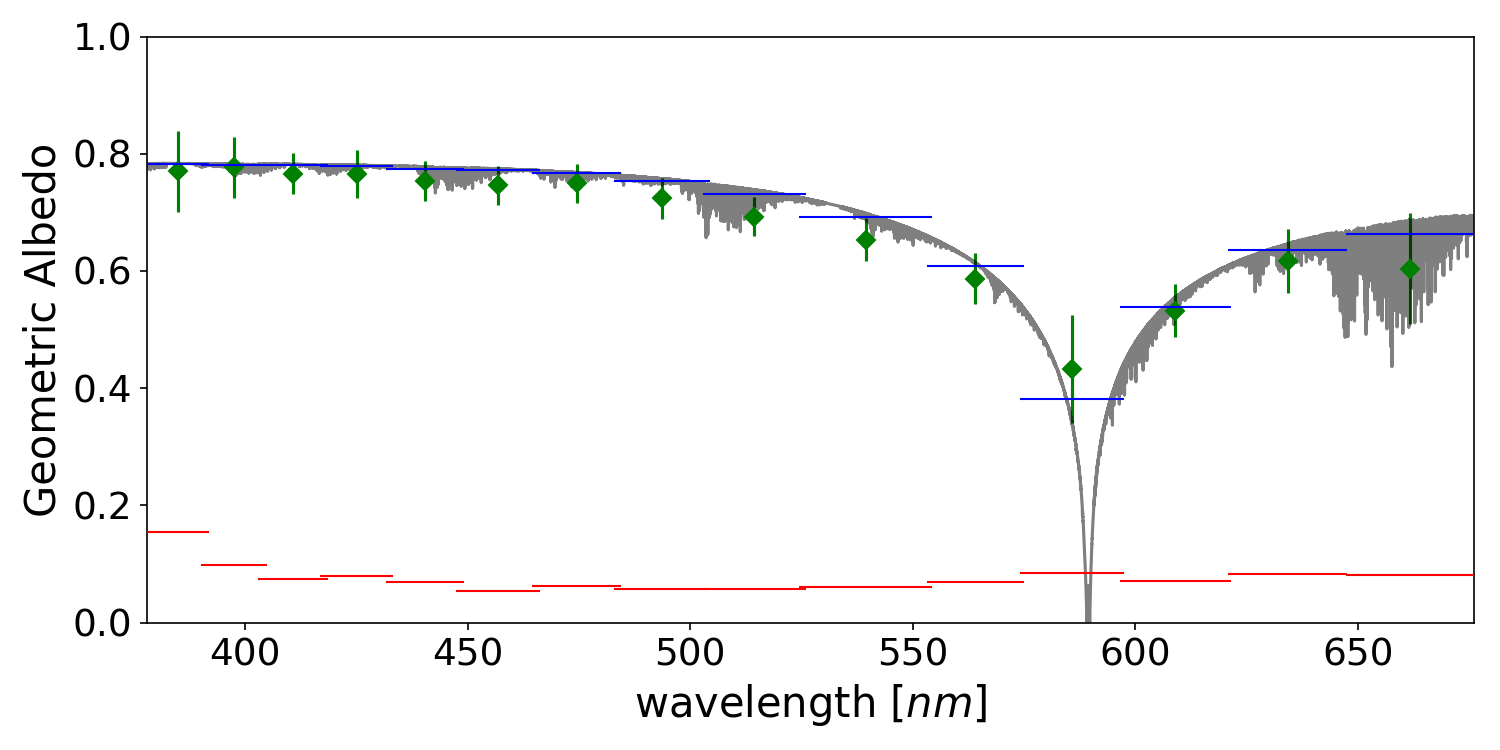}
	\caption{HD 209458 b; Model B ($\times 100$); $\chi^2 = 3.51$}
\end{subfigure}%
\begin{subfigure}[t]{0.5\hsize}
	\includegraphics[width=\hsize]{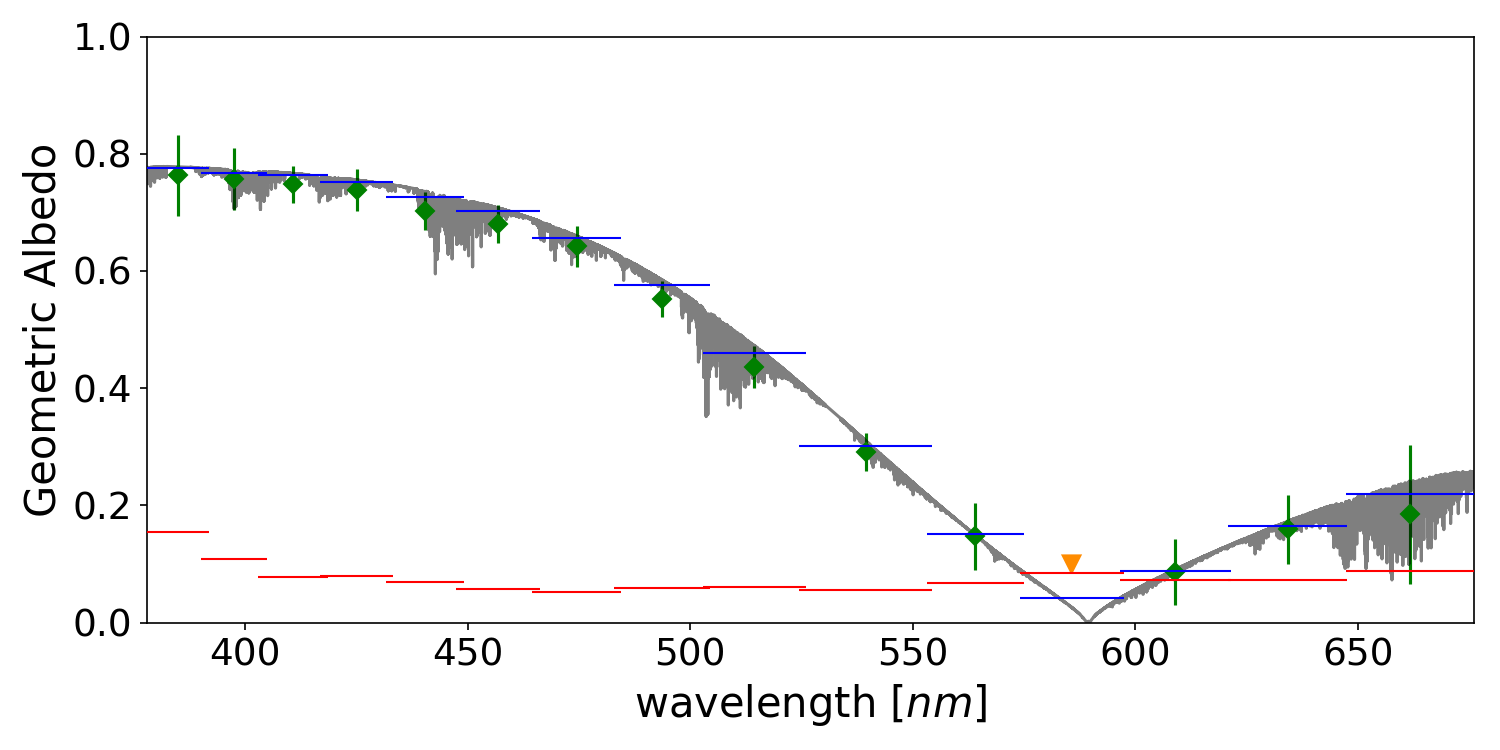}
	\caption{HD 209458 b; Model B ($\times 1$); $\chi^2 = 1.72$}
\end{subfigure}%

\caption{Distribution of the recovered albedo functions from the simulated ESPRESSO (MHR mode - 1UT) observations. For each wavelength bin: \textit{i)} the green dots represent the mean recovered albedo over the 100 simulated runs with error bars given by the 3 times the standard deviation; \textit{ii)}the blue horizontal lines represent the mean albedo of the simulated model over the bin; \textit{iii)} the red horizontal bar represents the $3\sigma$ detection limit of the albedo.}
	\label{fig:espresso_mhr}
	\end{figure*}

\clearpage\begin{figure*}
	\begin{subfigure}[t]{0.5\hsize}
	\includegraphics[width=\hsize]{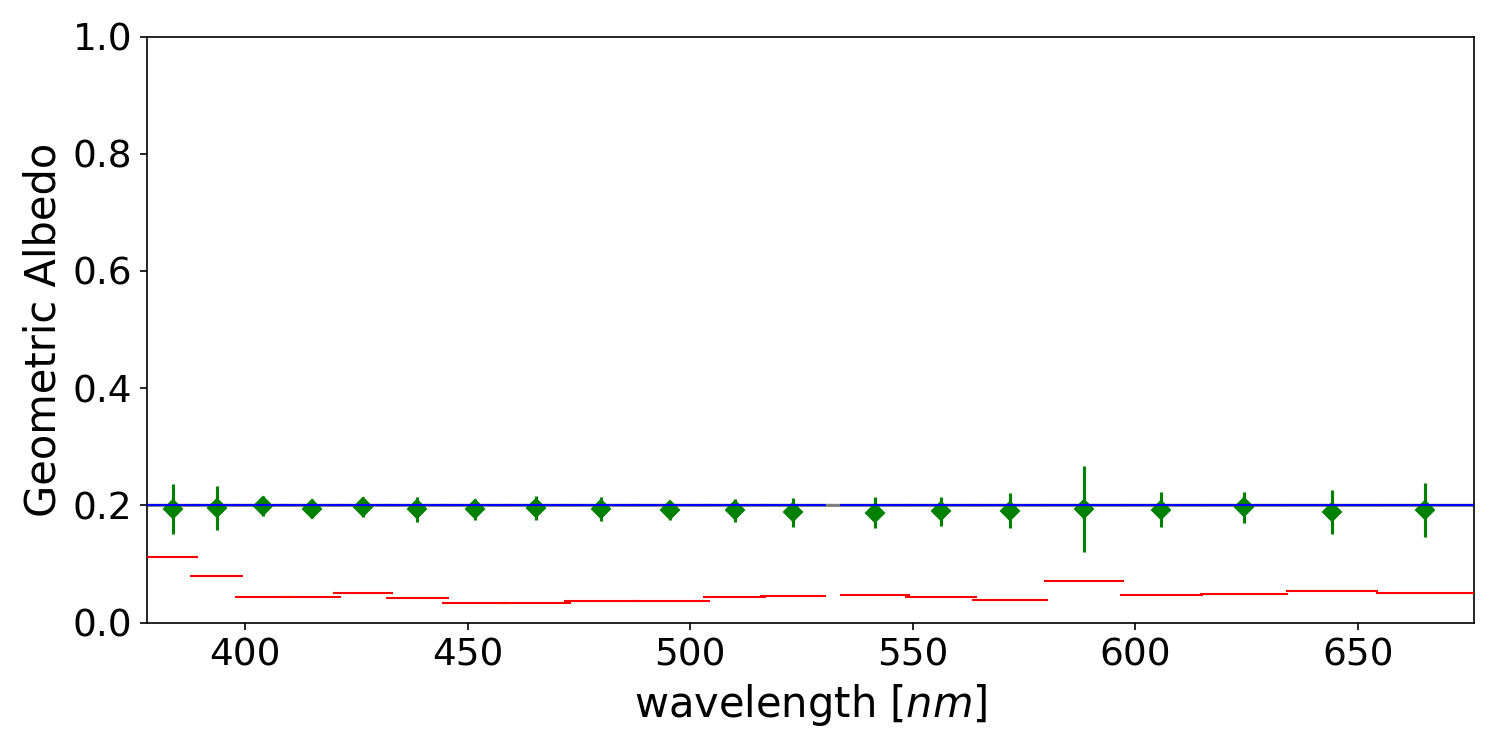}
	\caption{51 Peg b; $A_g$ = 0.2; $\chi^2 = 0.89$}
\end{subfigure}%
\begin{subfigure}[t]{0.5\hsize}
	\includegraphics[width=\hsize]{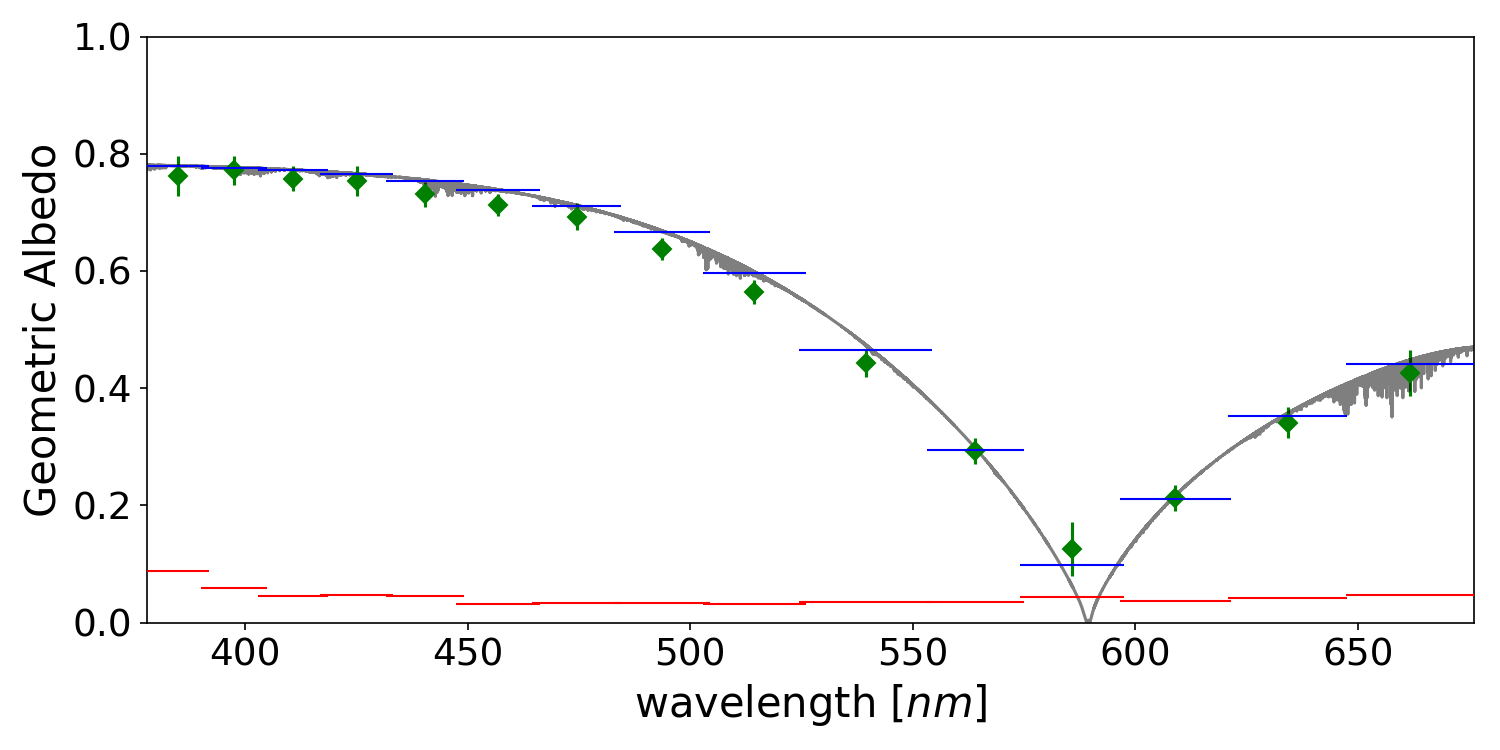}
	\caption{51 Peg b; Model A ($\times 100$); $\chi^2 = 6.75$}
\end{subfigure}%

\begin{subfigure}[t]{0.5\hsize}
	\includegraphics[width=\hsize]{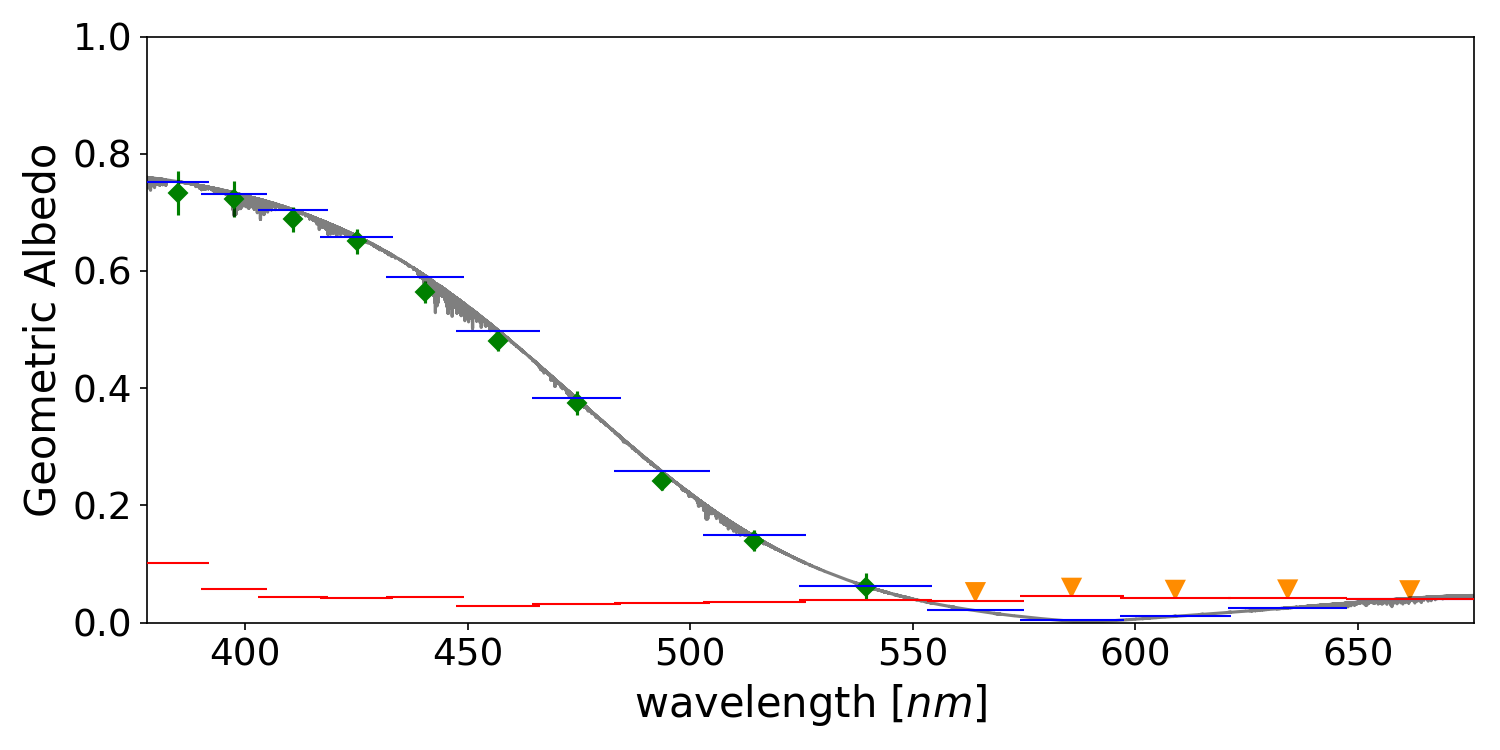}
	\caption{51 Peg b; Model A ($\times 1$); $\chi^2 = 3.45$}
\end{subfigure}%
\begin{subfigure}[t]{0.5\hsize}
	\includegraphics[width=\hsize]{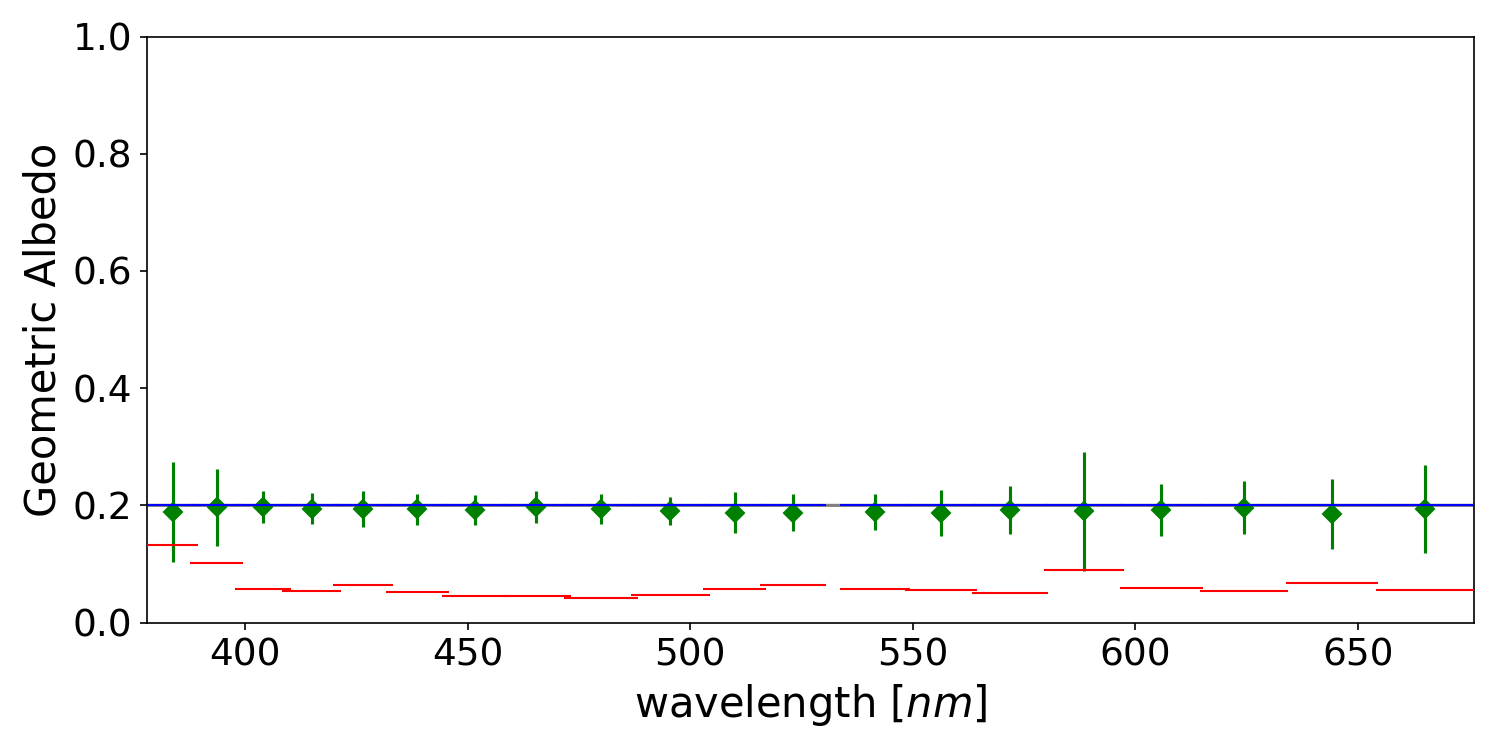}
	\caption{HD 209458 b; $A_g$ = 0.2; $\chi^2 = 0.54$}
\end{subfigure}%

\begin{subfigure}[t]{0.5\hsize}
	\includegraphics[width=\hsize]{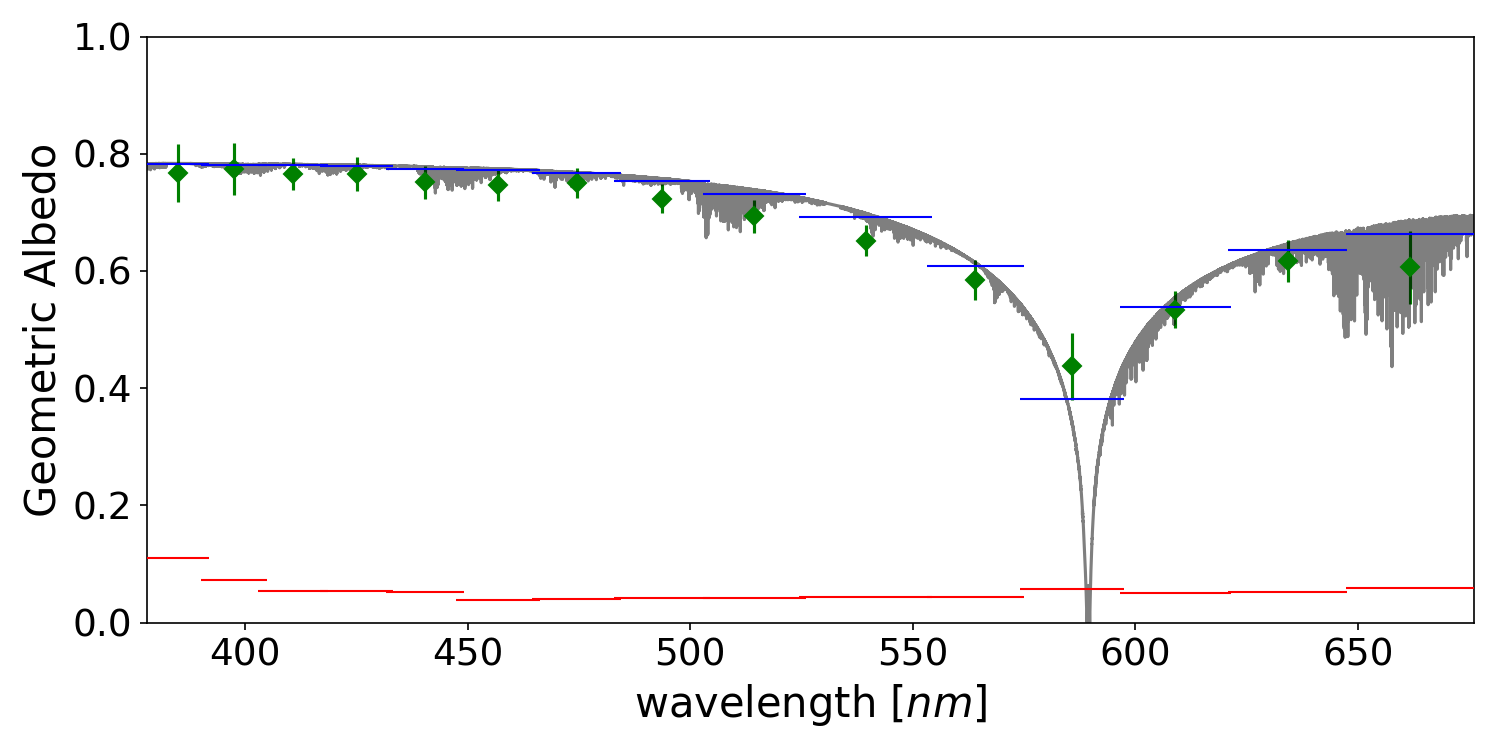}
	\caption{HD 209458 b; Model B ($\times 100$); $\chi^2 = 6.68$}
\end{subfigure}%
\begin{subfigure}[t]{0.5\hsize}
	\includegraphics[width=\hsize]{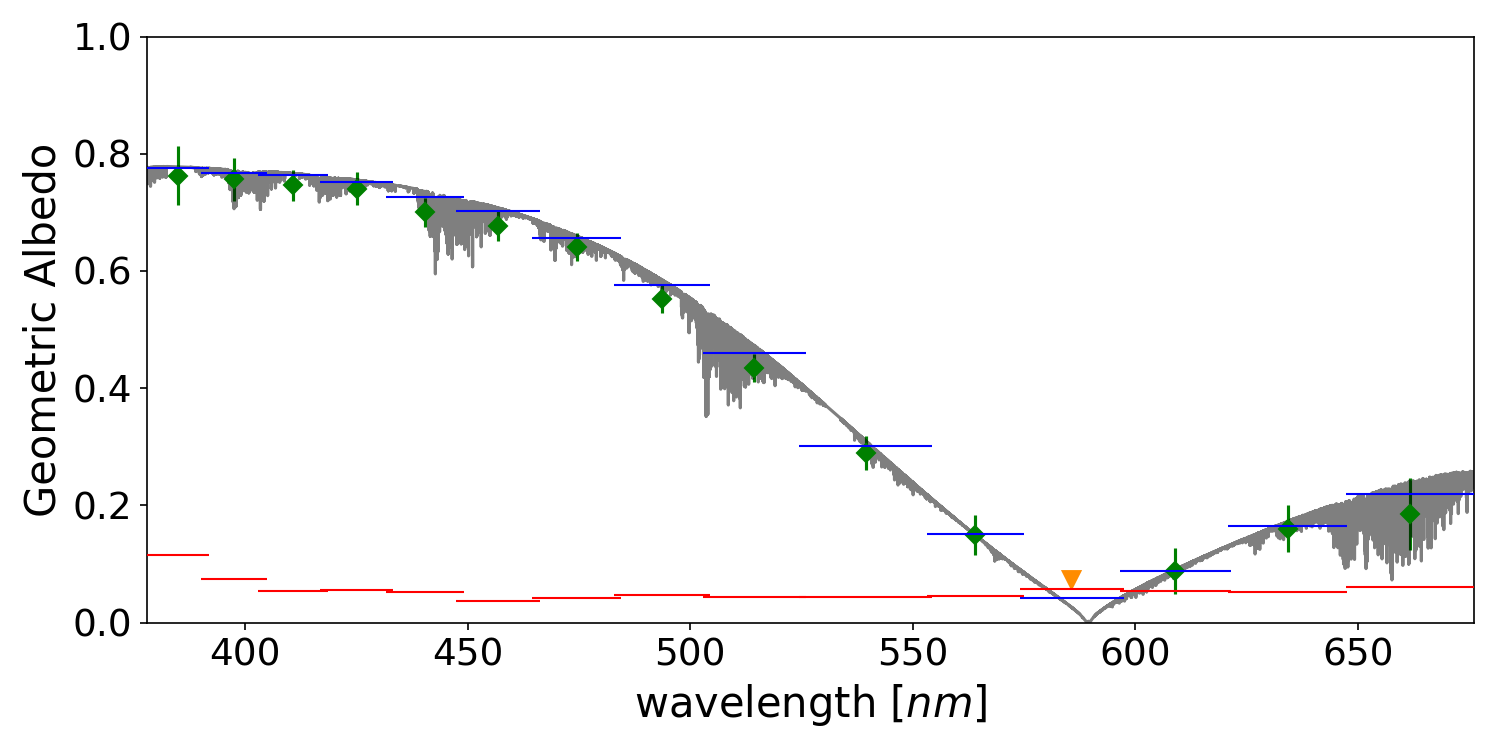}
	\caption{HD 209458 b; Model B ($\times 1$); $\chi^2 = 3.44$}
\end{subfigure}%

\caption{Distribution of the recovered albedo functions from the simulated ESPRESSO (MHR mode - 2UT) observations. For each wavelength bin: \textit{i)} the green dots represent the mean recovered albedo over the 100 simulated runs with error bars given by the 3 times the standard deviation; \textit{ii)}the blue horizontal lines represent the mean albedo of the simulated model over the bin; \textit{iii)} the red horizontal bar represents the 3$\sigma$ detection limit of the albedo.}
	\label{fig:espresso_mhr_2ut}
	\end{figure*}

\clearpage\begin{figure*}
	\begin{subfigure}[t]{0.5\hsize}
	\includegraphics[width=\hsize]{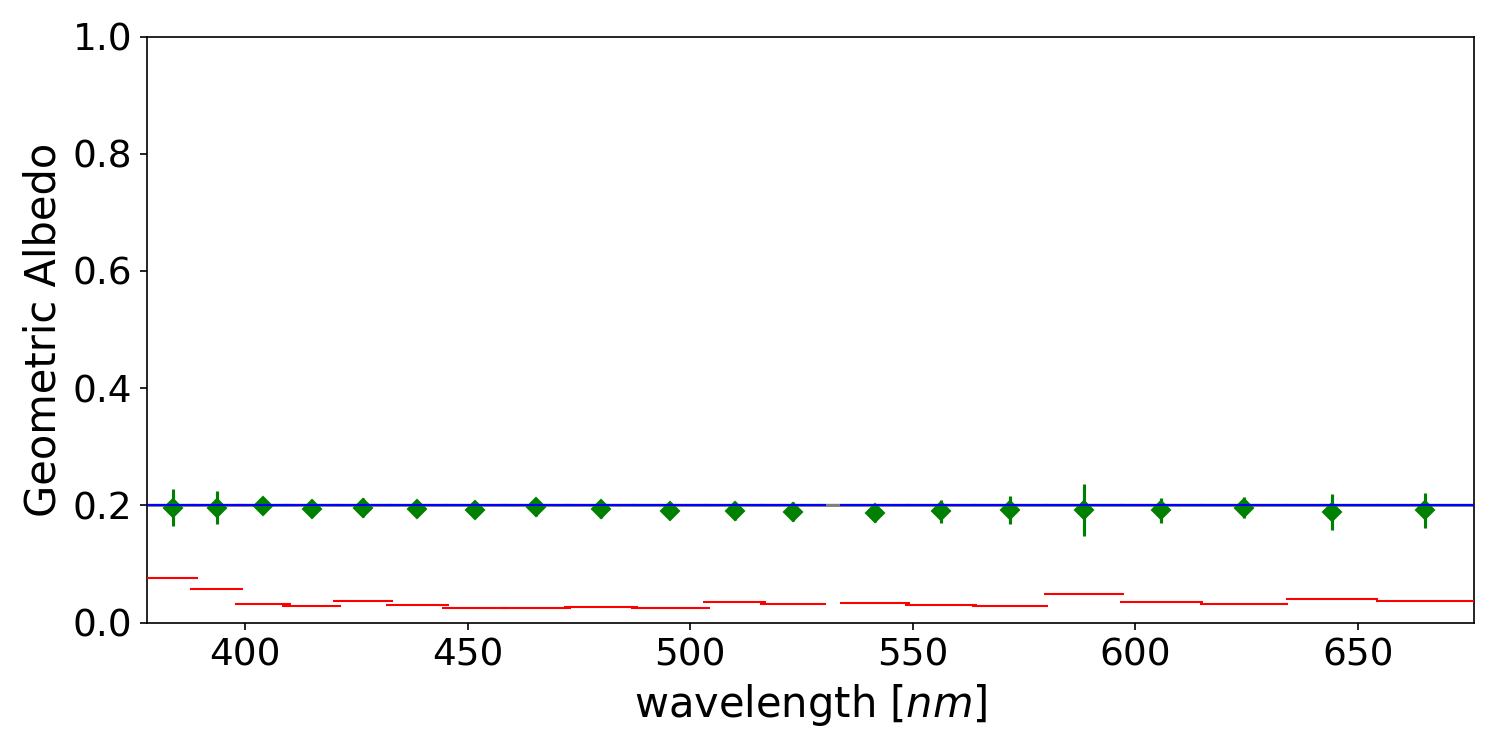}
	\caption{51 Peg b; $A_g$ = 0.2; $\chi^2 = 1.91$}
\end{subfigure}%
\begin{subfigure}[t]{0.5\hsize}
	\includegraphics[width=\hsize]{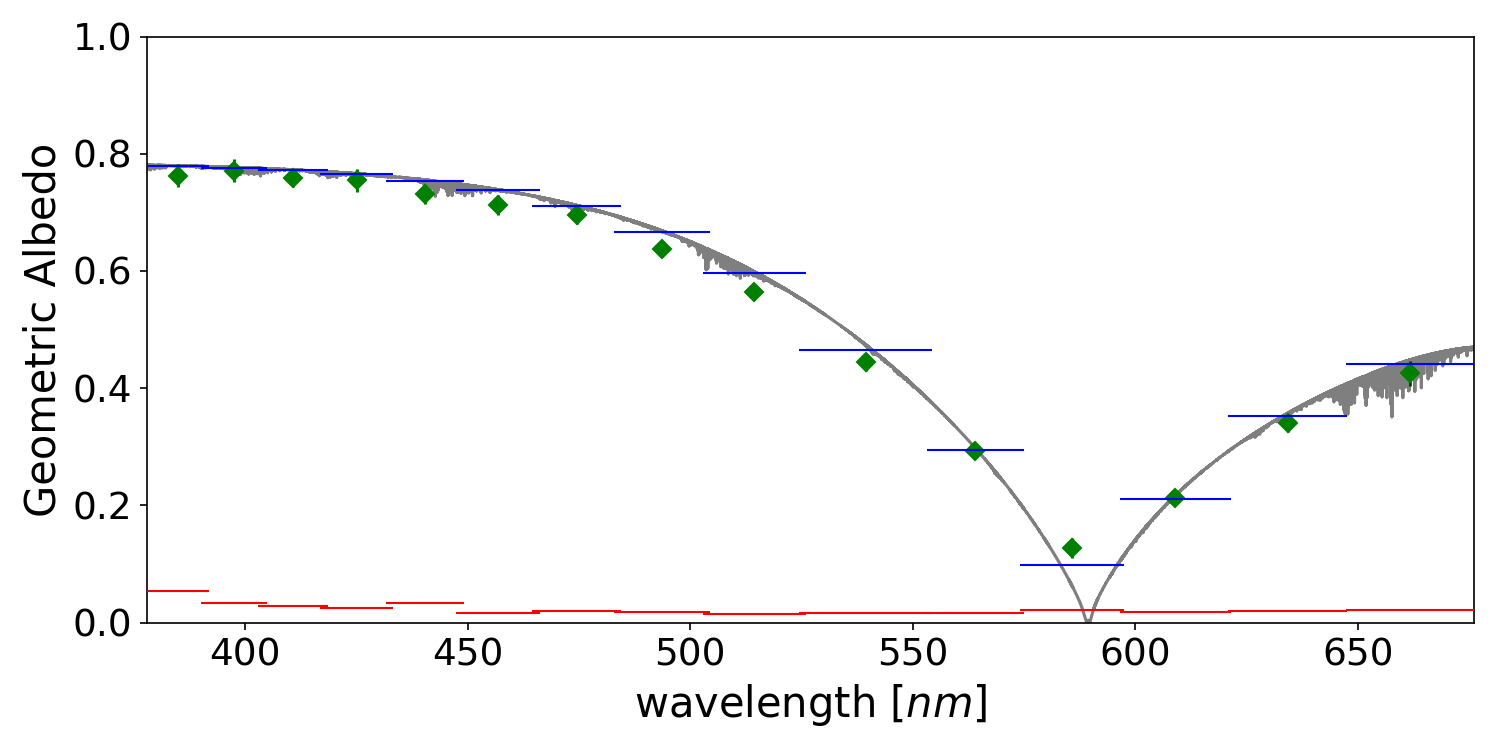}
	\caption{51 Peg b; Model A ($\times 100$); $\chi^2 = 14.87$}
\end{subfigure}%

\begin{subfigure}[t]{0.5\hsize}
	\includegraphics[width=\hsize]{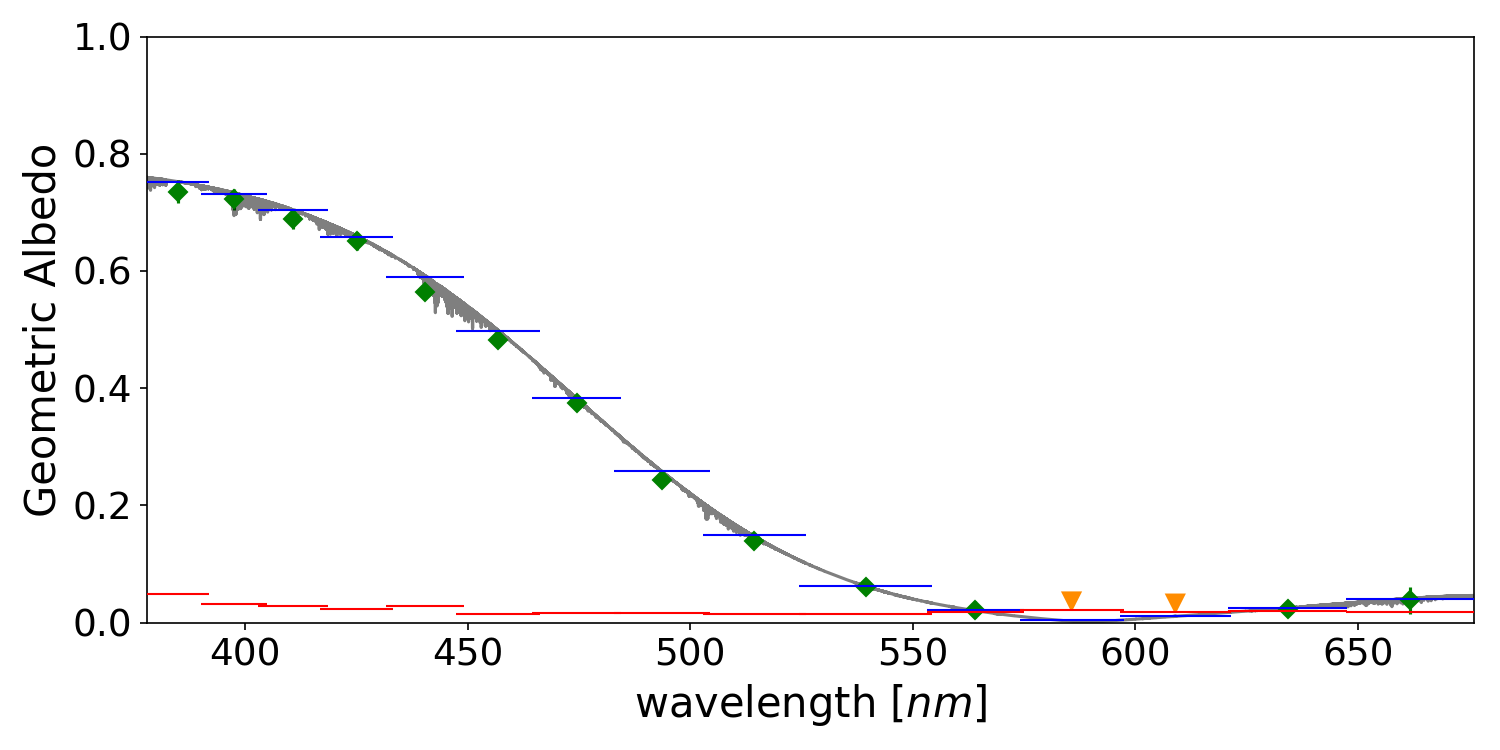}
	\caption{51 Peg b; Model A ($\times 1$); $\chi^2 = 8.11$}
\end{subfigure}%
\begin{subfigure}[t]{0.5\hsize}
	\includegraphics[width=\hsize]{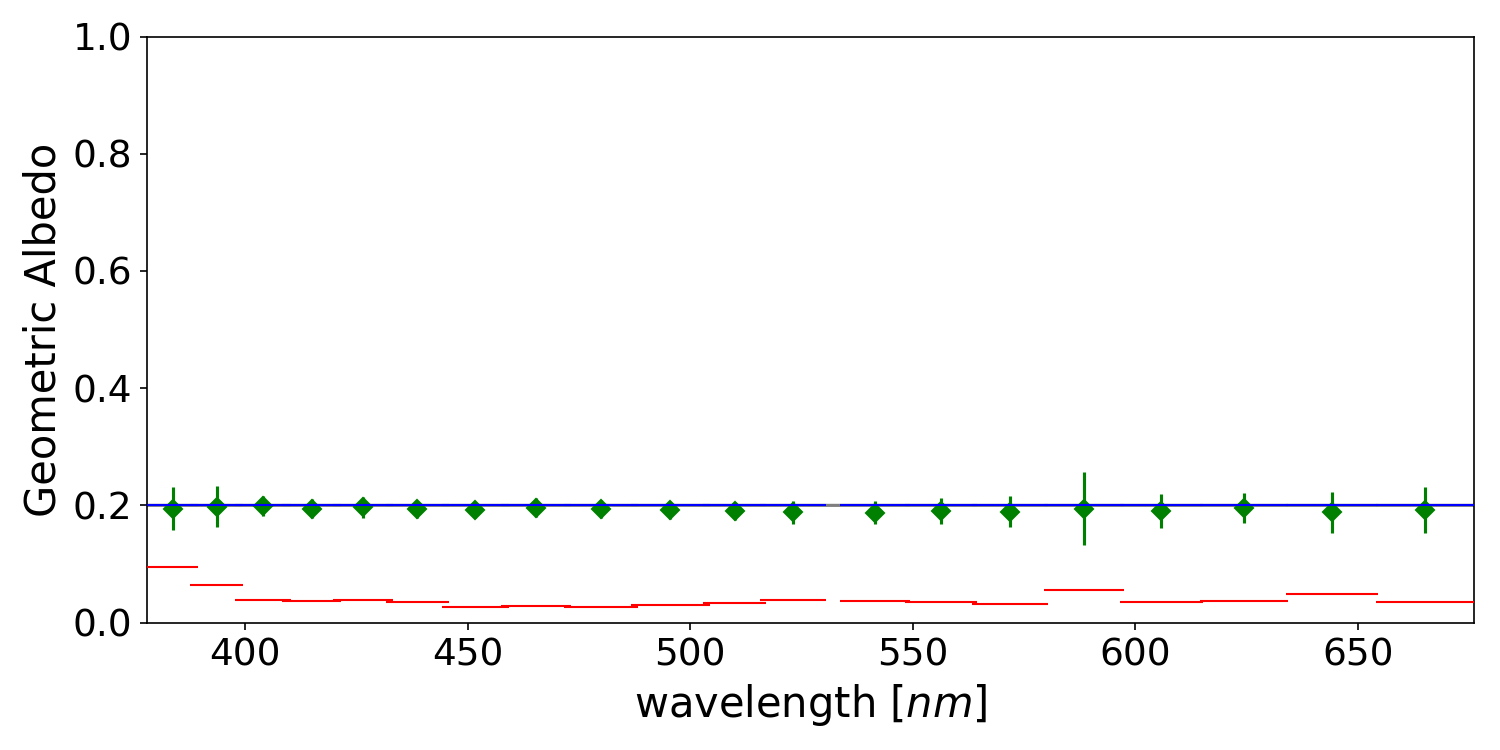}
	\caption{HD 209458 b; $A_g$ = 0.2; $\chi^2 = 1.42$}
\end{subfigure}%

\begin{subfigure}[t]{0.5\hsize}
	\includegraphics[width=\hsize]{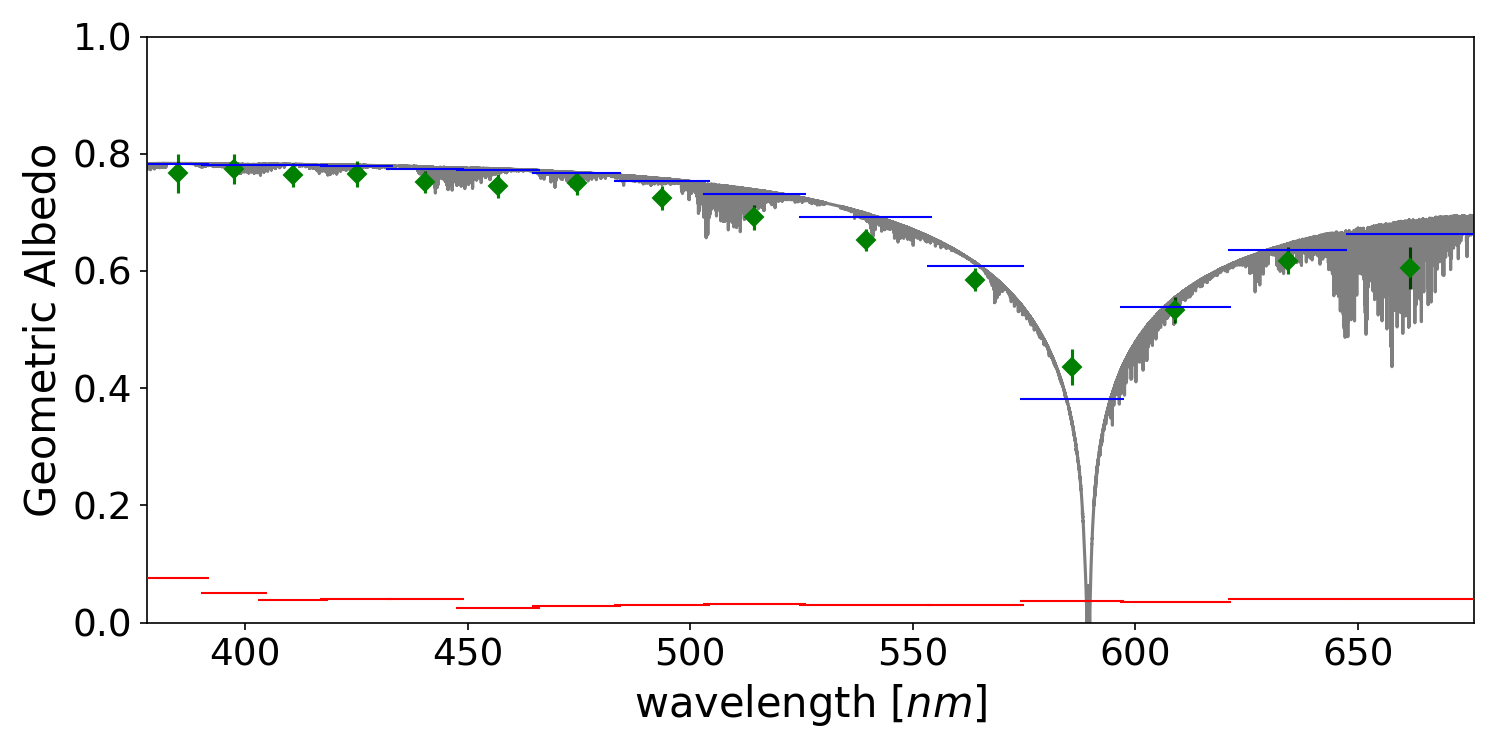}
	\caption{HD 209458 b; Model B ($\times 100$); $\chi^2 = 13.79$}
\end{subfigure}%
\begin{subfigure}[t]{0.5\hsize}
	\includegraphics[width=\hsize]{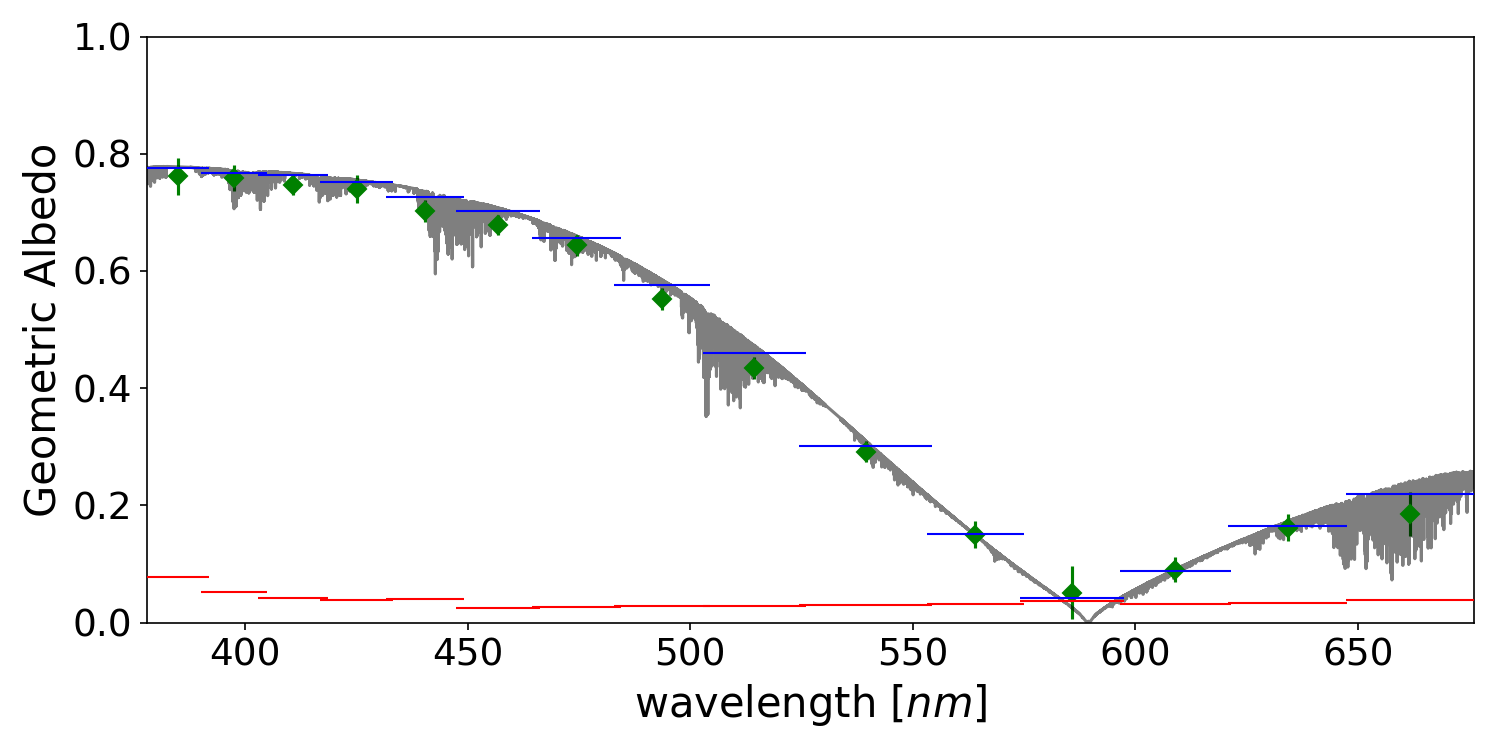}
	\caption{HD 209458 b; Model B ($\times 1$); $\chi^2 = 5.98$}
\end{subfigure}%

\caption{Distribution of the recovered albedo functions from the simulated ESPRESSO (MHR mode - 4UT) observations. For each wavelength bin: \textit{i)} the green dots represent the mean recovered albedo over the 100 simulated runs with error bars given by the 3 times the standard deviation; \textit{ii)}the blue horizontal lines represent the mean albedo of the simulated model over the bin; \textit{iii)} the red horizontal bar represents the $3\sigma$ detection limit of the albedo.}
	\label{fig:espresso_mhr_4ut}
	\end{figure*}

\clearpage\begin{figure*}
	\begin{subfigure}[t]{0.5\hsize}
	\includegraphics[width=\hsize]{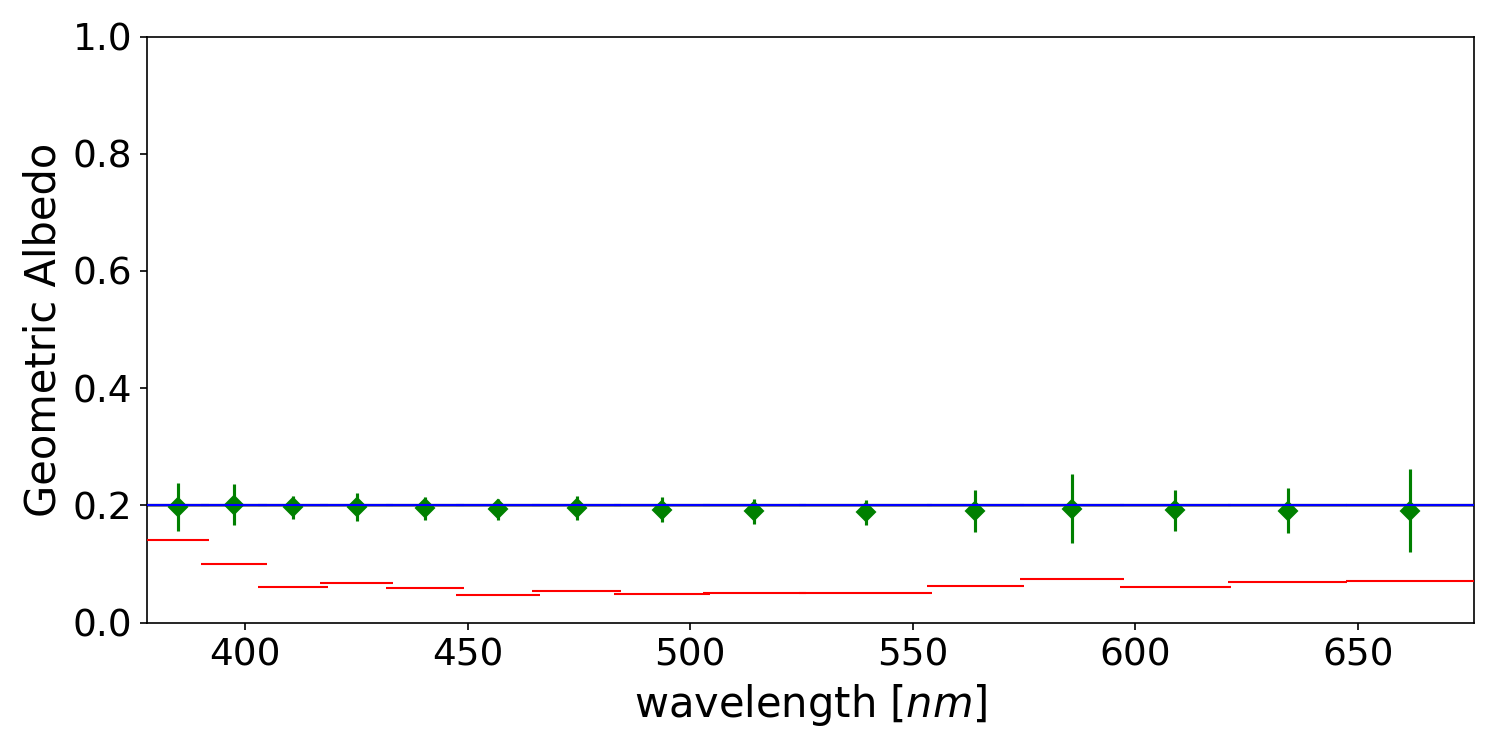}
	\caption{51 Peg b; $A_g$ = 0.2; $\chi^2 = 0.73$}
\end{subfigure}%
\begin{subfigure}[t]{0.5\hsize}
	\includegraphics[width=\hsize]{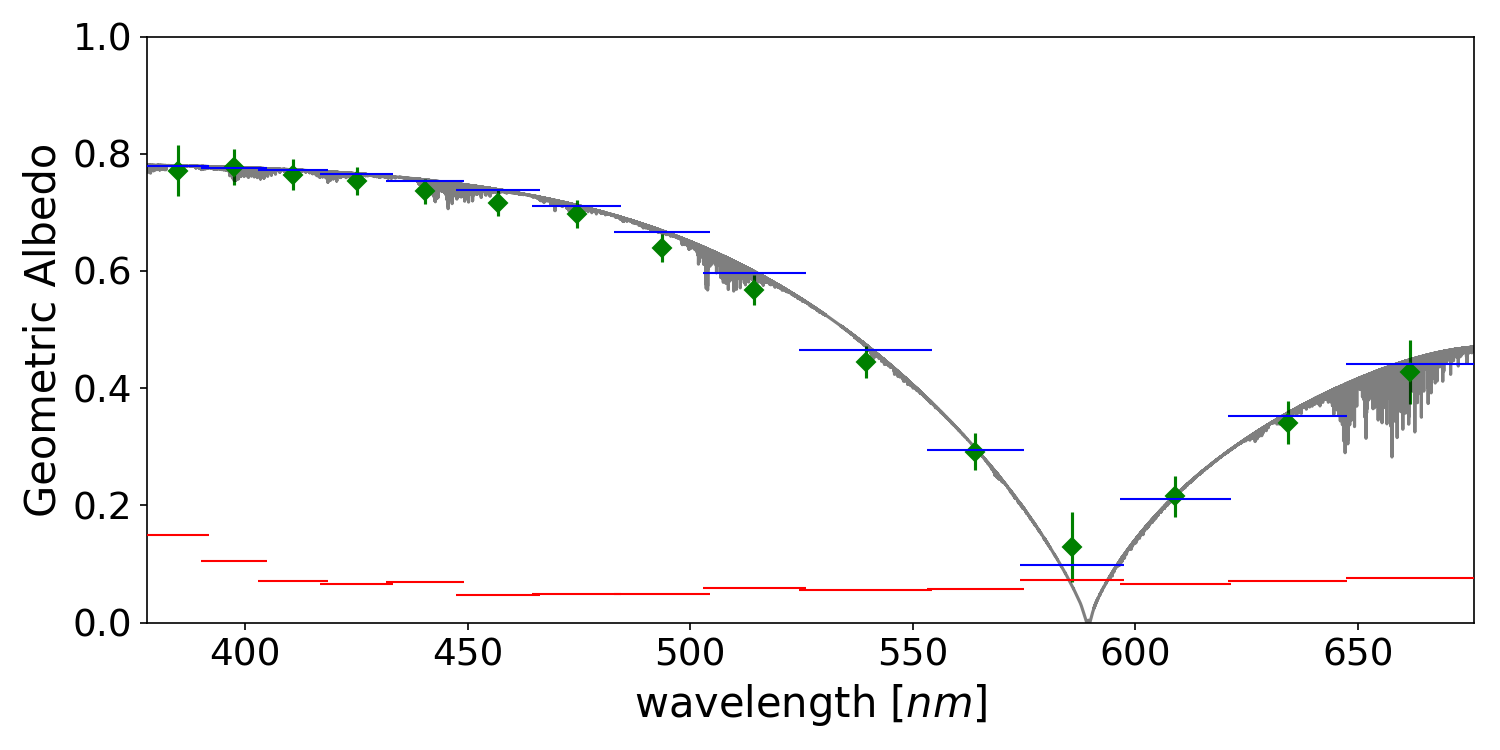}
	\caption{51 Peg b; Model A ($\times 100$); $\chi^2 = 3.6$}
\end{subfigure}%

\begin{subfigure}[t]{0.5\hsize}
	\includegraphics[width=\hsize]{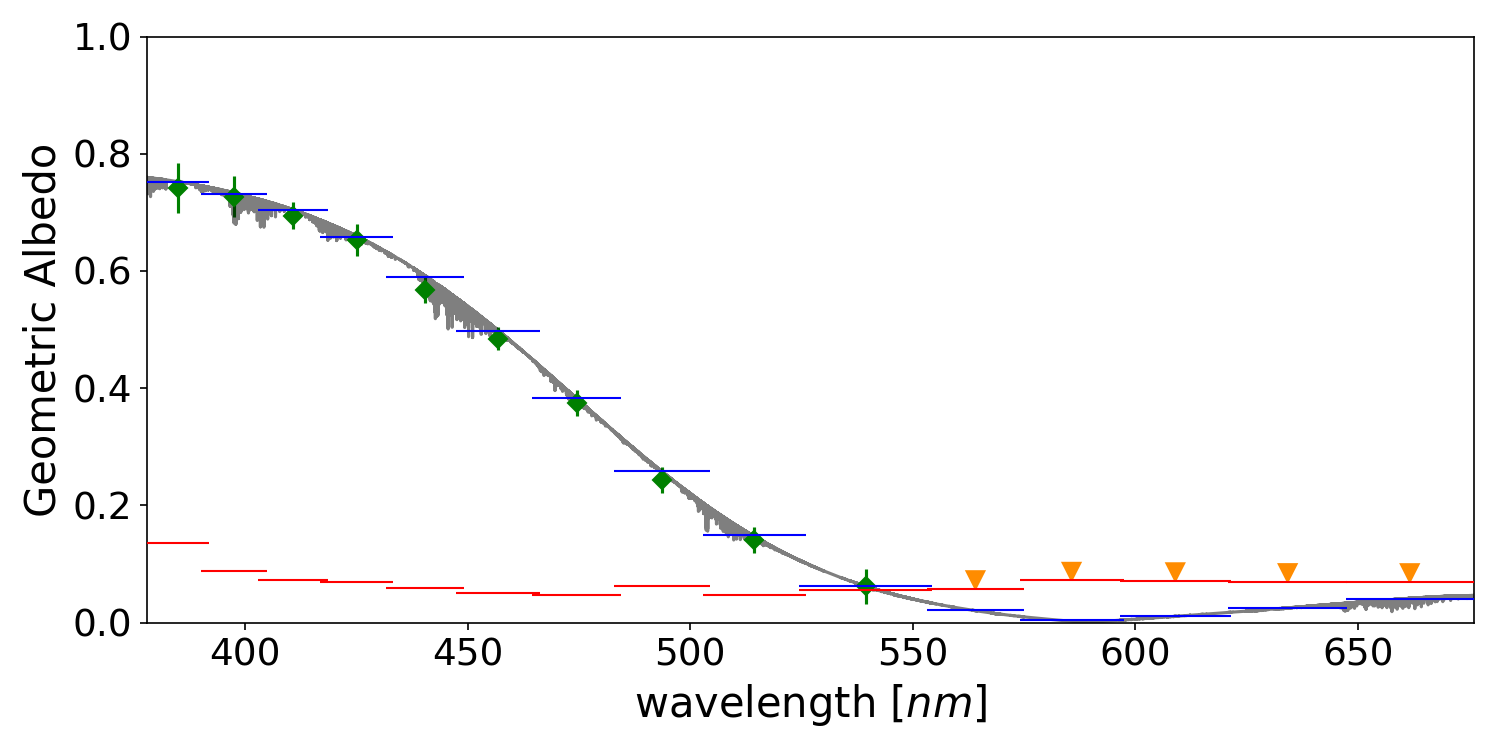}
	\caption{51 Peg b; Model A ($\times 1$); $\chi^2 = 1.61$}
\end{subfigure}%
\begin{subfigure}[t]{0.5\hsize}
	\includegraphics[width=\hsize]{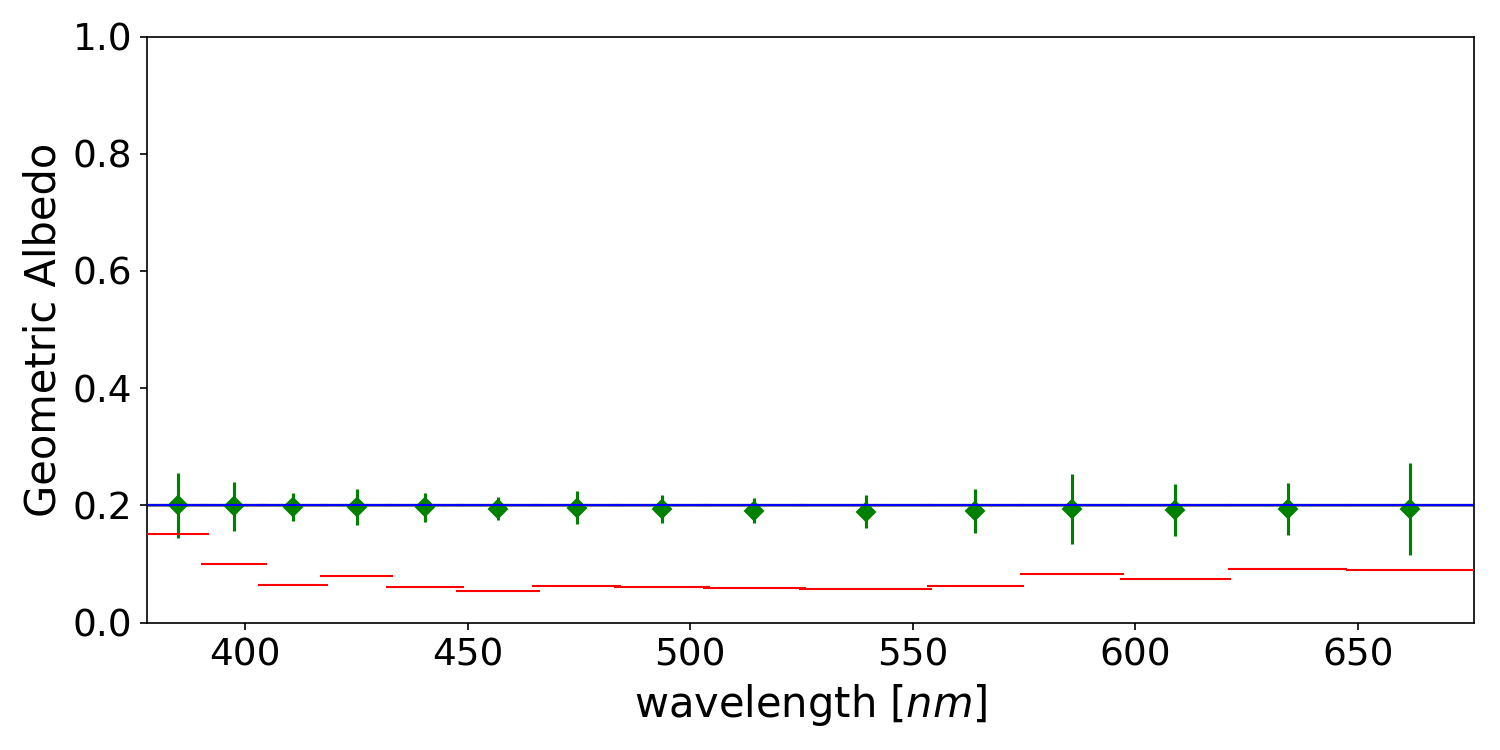}
	\caption{HD 209458 b; $A_g$ = 0.2; $\chi^2 = 0.41$}
\end{subfigure}%

\begin{subfigure}[t]{0.5\hsize}
	\includegraphics[width=\hsize]{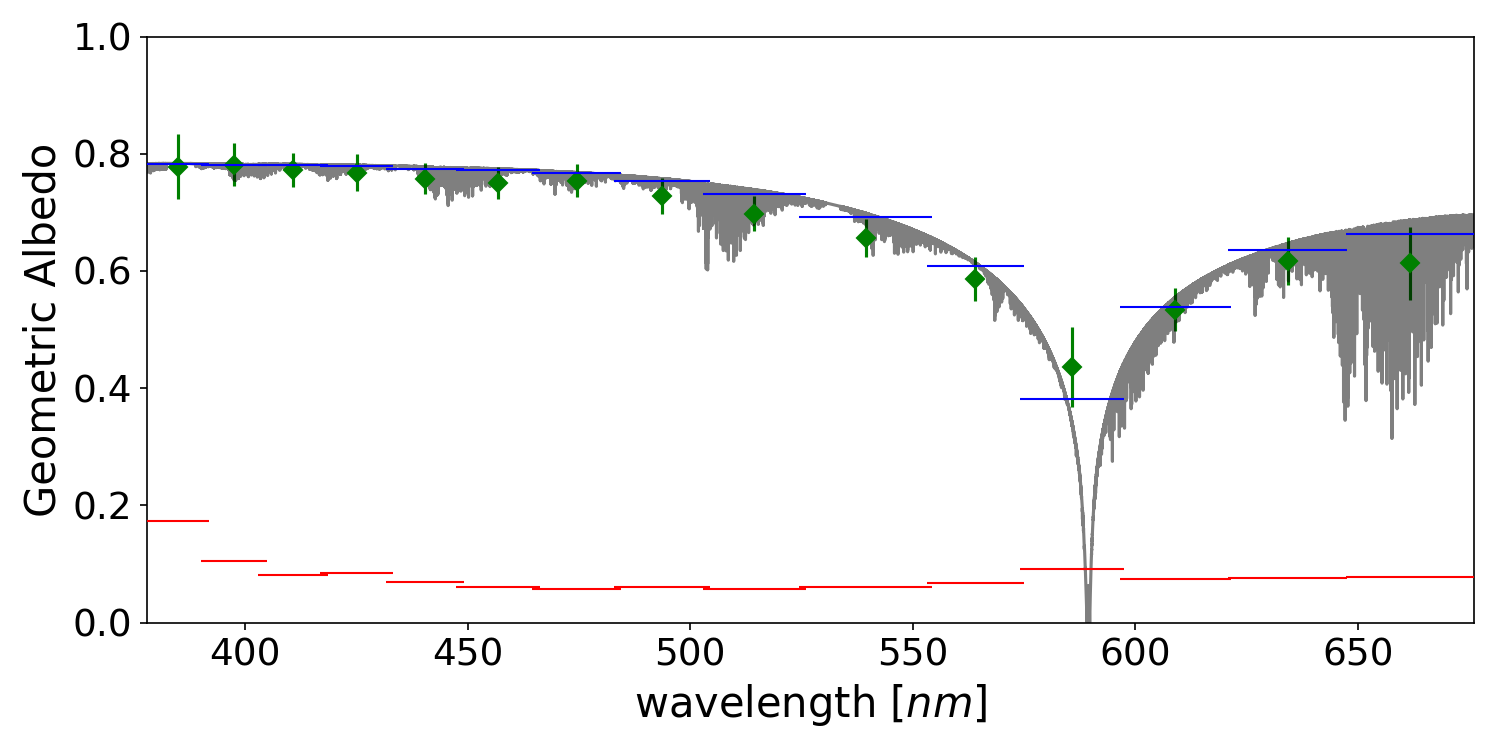}
	\caption{HD 209458 b; Model B ($\times 100$); $\chi^2 = 3.96$}
\end{subfigure}%
\begin{subfigure}[t]{0.5\hsize}
	\includegraphics[width=\hsize]{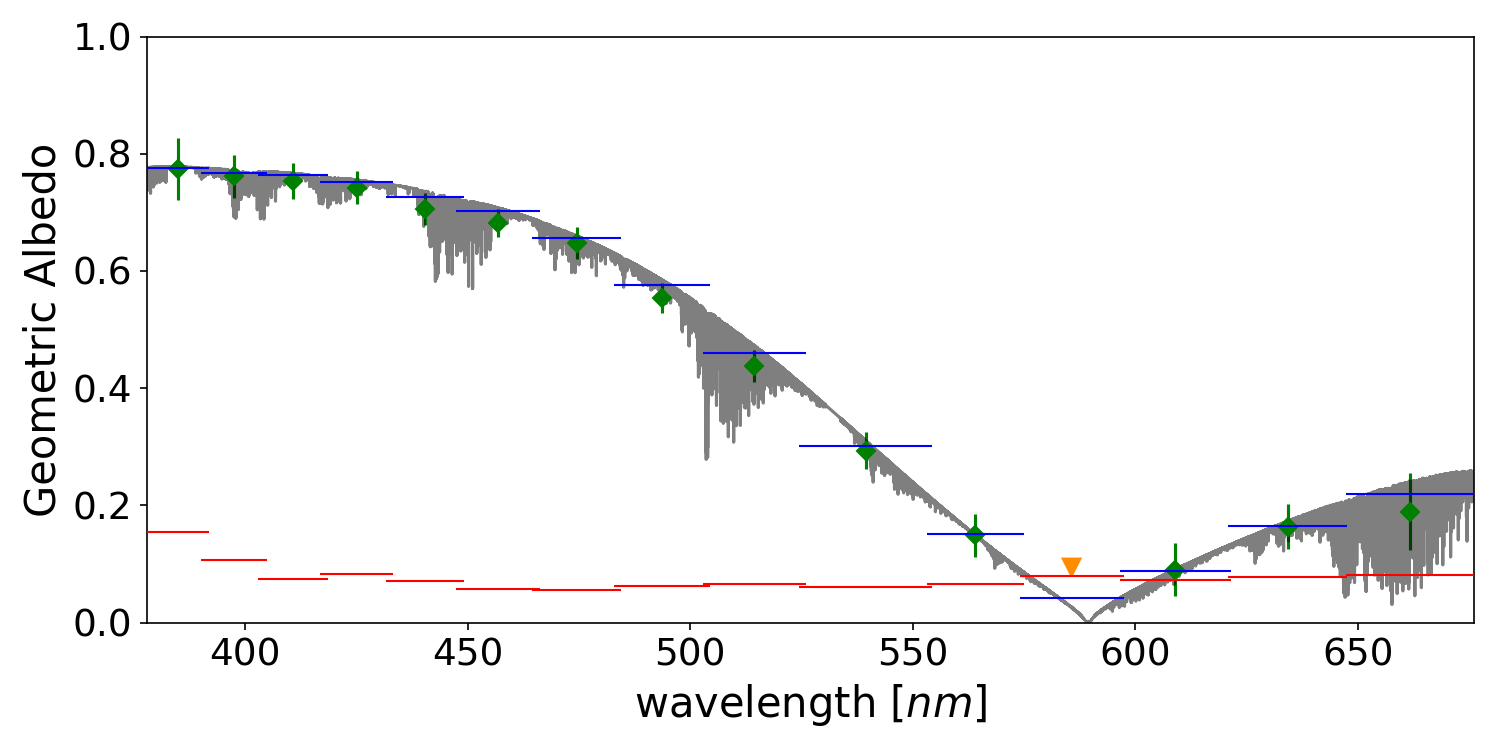}
	\caption{HD 209458 b; Model B ($\times 1$); $\chi^2 = 1.99$}
\end{subfigure}%

\caption{Distribution of the recovered albedo functions from the simulated ESPRESSO (UHR mode) observations. For each wavelength bin: \textit{i)} the green dots represent the mean recovered albedo over the 100 simulated runs with error bars given by the 3 times the standard deviation; \textit{ii)}the blue horizontal lines represent the mean albedo of the simulated model over the bin; \textit{iii)} the red horizontal bar represents the $3\sigma$ detection limit of the albedo.}
	\label{fig:espresso_uhr}
	\end{figure*}

\clearpage\begin{figure*}
	\begin{subfigure}[t]{0.5\hsize}
	\includegraphics[width=\hsize]{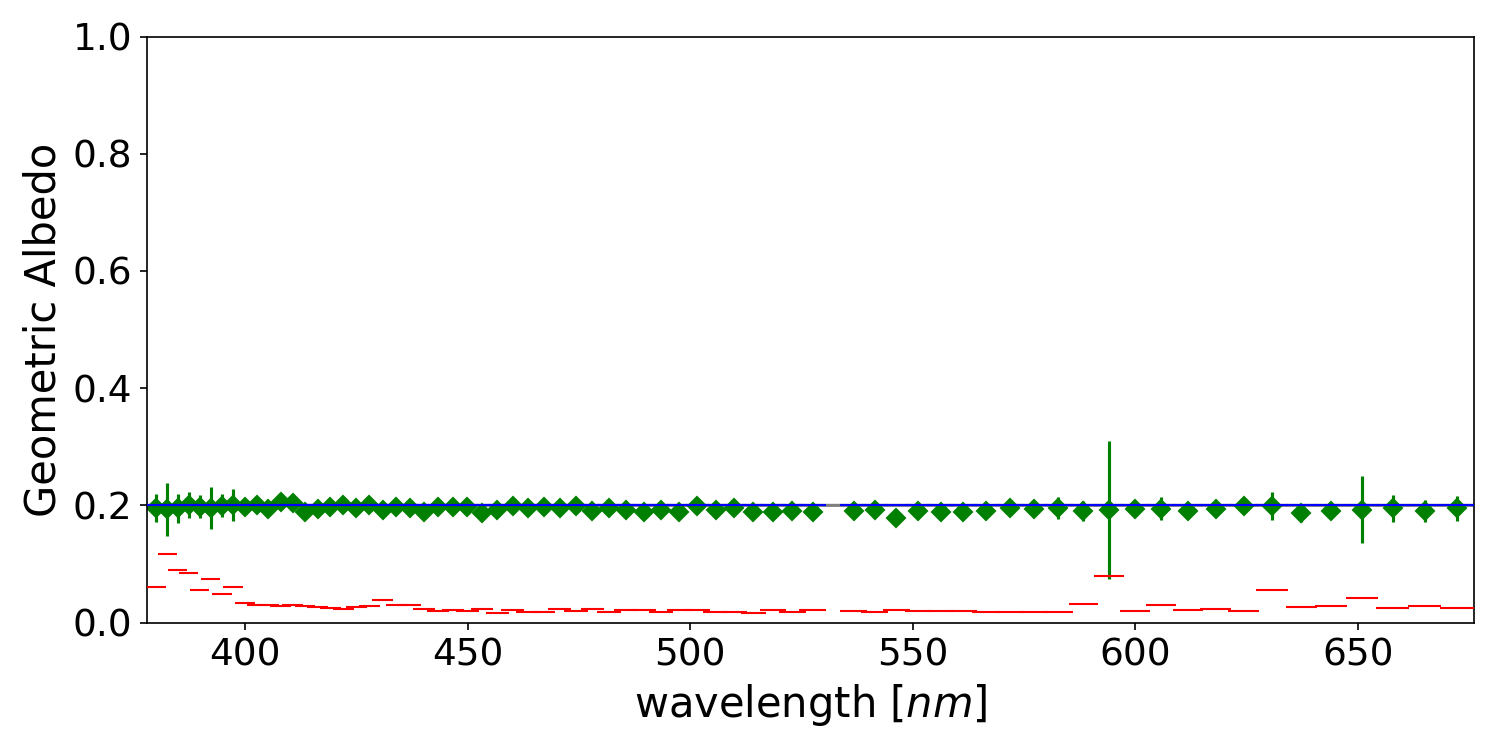}
	\caption{51 Peg b; $A_g$ = 0.2; $\chi^2 = 5.38$}
\end{subfigure}%
\begin{subfigure}[t]{0.5\hsize}
	\includegraphics[width=\hsize]{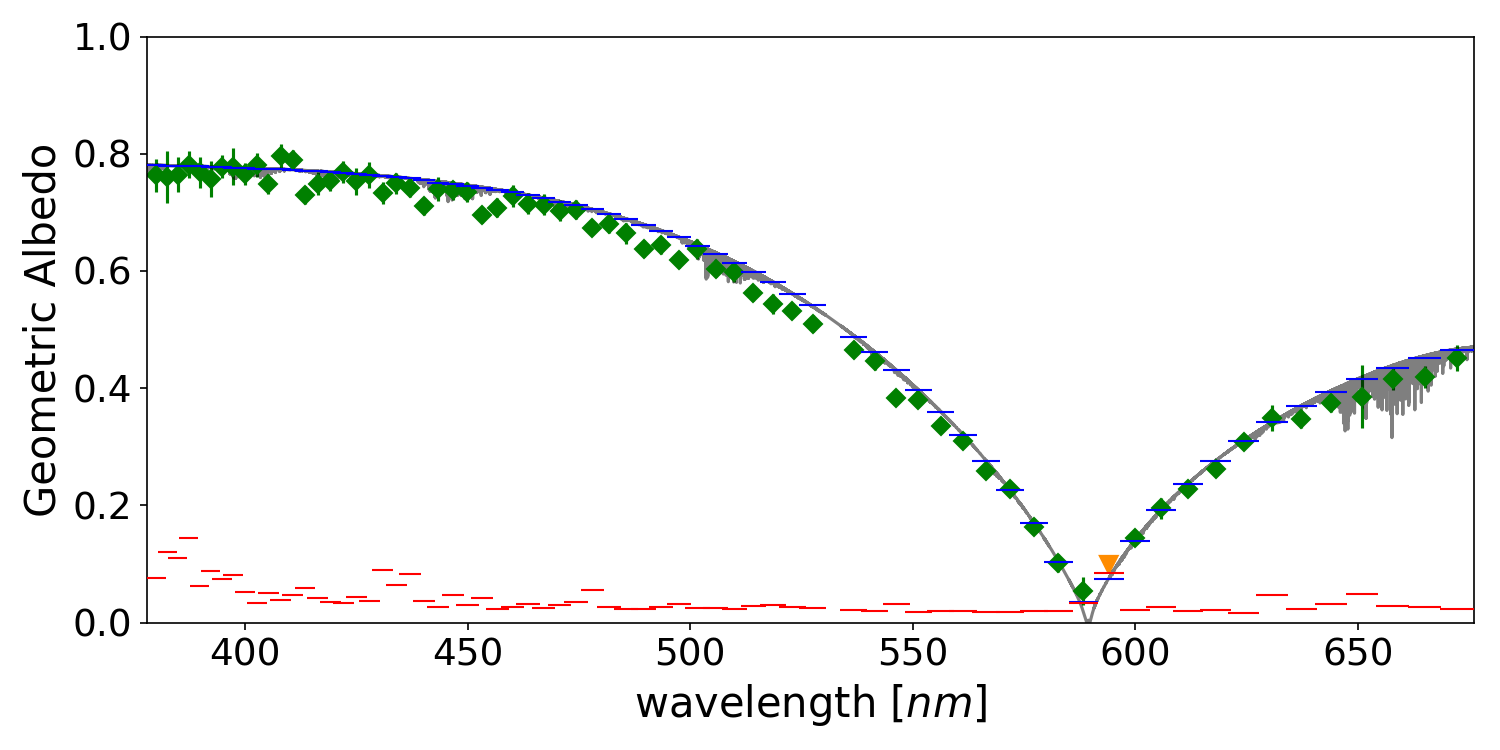}
	\caption{51 Peg b; Model A ($\times 100$); $\chi^2 = 19.77$}
\end{subfigure}%

\begin{subfigure}[t]{0.5\hsize}
	\includegraphics[width=\hsize]{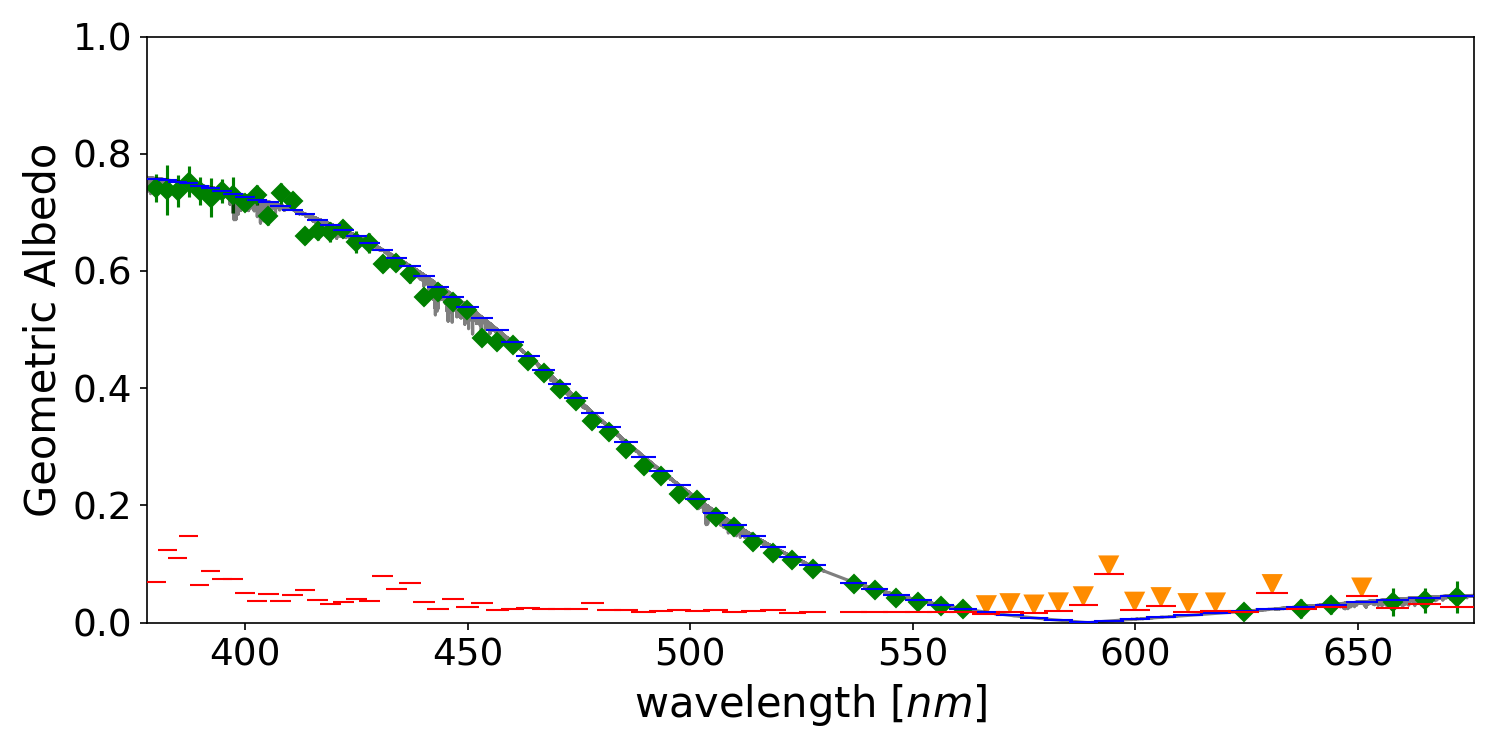}
	\caption{51 Peg b; Model A ($\times 1$); $\chi^2 = 7.84$}
\end{subfigure}%
\begin{subfigure}[t]{0.5\hsize}
	\includegraphics[width=\hsize]{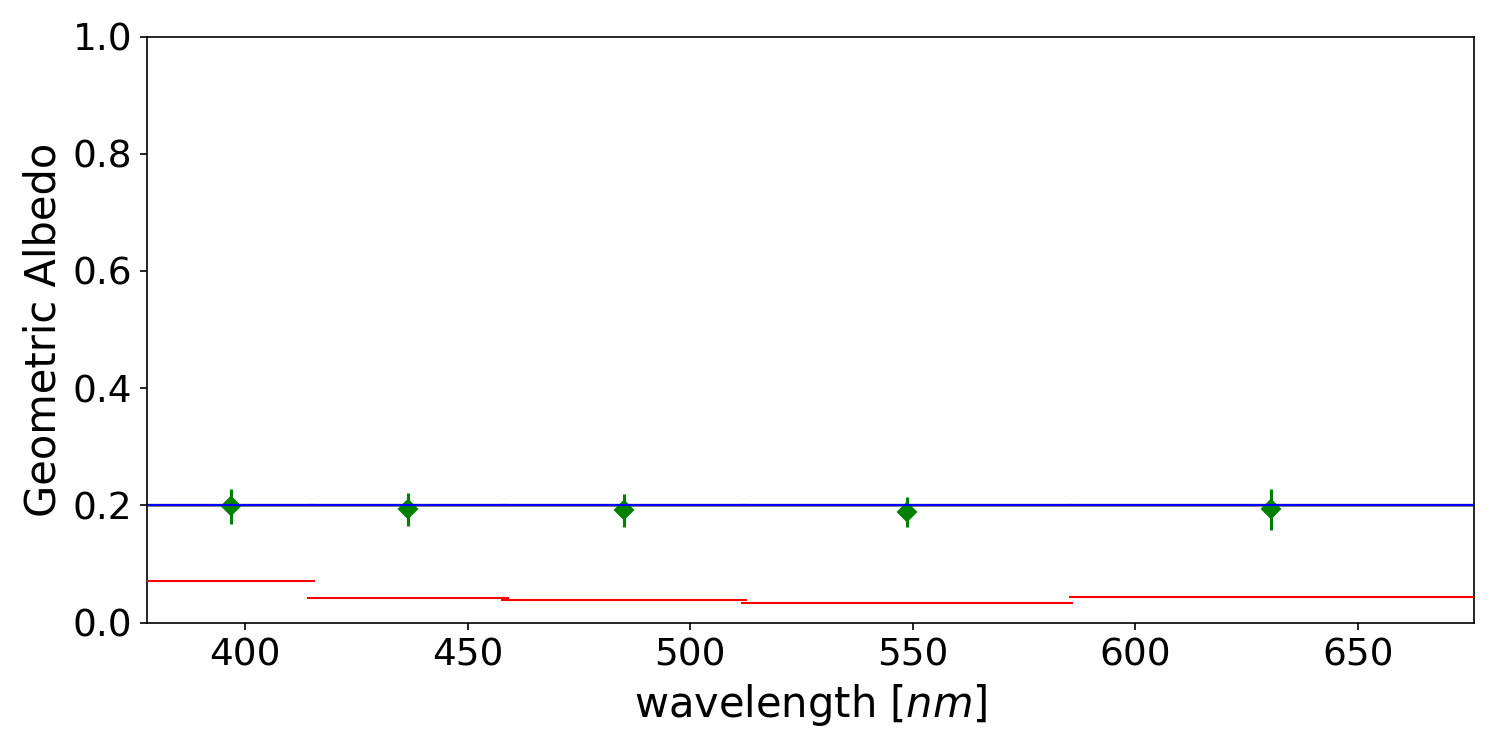}
	\caption{55 Cnc e; $A_g$ = 0.2; $\chi^2 = 0.67$}
\end{subfigure}%

\begin{subfigure}[t]{0.5\hsize}
	\includegraphics[width=\hsize]{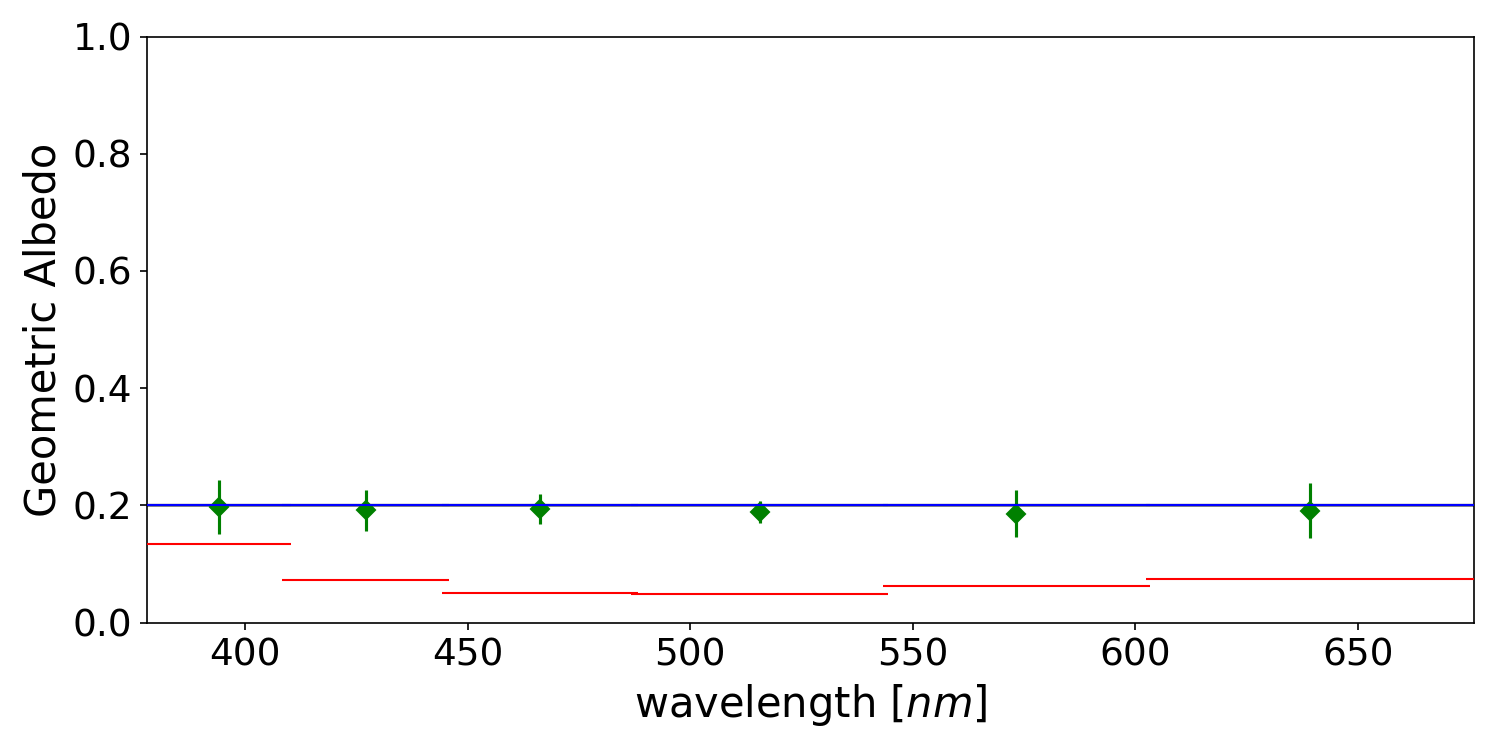}
	\caption{HD 109749 b; $A_g$ = 0.2; $\chi^2 = 1.05$}
\end{subfigure}%
\begin{subfigure}[t]{0.5\hsize}
	\includegraphics[width=\hsize]{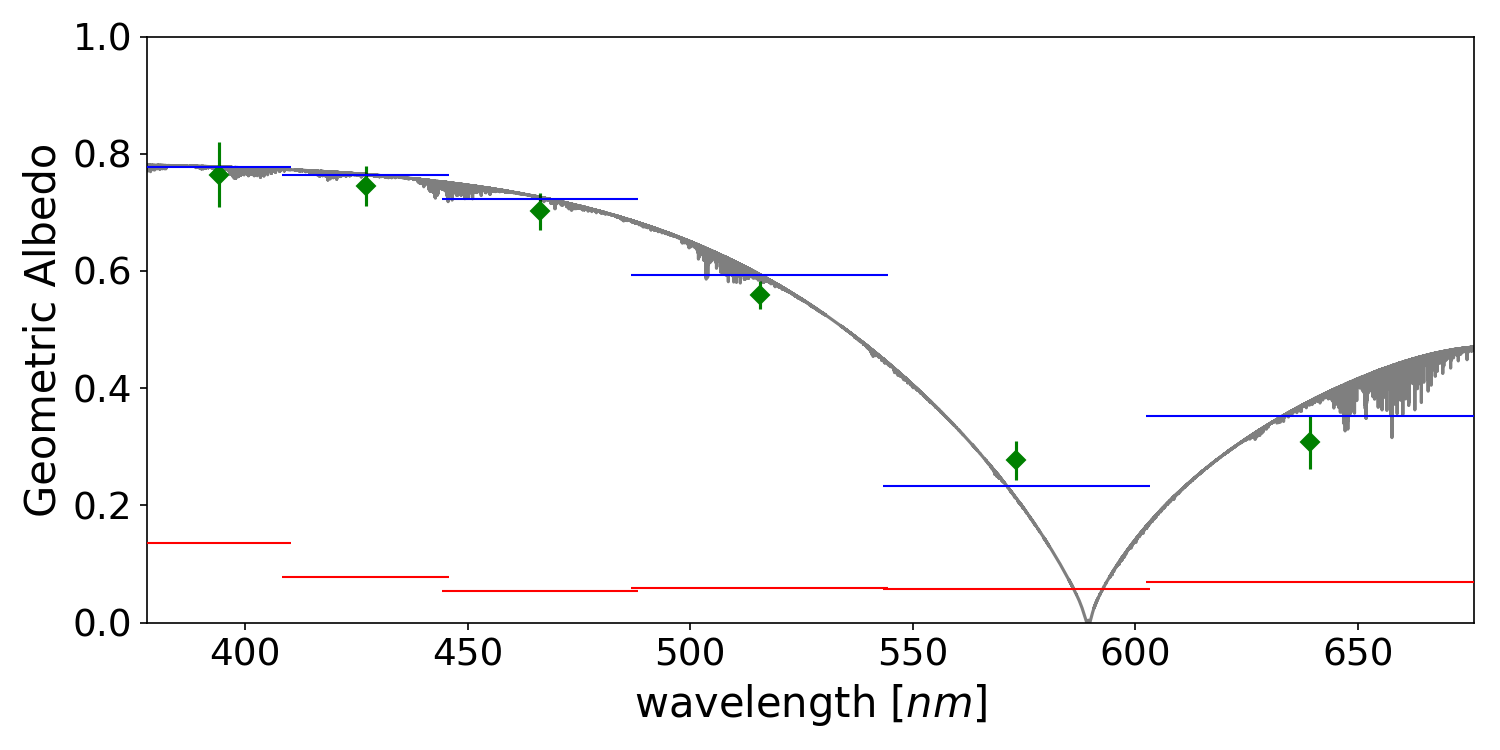}
	\caption{HD 109749 b; Model A ($\times 100$); $\chi^2 = 8.24$}
\end{subfigure}%

\begin{subfigure}[t]{0.5\hsize}
	\includegraphics[width=\hsize]{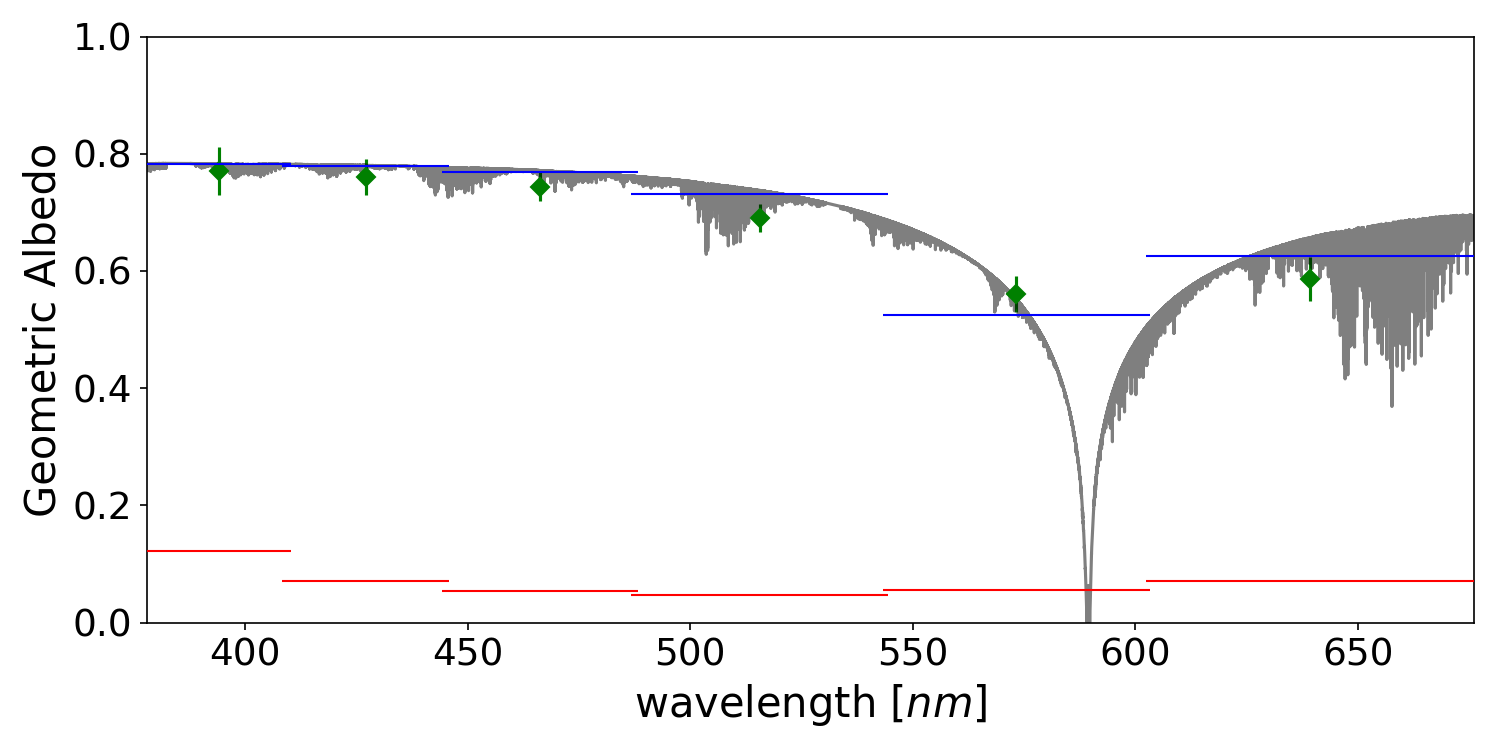}
	\caption{HD 109749 b; Model B ($\times 100$); $\chi^2 = 10.29$}
\end{subfigure}%
\begin{subfigure}[t]{0.5\hsize}
	\includegraphics[width=\hsize]{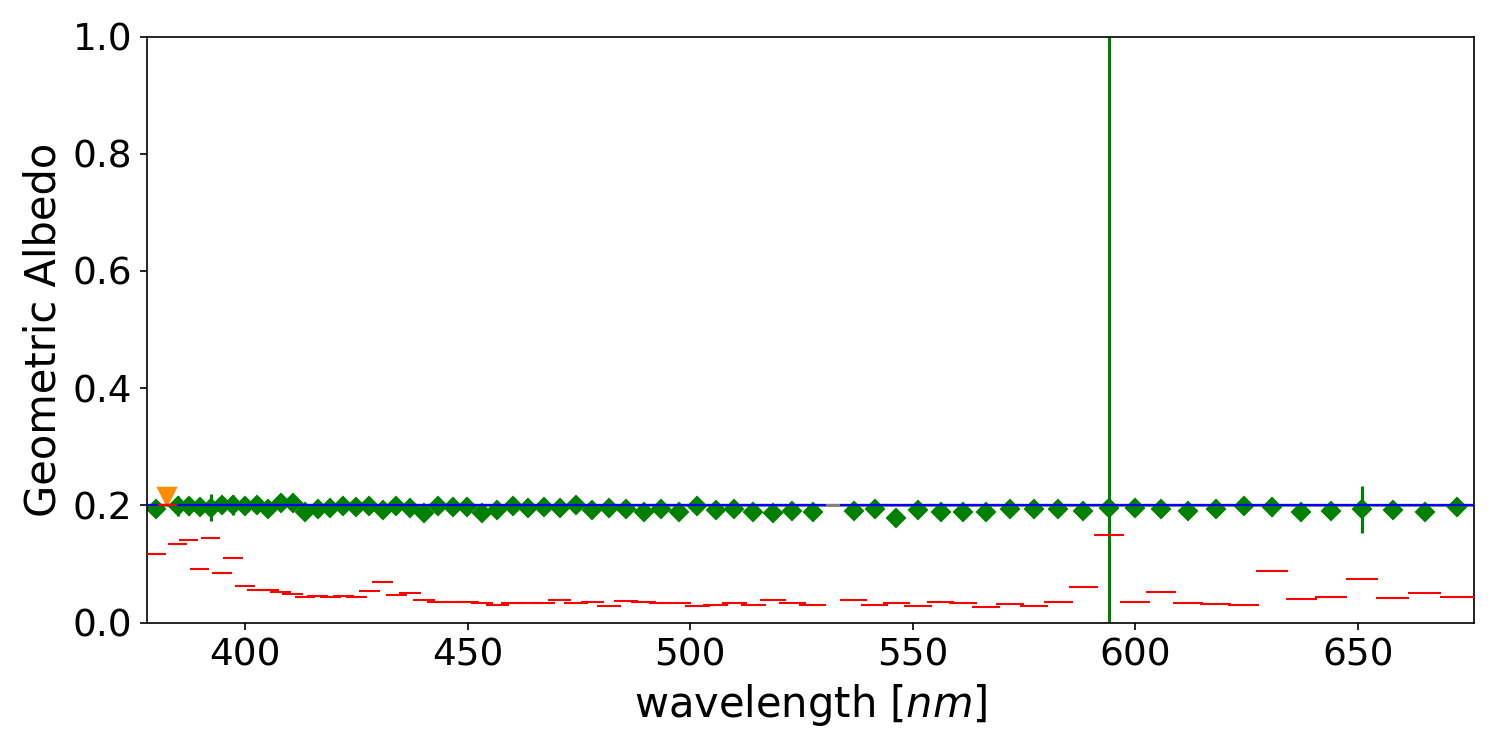}
	\caption{HD 209458 b; $A_g$ = 0.2; $\chi^2 = 2.04$}
\end{subfigure}%
\caption{Distribution of the recovered albedo functions from the simulated HIRES observations. For each wavelength bin: \textit{i)} the green dots represent the mean recovered albedo over the 100 simulated runs with error bars given by the 3 times the standard deviation; \textit{ii)}the blue horizontal lines represent the mean albedo of the simulated model over the bin; \textit{iii)} the red horizontal bar represents the $3\sigma$ detection limit of the albedo. {Note the large error bar on panel h, which results from the low number of spectral lines (3) in the numerical mask for that particular order (order \textit{103} of the HARPS spectrograph, with {590.7-nm <$\lambda$< 597.4-nm}).}}
\label{fig:hires}
\end{figure*}

\begin{figure*}
\ContinuedFloat
\begin{subfigure}[t]{0.5\hsize}
	\includegraphics[width=\hsize]{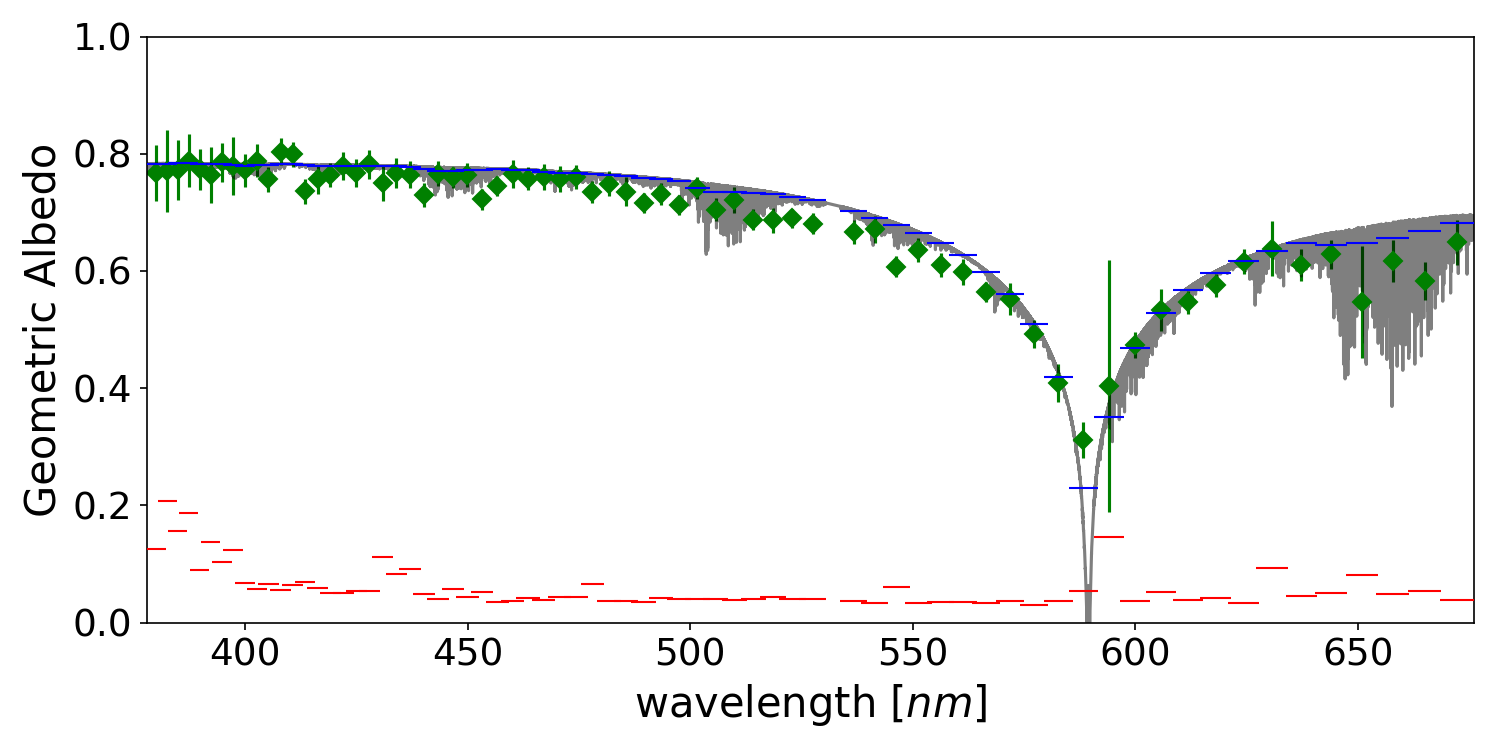}
	\caption{HD 209458 b; Model B ($\times 100$); $\chi^2 = 15.94$}
\end{subfigure}%
\begin{subfigure}[t]{0.5\hsize}
	\includegraphics[width=\hsize]{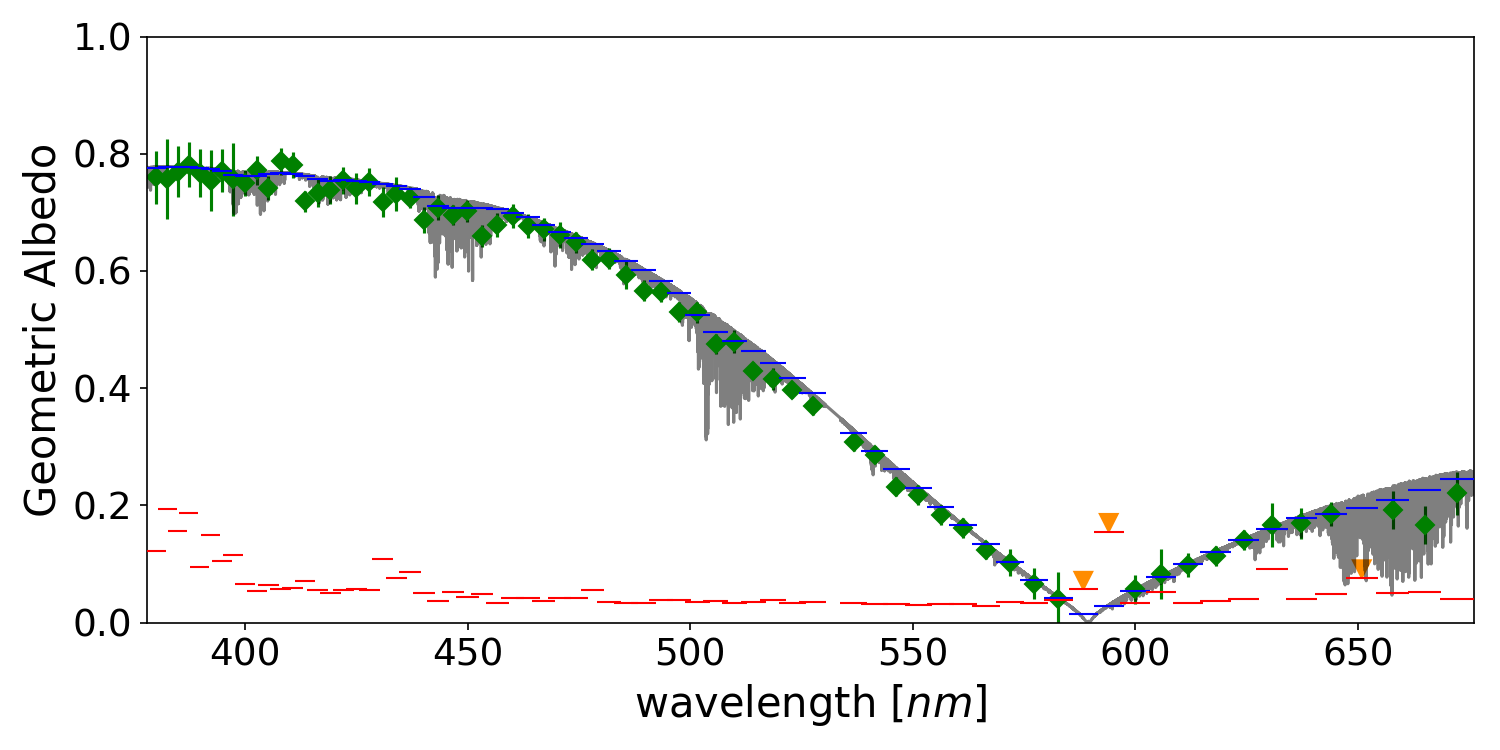}
	\caption{HD 209458 b; Model B ($\times 1$); $\chi^2 = 8.17$}
\end{subfigure}%

\begin{subfigure}[t]{0.5\hsize}
	\includegraphics[width=\hsize]{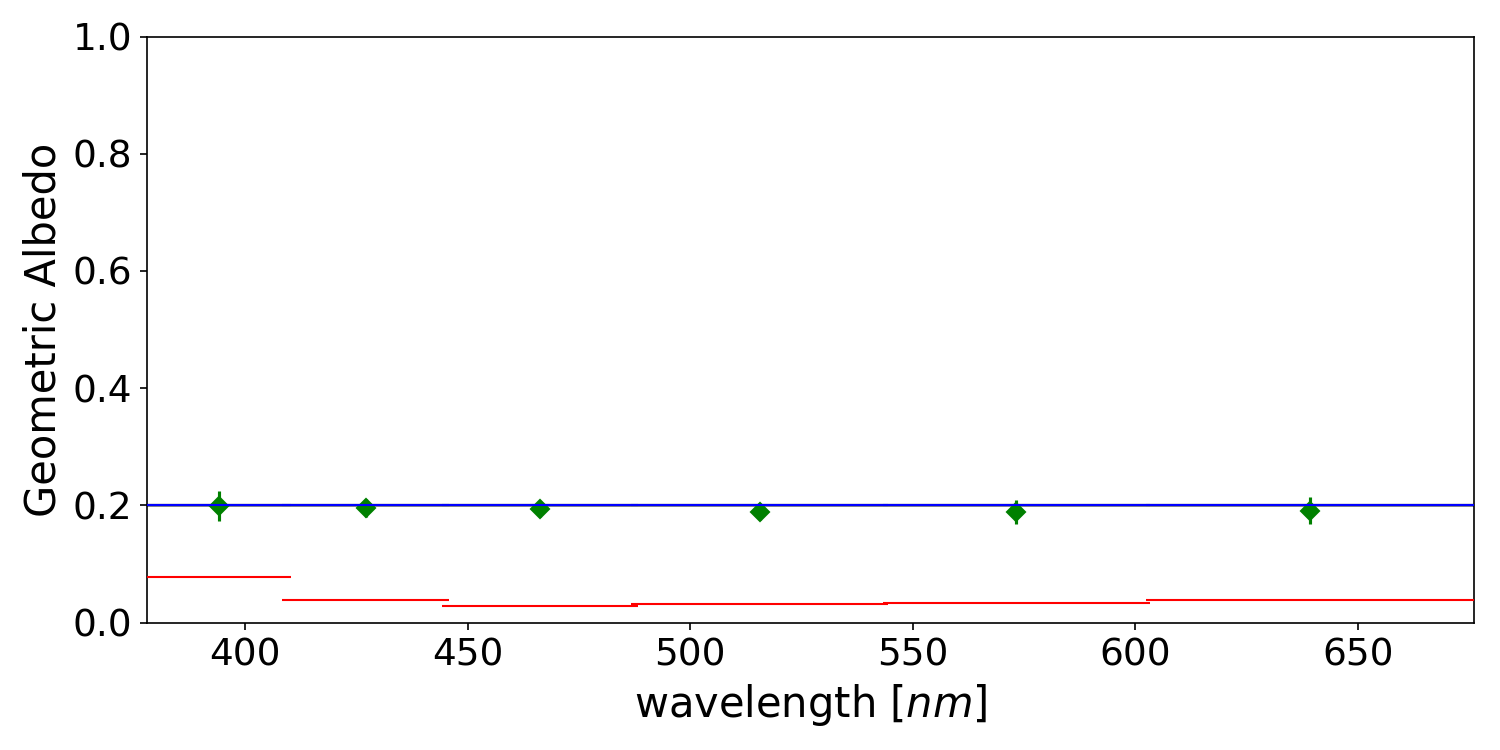}
	\caption{HD 76700 b; $A_g$ = 0.2; $\chi^2 = 2.13$}
\end{subfigure}%
\begin{subfigure}[t]{0.5\hsize}
	\includegraphics[width=\hsize]{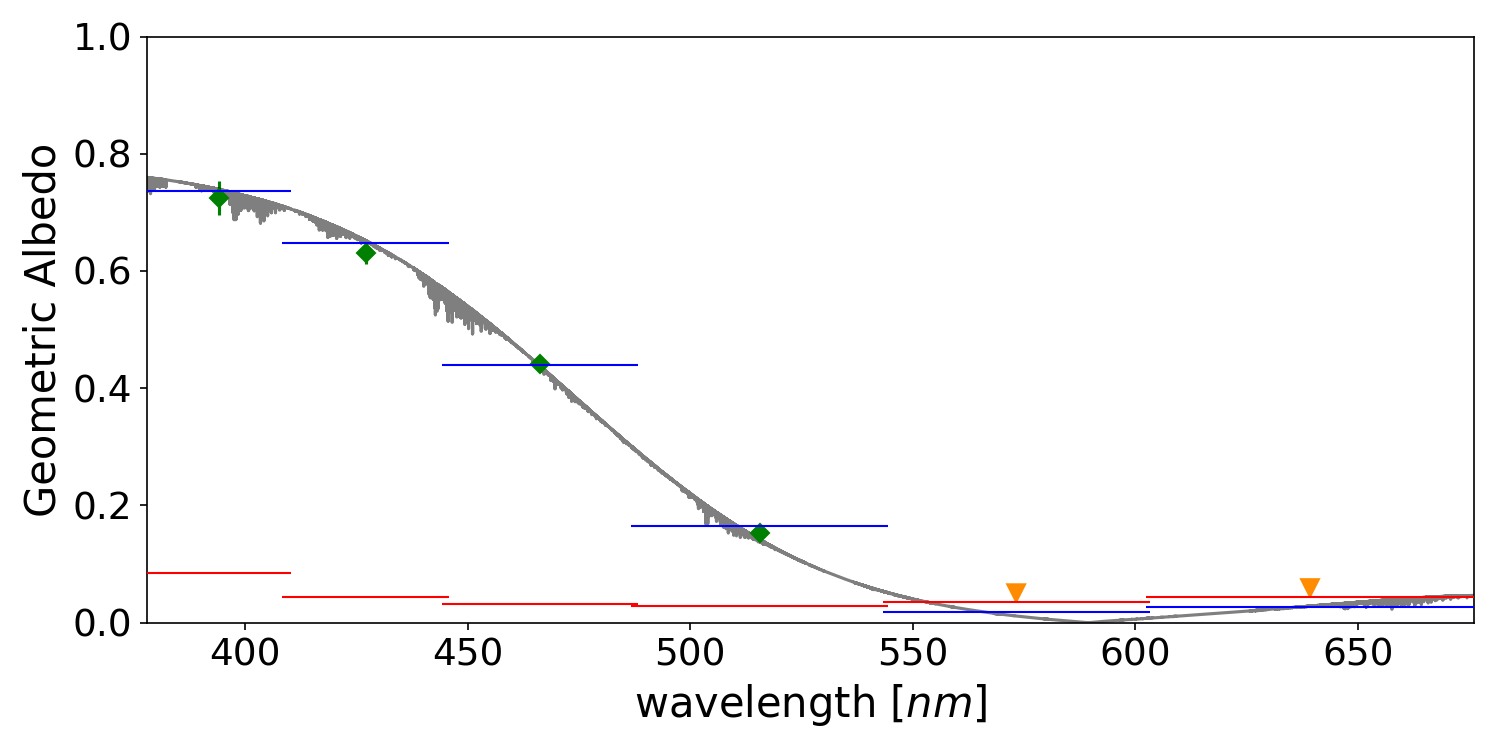}
	\caption{HD 76700 b; Model A ($\times 1$); $\chi^2 = 3.0$}
\end{subfigure}%

\begin{subfigure}[t]{0.5\hsize}
	\includegraphics[width=\hsize]{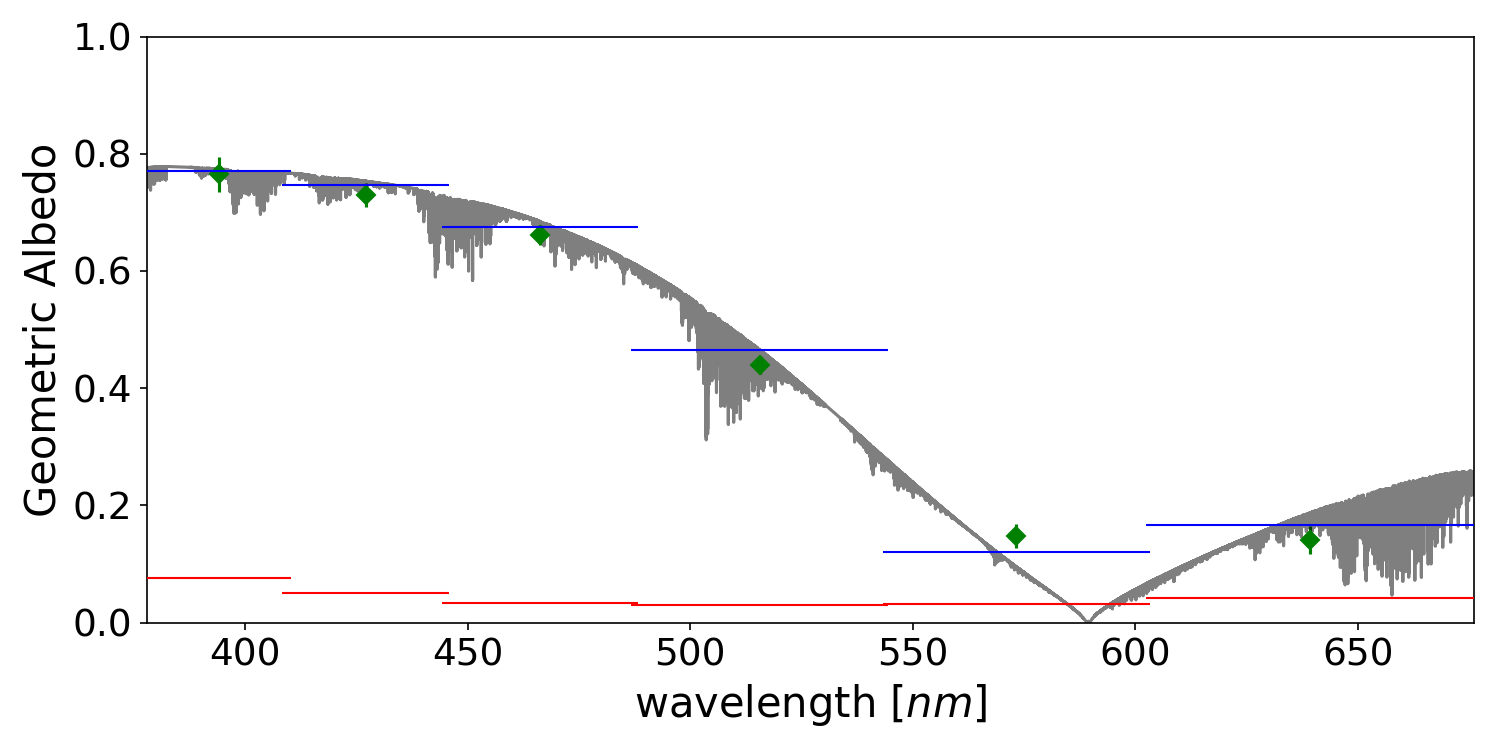}
	\caption{HD 76700 b; Model B ($\times 1$); $\chi^2 = 10.91$}
\end{subfigure}%
\contcaption{Distribution of the recovered albedo functions from the simulated HIRES observations. For each wavelength bin: \textit{i)} the green dots represent the mean recovered albedo over the 100 simulated runs with error bars given by the 3 times the standard deviation; \textit{ii)}the blue horizontal lines represent the mean albedo of the simulated model over the bin; \textit{iii)} the red horizontal bar represents the $3\sigma$ detection limit of the albedo.}

	\end{figure*}

\clearpage

%%%%%%%%%%%%%%%%%%%%%%%%%%%%%%%%%%%%%%%%%%%%%%%%%%



\bibliographystyle{mnras}
\bibliography{mnras_library} % if your bibtex file is called example.bib

%%%%%%%%%%%%%%%%%%%%%%%%%%%%%%%%%%%%%%%%%%%%%%%%%

%%%%%%%%%%%%%%%%% APPENDICES %%%%%%%%%%%%%%%%%%%%%

\appendix

\section{Optimal observation windows and detectability}
\label{app:optimalWindows}
{The reflected signal of the planet is small, and to detect it it is essential to choose orbital phases where that signal is close to its maximum.} As per equation Eq. \ref{eq:fluxRatioLambda}, this occurs at superior conjunction (or opposition in case of a transiting planet). Unfortunately at that point of the orbit, both the planet and star will have matching radial velocities and thus be difficult -- if not impossible -- to separate. Even with the precise stellar template built with the CCF technique, fluctuations on the stellar flux will leave residuals in the radial velocity domain at the stellar radial velocity residuals with amplitudes comparable or superior to the planetary signal. Thus, to maximize detectability of the planetary signal, it is necessary to choose orbital phases where 
\begin{inparaenum}
	\item[i)] the flux from the planet is as close to maximum as possible;
	\item[ii)] the planetary and stellar signals can be separated. 
\end{inparaenum}
In this work we have defined that the minimum distance in the radial velocity domain to be of 20 $\rm km \; s^{-1}$, which corresponds to about 3 times the FWHM of the stellar CCF. This value should be enough to avoid the region where the residuals from the removal of the stellar CCF are larger in amplitude than the planetary signal. 

We define detectability of the reflected signal from the planet as the S/N of the recovered planetary CCF over the spectral range of the selected instrument:
\begin{equation}
D\left(\phi\right) = \frac{A_{\rm{CCF}}}{\sigma_{\rm{noise}}}
\end{equation}
where $A_{\rm{CCF}}$ is the ratio of the amplitudes of the recovered planetary CCF by the stellar one (over the whole spectral range of the instrument) and $\sigma_{\rm{noise}}$ the noise of the final CCF. 

We computed the detectability of the planetary CCF from 51 Pegasi b after 10h worth of planetary observations with ESPRESSO in HR mode. Please note that although we are focusing on a single star+planet+instrument setup, the conclusions will be valid for all the simulations in this article. The top panel of figure \ref{fig:Detectability} represents the expected detectability as a function of the orbital phase, starting at inferior conjunction. We estimated $A_{\rm{CCF}}$ from Eq. \ref{eq:fluxRatioLambda} using the planetary parameters defined in Table \ref{tab:paramsPlanets} and assuming a Lambert phase function. The radial velocity of both the planet and star were also computed using planetary parameters defined in Table \ref{tab:paramsPlanets}. $\sigma_{\rm{noise}}$ was computed from the HARPS ETC for a 300s exposure with the stellar parameters defined in Table \ref{tab:paramsPlanets} and settings from Table \ref*{tab:paramsObservations} and extrapolated for 10h worth of observations, the telescope diameter and ESPRESSO's increase in sensitivity. The bottom panel of figure \ref{fig:Detectability} represents the expected detectability as a function of the distance between the planet and star in the radial velocity domain. The arrows represent the movement of the planetary signal along its orbit and located at orbital phases at 15 degrees intervals starting at inferior conjunction {($\phi = 0$)}. 

We have defined our optimal windows of observation as the orbital phases between 15 and 45 degrees on each side of superior conjunction {(i.e., orbital phases with $0.37 \lessapprox \phi \lessapprox 0.46$ or $0.54 \lessapprox \phi \lessapprox 0.63$, assuming that the inferior conjunction/transit phase corresponds to $\phi=0$ and one full orbit corresponds to $\Delta \phi = 1$)}. {These can be seen as the green areas in Figure \ref{fig:Detectability}}. In the case of 51 Pegasi b, this corresponds to orbital phases where the detectability of the planet varies between 75 per cent and 95 per cent of the maximum detectability (at superior conjunction). Please note that this is an ideal case where we assumed all observations to be at the same phase, hardly the case for real observation. However, these can be scheduled in to occur in the optimal windows defined above and as such the detectability should always be above 75 per cent of the maximum detectability. The red area represents the radial velocity range into which we assume the planetary and stellar CCF cannot be separated. 

\begin{figure}
	\centering
	\includegraphics[width=\hsize]{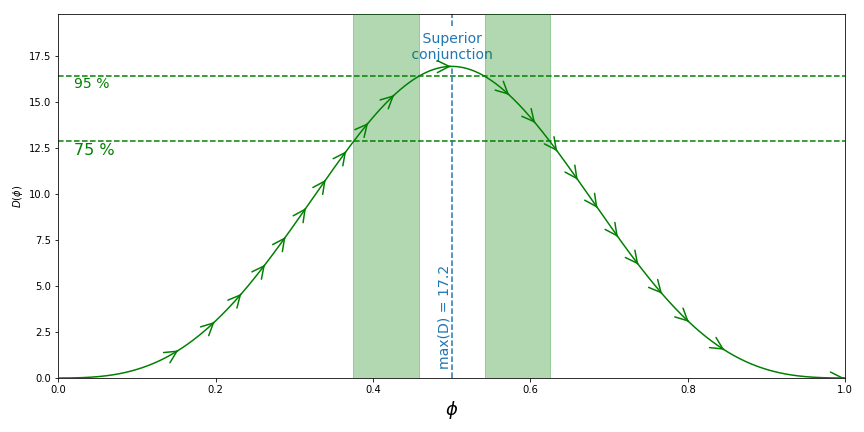}
	\includegraphics[width=\hsize]{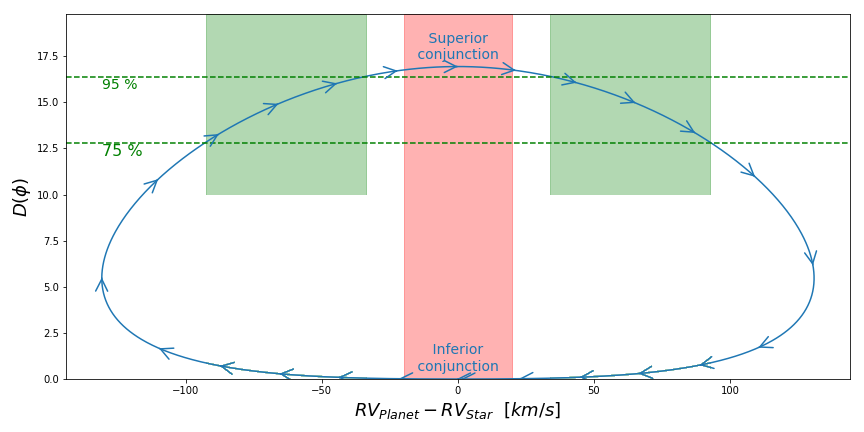}
	\caption{Detectability of the planetary signal from the 51 Pegasi system for 10h worth of observation of as a function: i) \textit{[Top]} of average orbital phase i) \textit{[Bottom]} the distance between the planet and its host in the radial velocity domain.}
	\label{fig:Detectability}
\end{figure}

%%%%%%%%%%%%%%%%%%%%%%%%%%%%%%%%%%%%%%%%%%%%%%%%%%

% Don't change these lines
\bsp	% typesetting comment
\label{lastpage}
\end{document}